\documentclass[11pt]{article}

\pdfoutput=1 
\usepackage{jheppub}

\usepackage{amsmath,amssymb,amsfonts}
\usepackage{hyperref}
\hypersetup{colorlinks,citecolor=blue,urlcolor=blue,linkcolor=blue}
\usepackage{graphicx}
\usepackage{xcolor}
\usepackage{enumitem}
\usepackage{caption}
\usepackage{subcaption}
\usepackage{comment}

\usepackage{nicefrac}
\usepackage{mathrsfs}
\usepackage{mdframed}
\usepackage{ulem}
\usepackage{dsfont}
\usepackage{units}
\usepackage[english]{babel}
\usepackage{float} 
\usepackage{braket}
\def\beq{\begin{equation}\begin{aligned}}
\def\eeq{\end{aligned}\end{equation}}

\begin{document}

\title{A Nonperturbative Toolkit for Quantum Gravity}
\author[1,2,3]{Vijay Balasubramanian}
\author[4]{Tom Yildirim}
\affiliation[1]{Department of Physics and Astronomy, University of Pennsylvania, Philadelphia, PA 19104, U.S.A.}
\affiliation[2]{Theoretische Natuurkunde, Vrije Universiteit Brussel and International Solvay Institutes, Pleinlaan 2, B-1050 Brussels, Belgium}
\affiliation[3]{Rudolf Peierls Centre for Theoretical Physics, University of Oxford,Oxford OX1 3PU, U.K.}
\affiliation[4]{Department of Physics, Keble Road, University of Oxford, Oxford, OX1 3RH, UK}
\emailAdd{vijay@physics.upenn.edu}
\emailAdd{tom.yildirim@physics.ox.ac.uk}

\abstract{We propose a method for demonstrating equivalences beyond the saddlepoint approximation between quantities in quantum gravity that are defined by the Euclidean path integral, without assumptions about holographic duality. The method involves three ingredients: (1) a way of resolving the identity with an overcomplete basis of microstates that is under semiclassical control, (2) a drastic simplification of the sum over topologies in the limit where the basis is infinitely overcomplete, and (3) a way of cutting and splicing geometries  to demonstrate equality between two different gravitational path integrals even if neither can be explicitly computed. We illustrate our methods by giving a general argument that the thermal partition function of quantum gravity with two boundaries factorises.   One implication of our results is that universes containing a horizon can sometimes be understood as superpositions of horizonless geometries entangled with a closed universe.}

\maketitle

\section{Introduction}

The Euclidean gravitational path integral has played a central tool in studies of  quantum gravity. For example, non-perturbative wormholes  in the path integral helped to explain how the entropy of Hawking radiation, ostensibly rising forever, eventually declines to zero consistently with unitarity \cite{Penington:2019kki,Almheiri:2019qdq}.  Likewise,
\cite{Balasubramanian:2022gmo,Balasubramanian:2022lnw,Climent:2024trz,Balasubramanian:2024yxk} constructed an infinite family of black hole microstates under good semiclassical control and showed that their overlaps receive nonpertubative corrections from wormholes. These overlaps led to a Hilbert space of precisely the right dimension to account for the  Beckenstein-Hawking entropy.

Strictly speaking, the Euclidean gravitational path integral, understood as a sum over topologies and metrics with a defined boundary condition and weighted by the exponential of the Einstein-Hilbert action, is not well defined by itself.  We will understand it as an effective description below some cutoff, completed at high energies by some more refined description,
perhaps string theory. However, the gravitational path integral does not seem to define a standard effective field theory. For example, some amplitudes that it computes vanish, even though the square of these amplitudes do not due to non-perturbative wormhole contributions.  This suggests an intepretation that the gravitational path integral computes coarse-grained information regarding some underlying fine-grained theory; see, for example, \cite{Saad:2019lba,Saad:2019pqd,Sasieta:2022ksu,Marolf:2020xie,deBoer:2023vsm,deBoer:2024mqg}.  Even understood this way, in many settings it is impossible to calculate explicitly, and we resort to approximations like summing over saddlepoints (extrema of the action) of the dominant topologies. Here, we propose a computational toolkit that allows us to extract exact relations in the underlying fine-grained theory despite 
 the coarse-gaining and the difficulty of doing explicit calculations.

The toolkit includes three components building on results in \cite{Iliesiu:2024cnh,Balasubramanian:2024yxk}. First, we propose a general way of resolving the identity in the quantum gravity Hilbert space with an overcomplete basis of microstates under semiclassical control.  Second, we show that inserting this resolution of the identity into the path integral drastically simplifies the sum over topologies in a limit where the basis is infinitely overcomplete. Third, in this limit, we develop a  method for conducting surgery on  geometries contributing to the path integral that allows us to demonstrate equality between gravitational path integrals for different quantities, even if neither can be calculated explicitly. As an application we demonstrate that in quantum gravity with two asymptotically locally AdS or flat boundaries the thermal partition function factorizes, even though the sum over geometries includes bulk wormholes.  This in turn implies that the trace of any function of the Hamiltonian factorizes.   Our computations make no assumptions about holographic duality. Along the way we show that states like the ones used in \cite{Balasubramanian:2022gmo,Balasubramanian:2022lnw,Climent:2024trz,Balasubramanian:2024yxk} to account for the entropy of black holes within different microcanonical windows actually span the complete gravity Hilbert space. 
As a consequence, a state corresponding to a black hole geometry can be written as a superposition of states that do not contain horizons, but are entangled with a compact universe.

Six sections follow.  In Sec.~\ref{sec:QTFTPI} we introduce some general features of the gravitational path integral.  In Secs.~\ref{sec:tool1}--\ref{sec:tool3} we develop our toolkit.  In Sec.~\ref{sec:Trzz} we apply the toolkit to factorization of the thermal partition function in AdS gravity with two boundaries.  We conclude in Sec.~\ref{sec:discussion} by discussing some implications of our formalism.

\section{The gravitational path integral} \label{sec:QTFTPI}

\begin{figure}[h]
    \centering
    \includegraphics[width=0.5\linewidth]{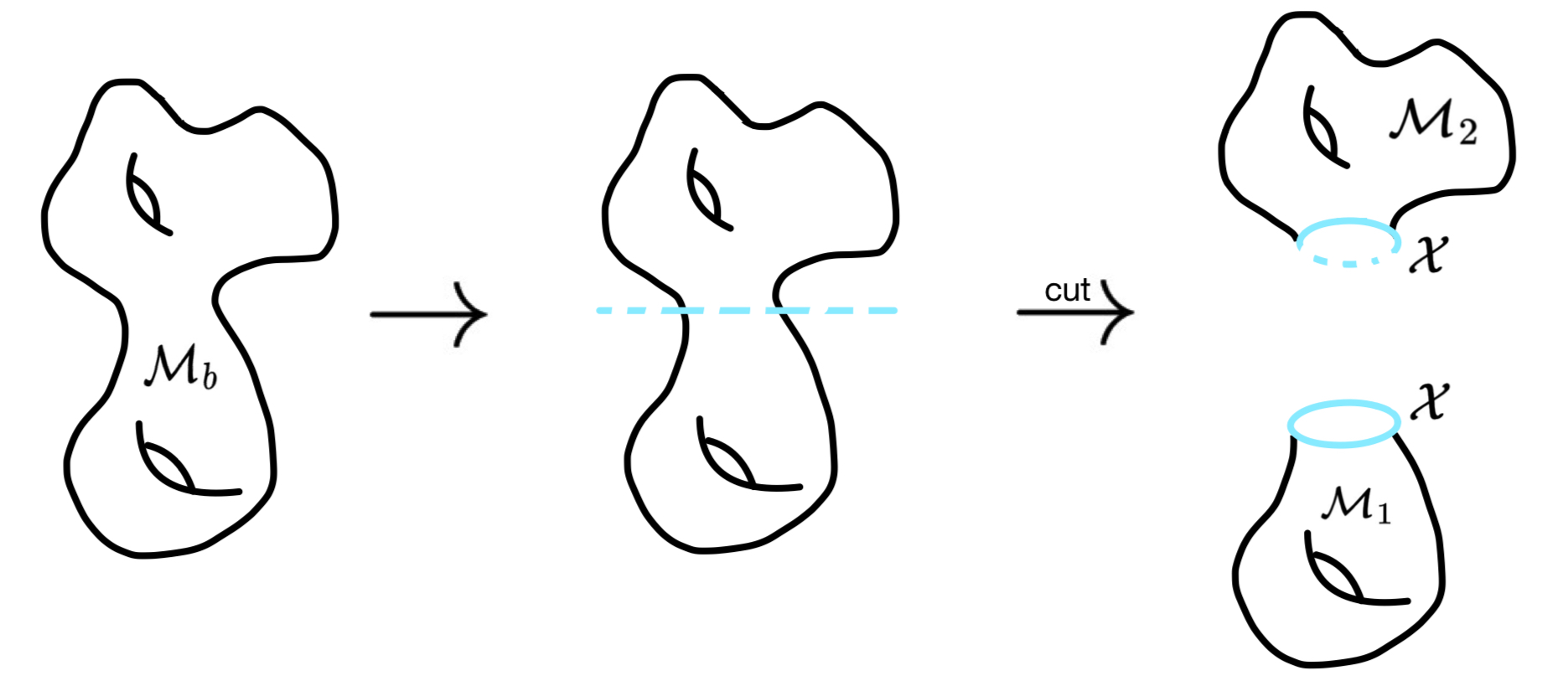}
    \caption{Cartoon of cutting a gravity path integral boundary condition to define a state.}
    \label{fig:cutpathint}
\end{figure}

The gravitational path integral is a map $\zeta[\Sigma_b] \rightarrow  \mathbb{C}$ from a set of boundary conditions for spacetime -- a boundary topology, metric, and matter sources, collectively denoted by $\Sigma_b$ -- to the complex numbers.  We are going to analyze the Euclidean gravitational path integral, and focus for concreteness on asymptotically local anti-de Sitter (AlAdS) gravity, although essentially identical analyses follow in the asymptotically flat case.  To calculate the path integral we must sum over all topologies and geometries ($g$), modulo diffeomorphisms, and matter fields ($\phi$), that are compatible with the asymptotic boundary conditions:
\begin{align}
\zeta[\Sigma_b] &= \int_{g,\phi \to \Sigma_b} \mathcal{D}g\mathcal \, \mathcal{D}\phi \, e^{-I_{bulk}[g,\phi]} \nonumber
\\
I_{bulk} &= -\frac{1}{16\pi G_{N}}\int_{\mathcal{M}} \sqrt{g}( R -2\Lambda + \mathcal{L}_{matter})  -\frac{1}{8\pi G_{N}}\int_{\partial\mathcal{M}= \mathcal{M}_b} \sqrt{h}K 
\label{eq:gravoverlap} 
\end{align}
where $\mathcal{M}$ is a bulk manifold filling in the boundary. We have used the Einstein-Hilbert action above, but could have instead used any self-consistent effective theory in any dimension, such as two-dimensional Jackiw-Teitelboim (JT) gravity.
We understand this path integral as an effective description below some cutoff, completed at high energies by some ultraviolet finite theory. We can  construct states of the theory by partitioning the  boundary  $\mathcal{M}_b =\mathcal{M}_1 \cup \mathcal{M}_2 $  so that $\partial\mathcal{M}_1= \partial\mathcal{M}_2=\mathcal{X}$ (see Fig.~\ref{fig:cutpathint}). The path integral then produces a functional of the fields and geometry on the cut $\mathcal{X}$, and formally defines states  $|\mathcal{M}_{1,2}\rangle \in \mathcal{H}_{\mathcal{X}}$ in a Hilbert space $\mathcal{H}_{\mathcal{X}}$ such that $\langle \mathcal{M}_{2}|\mathcal{M}_{1}\rangle=\zeta[\Sigma_b]$. In general, the path integral cannot be performed exactly, and so we resort to saddlepoint approximations to evaluate overlaps of states constructed in this way. Note that defining a state by partitioning boundary data is a standard way of producing states in the \textit{bulk} quantum gravitational Hilbert space. For instance, such methods were used in \cite{Iliesiu:2024cnh} and \cite{Marolf:2020xie}  to construct  states in JT gravity and a topological toy model respectively, and in this work we will apply this method in higher dimensional Einstein-Hilbert gravity. Crucially, we do not assume any kind of dual boundary CFT construction.  It is clear our states live in the bulk Hilbert space as we can, in principle, compute the wavefunctionals $\Psi_{\mathcal{M}_{1}}(h,\phi) \equiv\braket{h,\phi|\mathcal{M}_{1}}$ between our states and states defined by some bulk time-slice  data $h_{\mu \nu},\phi$ by performing the path integral subject to these joint boundary conditions.

For a given cut $\mathcal{X}$ one option for producing different states is to fix the boundary manifold $\mathcal{M}_b$ and to create states $\ket{i}$ by inserting matter operators $\mathcal{O}_i$ on $\mathcal{M}_{b}$ (see, e.g., \cite{Marolf:2017kvq,Marolf:2020xie,Colafranceschi:2023moh}). The overlap of such states, $\braket{i|j}$ is computed by evaluating the path integral with $\mathcal{O}_{i,j}$ inserted on $\mathcal{M}_{1,2}$.  Likewise we can calculate the square of the overlap $\braket{i|j}\braket{j|i}$ by inserting $\mathcal{O}_{i,j}$ on two copies of the boundary manifold and summing over all Euclidean bulk geometries and topologies that fill in these boundary conditions.  Recent work has shown that path integral for square of the overlap can have additional contributions beyond a product of the contributions to $\braket{i|j}$ -- there can be connected topologies, i.e., wormholes, that interpolate between the two boundaries.  These non-perturbative contributions, which are akin to instantons in quantum field theory, can produce an apparently paradoxical result where 
\begin{equation}
\braket{i|j}= \delta_{ij} ~~~~\mathrm{while}~~~
\braket{i|j}\braket{j|i}= \delta^2_{ij} + Z_{WH} \, .
\label{eq:coarseinner}
\end{equation}
where $Z_{WH}$ is the wormhole contribution, at least in the semiclassical saddlepoint approximation of the gravitational path integral.
Following 
\cite{Saad:2019lba,Saad:2019pqd,Sasieta:2022ksu,Marolf:2020xie,deBoer:2023vsm,deBoer:2024mqg,Penington:2019kki,Almheiri:2019qdq,Balasubramanian:2022gmo,Balasubramanian:2022lnw,Climent:2024trz,Balasubramanian:2024yxk,Marolf:2024jze} we interpret this situation as arising because the gravitational path integral, treated as an effective theory below some cutoff, and certainly after taking a saddlepoint approximation, calculates coarse-grained observables that compute some kind of average over corresponding quantities in an underlying more fundamental theory that we refer to as the fine-grained theory. To reflect this interpretation we will write an overbar over quantities computed using the gravitational path integral, for example,
\begin{equation}
\overline{\braket{i_1|j_i}\braket{i_2|j_2} \cdots \braket{i_n|j_n}}  \, .
\end{equation}
Thus, it can transpire that $\overline{\braket{i|j}}$ vanishes because of cancellation of random phases in the coarse-grained computation of overlaps of  states in the fine-grained theory, while these phases cancel in the magnitude squared $\overline{\braket{i|j}\braket{j|i}}$.  If so, the square root of the latter quantity quantifies the actual magnitude of the overlap in the fine-grained theory.

\begin{figure}[h]
    \begin{subfigure}[b]{0.2\linewidth}
        \centering
        \includegraphics[width=\linewidth]{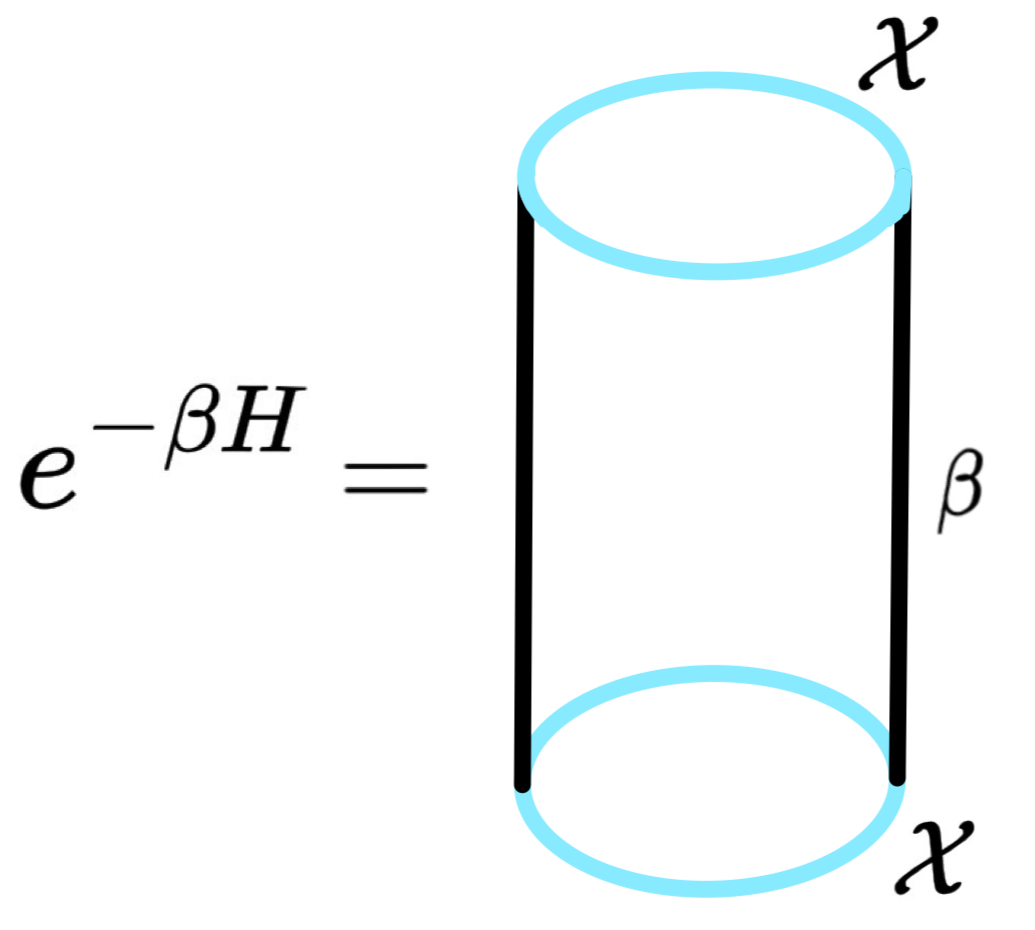}
      \caption{}
        \label{fig:hamiltonian}
    \end{subfigure}
    \begin{subfigure}[b]{0.2\linewidth}
        \centering
        \includegraphics[width=\linewidth]{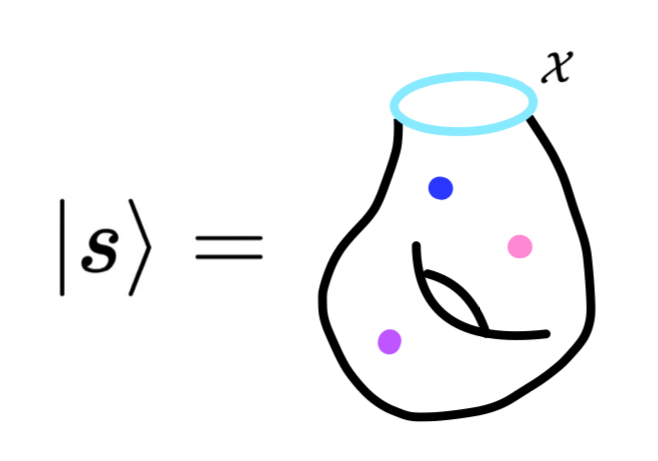}
        \caption{}
        \label{fig:state}
    \end{subfigure}
    \hfill
    \begin{subfigure}[b]{0.2\linewidth}
        \centering
        \includegraphics[width=\linewidth]{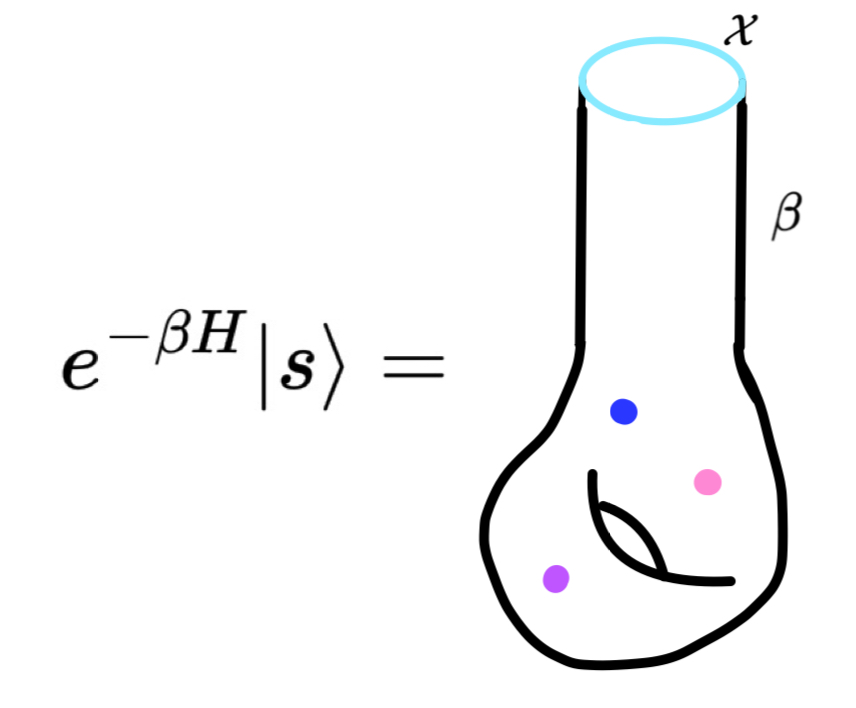}
        \caption{}
        \label{fig:timeEstate}
     \end{subfigure}
     \hfill
     \begin{subfigure}[b]{0.2\linewidth}
        \centering
        \includegraphics[width=\linewidth]{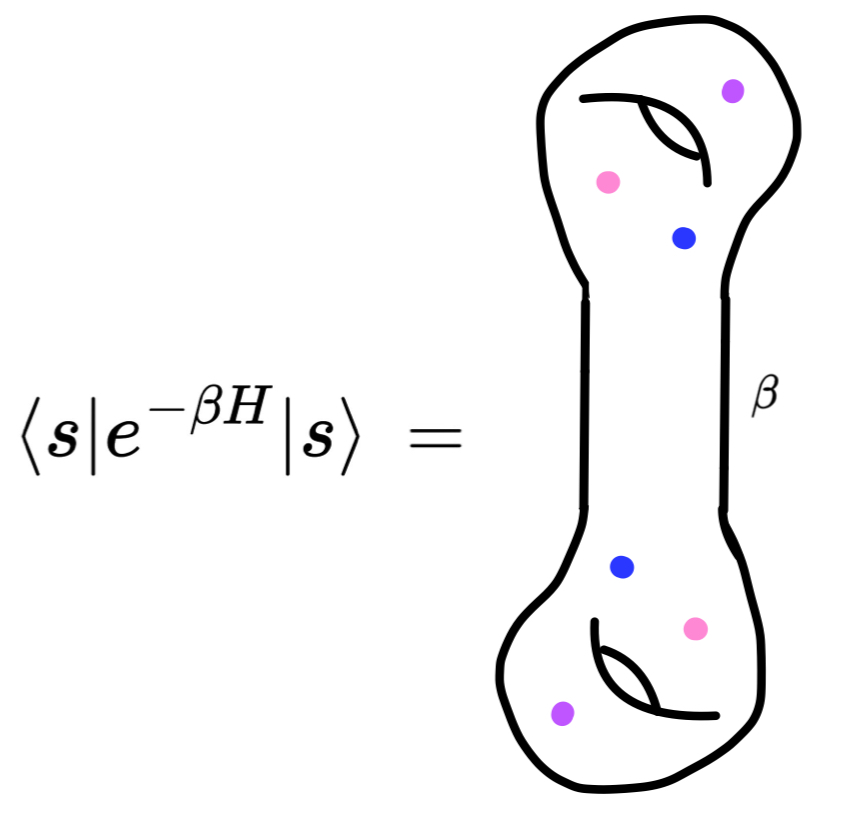}
        \caption{}
        \label{fig:timeEoverlap}
     \end{subfigure}
    \caption{Action of $e^{-\beta H}$ on states.  ({\bf a}) Action of the Hamiltonian. ({\bf b}) Example state. ({\bf c}) Action of the Hamiltionian on a state. ({\bf d}) Survival amplitude after Euclidean time evolution. }
\end{figure}

\paragraph{Action of the Hamiltonian} \label{hamiltonian}
Next, we define the action of the Hamiltonian on our states. The action of $e^{-\beta H}$ is to glue an asymptotic boundary ``cylinder"  $\mathcal{X} \times \mathbb{I_{\beta}}$ to the cut (Figs.~\ref{fig:hamiltonian},  \ref{fig:state} and \ref{fig:timeEstate}). In other words, it lengthens the cut open boundary by a Euclidean boundary time $\beta$. This is motivated by ADM formalism for Einstein gravity, where the bulk Hamiltonian generates boundary time  evolution. The so-called survival amplitude $\langle\mathbf{s} | e^{-\beta H}|\mathbf{s}\rangle$  is therefore constructed by sewing the $ |\mathbf{s} \rangle$ bra and ket together via a boundary cylinder of length $\beta$ resulting in the closed boundary manifold $\mathcal{M}_{\mathbf{s}}(\beta)$ (Fig.~\ref{fig:timeEoverlap}). 
The Hamiltonian can alternatively viewed as creating a state on two copies of $\mathcal{X}$ -- pictorially, we imagine bending the cylinder in Fig.~\ref{fig:hamiltonian} around into a half-torus.    We denote the state constructed in this manner by the operator $e^{-\frac{\beta H}{2}}$ as $|\beta\rangle$ (note the $1/2$ in the exponent).  
In QFT, a similar half-torus path integral prepares the Thermofield Double (TFD) state  $|TFD(\beta)\rangle = \sum_{n} e^{\frac{-\beta E_{n}}{2}} |E_n\rangle^* _{L}|E_n\rangle_R$ defined on two copies of the theory. 
In what follows we refer to the norm  
\begin{equation}
\braket{\beta|\beta}\equiv Z(\beta)
\label{eq:Zpartitionfunction}
\end{equation}
as the ``Z partition function". We introduce this terminology as it is not clear from the gravitational Hilbert space perspective presented here why this path integral should compute the thermal partition sum of the theory. In \cite{Balasubramanian:2025hns} we use the toolkit developed here to show that this is indeed the case, once non-perturbative wormhole corrections to the thermal trace are taken into account.

\paragraph{Traces and the identity}Consider a set of $\kappa$ states $\{|i \rangle\} $ constructed as described above by cutting open the path integral with inserted operators. The span of these states, $\mathcal{H}_{\kappa} \equiv span\{ |i\rangle , i= i,2, ... \kappa \}$ is a subspace of the full Hilbert space. As we discussed, even if these states are orthogonal in the low-energy effective theory, they can have overlaps in the fine-grained theory, arising from nonperturbative effects. We characterize these overlaps by the Gram matrix $G_{ij} \equiv \braket{i|j}$.  Because of these overlaps the dimension of $\mathcal{H}_{\kappa}$ can be less than $\kappa$, the number of states in $\{|i \rangle\} $, i.e., the latter can be overcomplete as a basis for $\mathcal{H}_{\kappa}$.   The Gram matrix can be used to construct an orthonormal basis $\{|v_\gamma\rangle\}$ from an overcomplete one:
\beq \label{Shellbasis}
|v_\gamma\rangle= G^{-1/2}_{ji}U_{i\gamma}|j\rangle
\eeq
where $U$ is a unitary transformation that diagonalizes $G$ ($G=UDU^{\dagger}$ with $D$ diagonal) (see Appendix~\ref{LA} and Appendix~A of \cite{Balasubramanian:2024yxk} for details).    Here and in what follows we will use the convention that repeated indices are summed. 
Following \cite{Boruch:2024kvv,Balasubramanian:2024yxk} the inverse of the Gram matrix is defined via analytic continuation of $G^{n}_{ij}$ from positive integers $n$ such that 
\beq \label{eq:shellTr}
 G^{-1}_{ij} \equiv \lim_{n\to -1} G^{n}_{ij}.
\eeq
This generalized inverse is well-defined as it inverts the nonzero eigenvalues of $G$ and leaves the zero eigenvalues untouched (see Appendix \ref{LA}).  

The matrix elements of operators in the orthonormal basis can be expressed as
\beq \label{reconstruct}
\bra{v_\gamma}O\ket{v_\xi} = G^{-1/2}_{ki}G^{-1/2}_{jl}U^{-1}_{\gamma k} U_{l \xi} \langle i|O|j\rangle.
\eeq
So the trace over $\mathcal{H}_{\kappa}$ is independent of $U$,
\beq \label{eq:trace}
Tr_{\mathcal{H}_{\kappa}}(O) = G^{-1}_{ij}\langle j|O|i\rangle \, .
\eeq
Thus, even if $\{\ket{i}\}$ is overcomplete as a basis for the span, we can use these states and the associated Gram matrix to compute traces of operators.  Another useful quantity is the resolution of the identity on $\mathcal{H}_{\kappa}$ 
\beq \label{eq:id}
\mathds{1}_{\mathcal{H}_{\kappa}} = \sum_{\gamma}|v_\gamma\rangle\langle v_\gamma| = G^{-1}_{ij} |i\rangle\langle j|
\eeq
in terms of which the dimension of $\mathcal{H}_{\kappa}$ is
\beq
\dim(\mathcal{H}_{\kappa}) = Tr_{\mathcal{H}_{\kappa}}(\mathds{1})= G^{-1}_{ij} \braket{j|i} = 
\lim_{n\to -1} G^n_{ij} \braket{j|i} = \lim_{n\to -1}  \mathrm{Tr}(G^{n+1})
\label{eq:dimHk}
\eeq
where used the definition   (\ref{eq:shellTr}).
Eq.~\ref{eq:dimHk} provides an expression for the dimension of  the $\mathcal{H}_{\kappa}$ subspace of the fine-grained theory. In practice we will use the gravitational path integral to calculate the dimension as 
\begin{equation}\label{eq:tracedimcount}
\overline{\dim(\mathcal{H}_{\kappa})} =\lim_{n\to -1}  \overline{\mathrm{Tr}(G^{n+1})} \, .
\end{equation}

Below we will show that certain classes of  states constructed in this way by the Euclidean path integral can form a basis of the complete Hilbert space of the theory.

\begin{figure}[H]
   \centering
     \begin{subfigure}{0.6\linewidth}
      \centering
     \includegraphics[width=\linewidth]{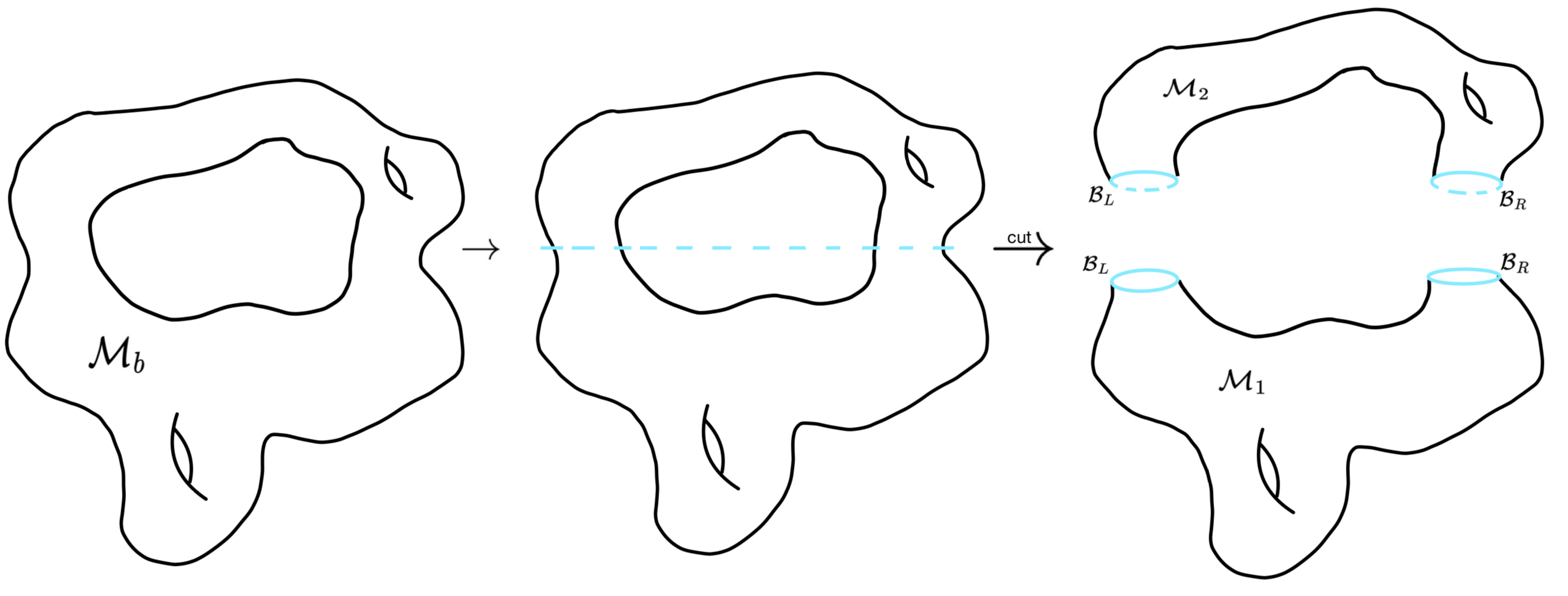}
     \caption{Construction of an example state in $\mathcal{H}_{LR}$.}
     \label{fig:singleBcut}
     \end{subfigure}
    \begin{subfigure}{0.7\linewidth}
    \centering
    \includegraphics[width=\linewidth]{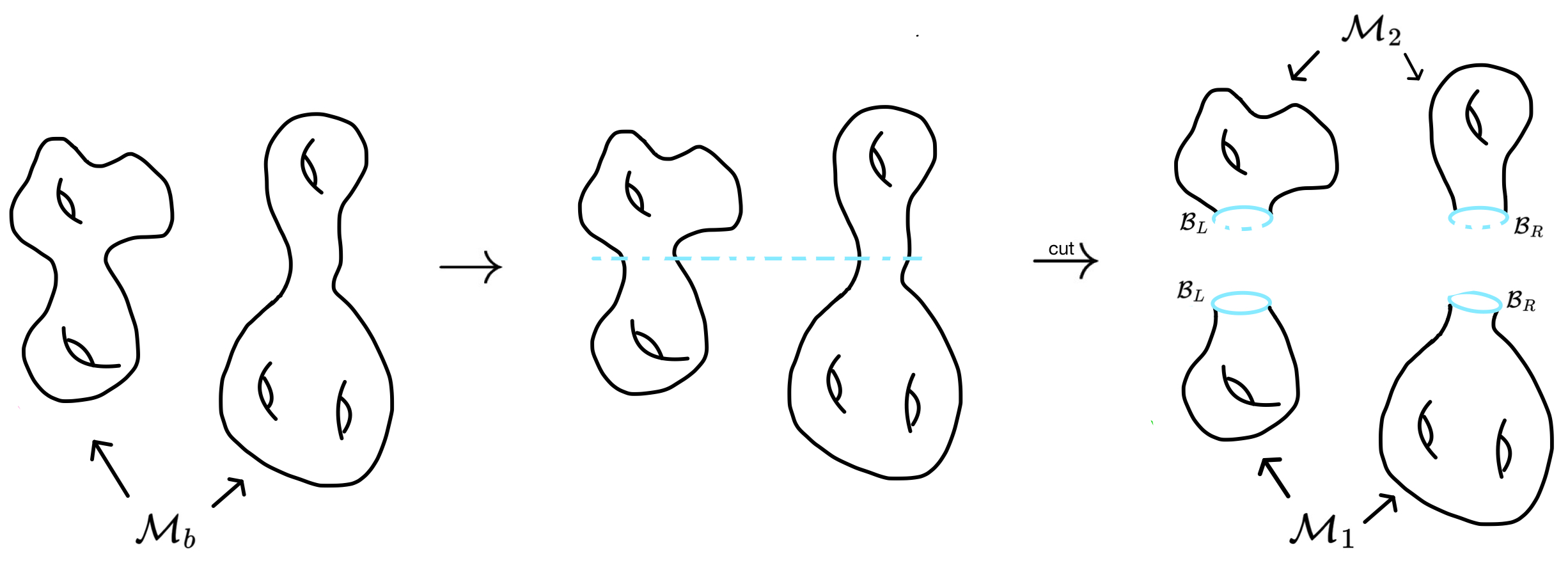}
    \caption{Construction of an example state in $\mathcal{H}_{\mathcal{B}_L}\otimes \mathcal{H}_{\mathcal{B}_R}$.}      \label{fig:tensorcut}
 \end{subfigure}
 \caption{Two distinct ways of preparing states in $\mathcal{H}_{L \cup R}$. ({\bf a})  Construction of an example state in $\mathcal{H}_{LR}$.  ({\bf b})  Construction of an example state in $\mathcal{H}_{\mathcal{B}_L}\otimes \mathcal{H}_{\mathcal{B}_R}$.}
 \end{figure}

\paragraph{The two-sided Hilbert space}\label{prob2}
Consider quantum gravity with an asymptotic boundary having two disconnected components.  We can build states for this theory by cutting open the Euclidean path integral on a slice with two disconnected boundary components $\mathcal{X}= \mathcal{B}_L \cup \mathcal{B}_R$, as after Lorentzian continuation each of these components continues to a separate asymptotic Lorentzian boundary. We will call this Hilbert space $\mathcal{H}_{L \cup R}$. There are two ways of preparing states in $\mathcal{H}_{L \cup R}$. First, we could fill in and cut open a single connected boundary manifold, as in Fig.~\ref{fig:singleBcut}. For example, the state $\ket{\beta}$ is of this kind. Let us call the resulting Hilbert $\mathcal{H}_{LR}$. Note that while the two components of the \textit{boundary} cut are disconnected, there can exist bulk geometries satisfying these boundary conditions  which connect the boundaries via a single bulk.  Alternatively we could cut open two disconnected closed boundary manifolds $\mathcal{M}_L$ and $\mathcal{M}_R$ with cuts $\mathcal{B}_L$ and $\mathcal{B}_R$. As the two boundary manifolds are disconnected, this construction prepares states in a tensor product Hilbert space which we call $\mathcal{H}_{\mathcal{B}_L}\otimes \mathcal{H}_{\mathcal{B}_R}$ (Fig.~\ref{fig:tensorcut}).

We might expect that the full Hilbert space would be an (internal) direct sum $\mathcal{H}_{LR} \oplus \mathcal{H}_{\mathcal{B}_L}\otimes \mathcal{H}_{\mathcal{B}_R}$ or perhaps $\mathcal{H}_{L \cup R}=\{u+v\;|\;u\in \mathcal{H}_{LR},\, v\in \mathcal{H}_{\mathcal{B}_L}\otimes \mathcal{H}_{\mathcal{B}_R}\}$.  But holography with AdS boundary conditions requires that the full Hilbert space with disconnected boundaries has a tensor product structure because the dual CFT Hilbert space does.  In other words, we expect that, despite appearances, 
\beq \label{eq:facprob}
\mathcal{H}_{L \cup R}= \mathcal{H}_{\mathcal{B}_L}\otimes \mathcal{H}_{\mathcal{B}_R}=\mathcal{H}_{LR}.
\eeq
In this paper we will take this equality for granted and study the structure of $\mathcal{H}_{LR}$. In  \cite{Balasubramanian:2025zey} we show that the factorisation (\ref{eq:facprob}) results from the existence of novel Euclidean wormhole saddles connecting these various boundary conditions.

\paragraph{Levels of approximation}\label{approximations}
In this paper are interested in the task of establishing whether two quantities $\mathcal{A}$ and $\mathcal{B}$ in the fine-grained theory are equal $\mathcal{A}\stackrel{?}{=}\mathcal{B}$. 
As we have discussed, one challenge to showing such a thing is that the gravitational path integral seems to compute coarse grained averages in a probability distribution whose precise nature is not known.  From this perspective, the separate path integrals  computing $\mathcal{A}$ and $\mathcal{B}$ can at best show that $\overline{\mathcal{A}}=\overline{\mathcal{B}}$ where the overline indicates an average over the implicit distribution.  To additionally show that $\mathcal{A}=\mathcal{B}$ for each configuration in the support of the distribution, we have to additionally demonstrate that 
$\overline{(\mathcal{A}-\mathcal{B})^2}=0$.   To calculate $\overline{\mathcal{A}}$, $\overline{\mathcal{B}}$ and $\overline{(\mathcal{A}-\mathcal{B})^2}$ we have to evaluate the path integrals computing these quantities.  This is challenging in its own right, and can be carried out at various levels of precision:
\begin{enumerate}[align=left, labelwidth=*, labelsep=1em, leftmargin=*]
\item[\textbf{Level 1}] Sum over saddlepoints consistent with boundary conditions.   The sum includes all topologies on which we can place a metric that solves the equations of motion. Sometimes we approximate further by only keeping the  saddlepoint that dominates the sum.
\item[\textbf{Level 2}] Sum over all geometries, not only ones that solve the equations of motion, associated to topologies that admit saddlepoints.   This includes all perturbative fluctuations.
\item[\textbf{Level 3}]  Sum over all topologies and geometries, even for topologies that do not admit saddlepoint geometries.  In higher dimensions we may not have a complete classification of the possibilities, and thus have to restrict the sum to whatever is known.
\end{enumerate}
An ultraviolet finite completion in something like string theory could require additional ingredients that are not captured at all by the effective description via the gravitational path integral.  We will have nothing to say about the effects of such ingredients.  As we proceed below, we will indicate the level of approximation at which we are working.

\section{Tool 1: A convenient set of states} \label{sec:tool1}
In this section we will show how to construct a candidate overcomplete basis for $\mathcal{H}_{LR}$ that is under semiclassical control.

\subsection{Construction of the shell states} \label{sec:shellstates}
\begin{figure}[h]

        \begin{subfigure}[b]{.43\linewidth}
            \centering
            \includegraphics[width=0.9\linewidth]{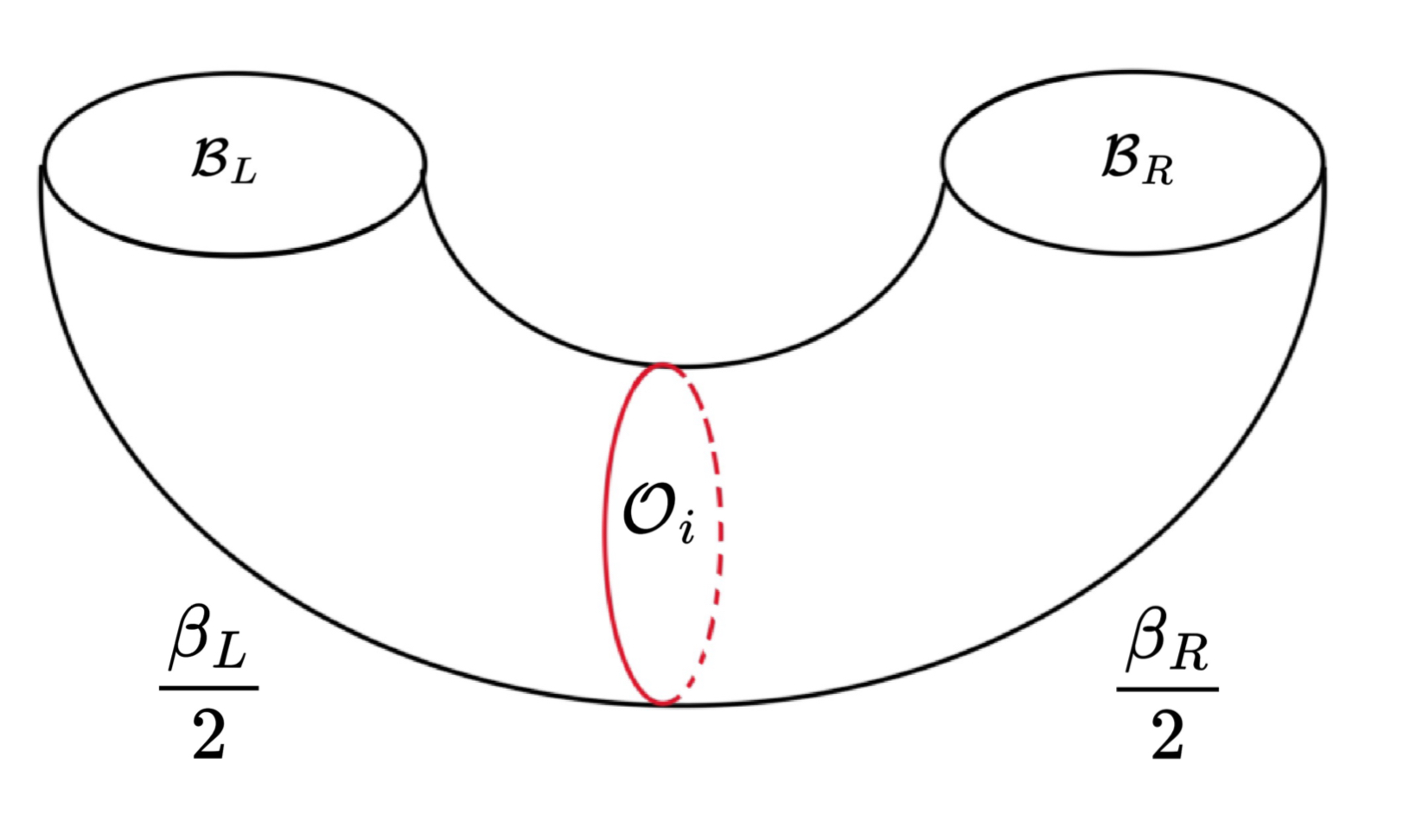}
            \caption{}
        \end{subfigure}
 \hfill
        \begin{subfigure}[b]{.5\linewidth}
            \centering  
            \includegraphics[width=0.9\linewidth]{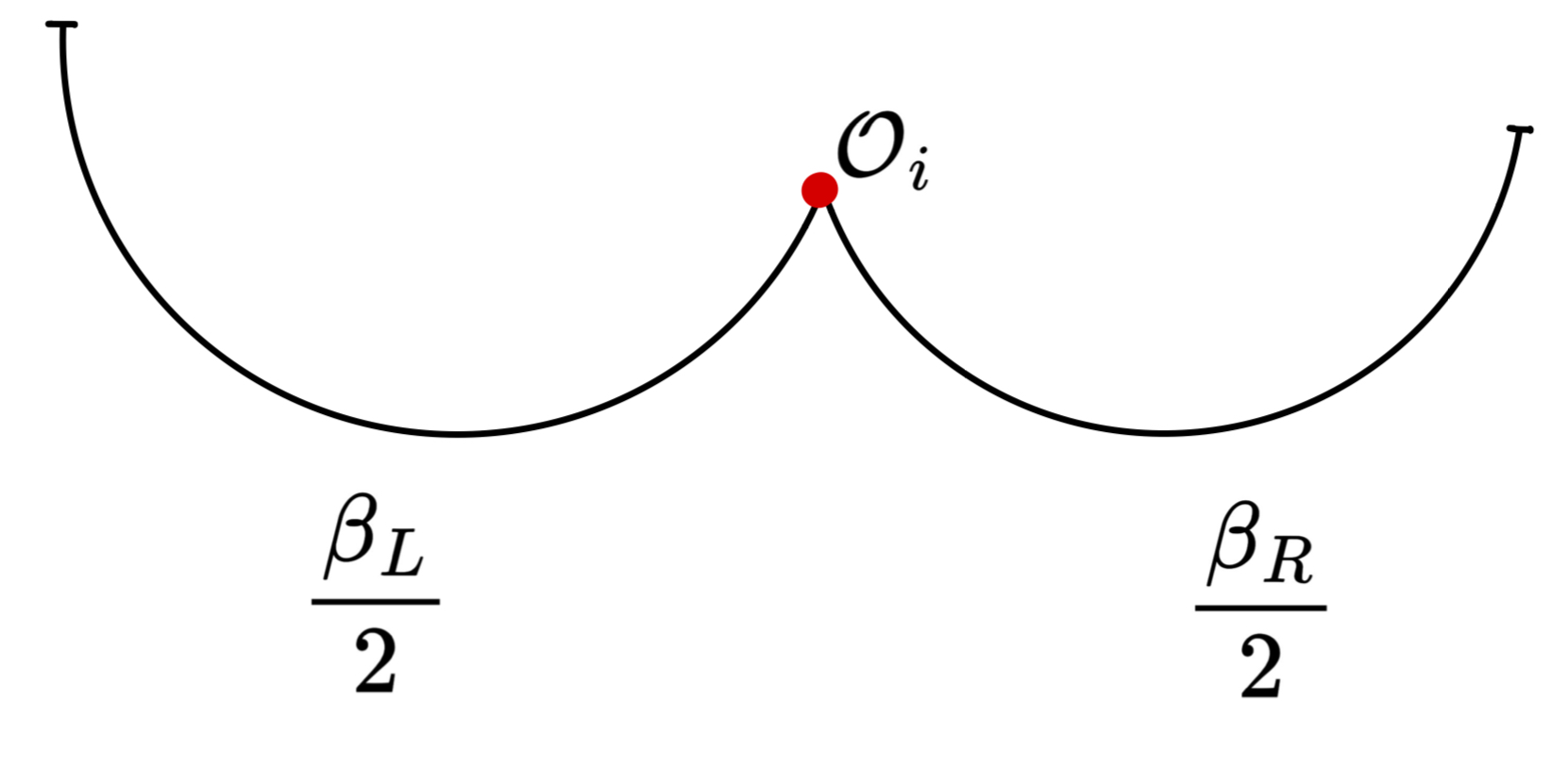}
            \caption{}
        \end{subfigure}
    \caption{Asymptotic boundary condition for the gravity path integral defining the shell state.  ({\bf a}) Cut-open Euclidean boundary with topology $\mathbb{I}_{\frac{\beta_L+\beta_R}{2}}\times\mathbb{S}^{d-1}$ for preparation of the shell states. The shell operator $\mathcal{O}_{i}$  is pictured in red. In AdS/CFT we can also perform the path integral in the boundary CFT with insertion of a $\mathbb{S}^{d-1}$ symmetric operator dual to the shell.  ({\bf b}) Euclidean boundary with the $\mathbb{S}^{d-1}$ suppressed. We adopt this convention for the rest of the paper. Here $\beta_{L,R}/2$ are  Euclidean ``preparation times''. }
    \label{fig:shell_b.c}
\end{figure}

Our candidate basis for the two-sided Hilbert space  consists of the shell states studied in \cite{Sasieta:2022ksu,Balasubramanian:2022gmo,Balasubramanian:2022lnw,Antonini:2023hdh}; see \cite{Balasubramanian:2024rek} for a comprehensive discussion. We  define these states by cutting open the Euclidean gravitational path integral.  The resulting  boundary $\mathbb{I}\times \mathbb{S}^{d-1}$ has a cut with two boundary components $\mathcal{B}_L=\mathcal{B}_R$, each of topology $\mathbb{S}^{d-1}$ (Fig.~\ref{fig:shell_b.c}).  We also insert a $\mathbb{S}^{d-1}$ symmetric heavy ($m \sim \mathit{O}(1/G_{N})$) dust shell operator  $\mathcal{O}_{S}$ separated by a boundary length $\frac{\beta_L}{2}$ and $\frac{\beta_R}{2}$ from the $\mathcal{B}_{L,R}$ cuts.  Following \cite{Balasubramanian:2022gmo} we can obtain an infinite family of shell states in this way by varying the mass $m_i$ 
of the shell operators. We call the resulting state $|i\rangle$.

The construction of these states is intuitive within AdS$_{d+1}$/CFT$_{d}$ but applies to any dimension and cosmology with a time-like asymptotic boundary. In the AdS/CFT setting, the CFT  lives on the boundary geometry.  We define states on two copies of the CFT with Hamiltonians ${H}^{CFT}_{L}={H}^{CFT}_{R}$ by inserting an operator $\mathcal{O}$ in the CFT  path integral and evolving with the $L$ and $R$ Hamiltonian. Via the usual state-operator map this yields a state $|\mathcal{O}\rangle =|e^{-{\beta_{L}{H}_{L}}}\mathcal{O}e^{-{\beta_{R}{H}_{R}}}\rangle $ on  $\mathcal{H}^{CFT}_{L} \otimes\mathcal{H}^{CFT}_{R}$ (Fig.~\ref{fig:shell_b.c}). Taking $|E_n\rangle$ to label energy eigenstates for one copy of the CFT, such a state in the energy basis is given by:
\beq \label{eq:pets}
|\mathcal{O}\rangle = \sum_{n,m}e^{-(\beta_{L}E_{n} +\beta_{R}E_{m})}\mathcal{O}_{mn}|E_n\rangle_L|E_m\rangle_R
\eeq
and thus corresponds to a partially entangled state (PETS) \cite{Goel:2018ubv}. The CFT shell states are defined by a choice of  operator $\mathcal{O}$ that is holographically dual to the shell operator insertion on the asymptotic boundary of AdS$_{d+1}$. In the CFT this is achieved by inserting  products of a scalar operator $\mathcal{O}_{\Delta}$ of $\mathit{O}(1)$ scaling dimension $\Delta$. In order to make the states sufficiently heavy $m \sim \mathit{O}(1/G_{N})$, we must insert $n \sim \ell^{d-1}/\mathrm{G_N}$  such operators ($\ell$  is the AdS radius) at spherically symmetric points on the $\mathbb{S}^{d-1}$, resulting in the operator $\mathcal{O}_{S}$. 

To be clear, a particular microstate of the gravity theory is constructed by specifying the microscopic details of the operator insertions that make up the heavy shell. There are many such exact microstates which share the same semiclassical description as a spherically symmetric heavy dust shell in the bulk. As discussed Sec.~\ref{sec:QTFTPI} above, the semiclassical gravitational path integral in the manner used here appears to compute a coarse-graining over these fine-grained details. Thus while  we have some particular microscopic realization in mind for each of our shell states, we can, remarkably,  learn much about the underlying theory without specifying these details.

\begin{figure}[h]
    \centering
    \includegraphics[width=0.3\linewidth]{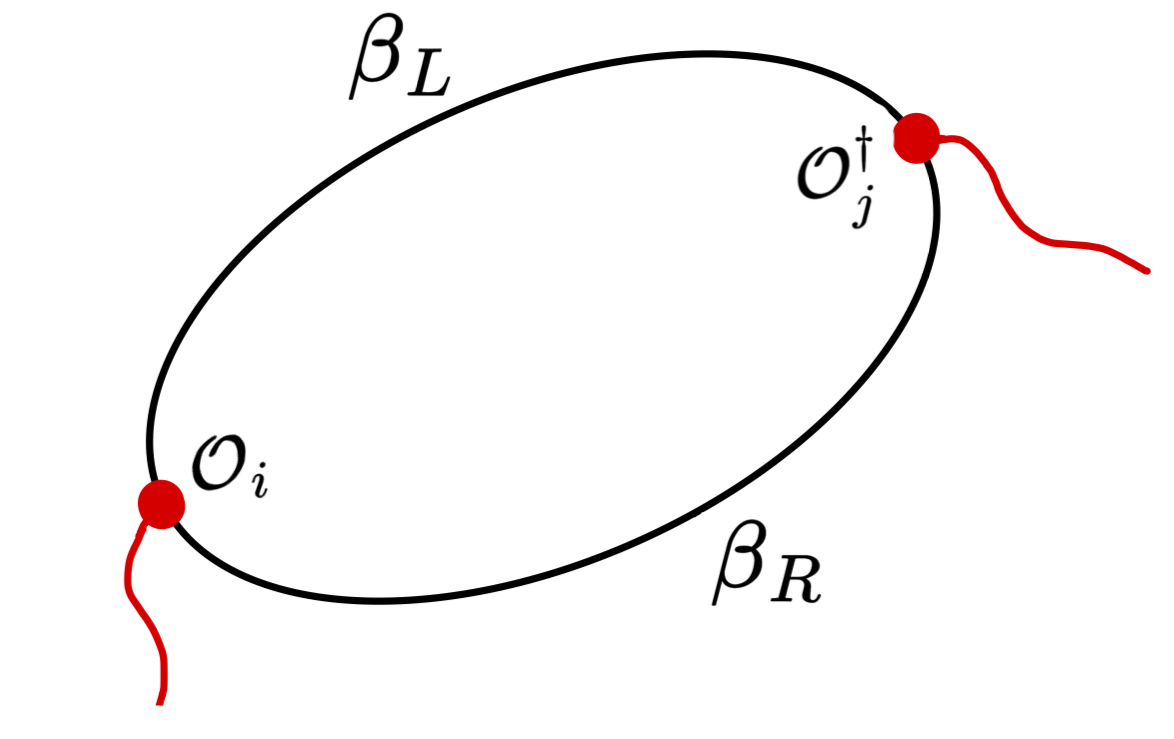}
    \caption{Shell asymptotic boundary condition for $\overline{\langle j|i\rangle} $, consisting of the operator insertions $\mathcal{O}_{i}$ and  $\mathcal{O}^{\dagger}_{j}$ separated by asymptotic time extent $\beta_{L}$ and $\beta_{R}$ respectively. The red lines represent the shells propagating into the bulk.}
    \label{fig:shell_bdry}
\end{figure}

\begin{figure}[h]
            \centering
            \includegraphics[width=1\linewidth]{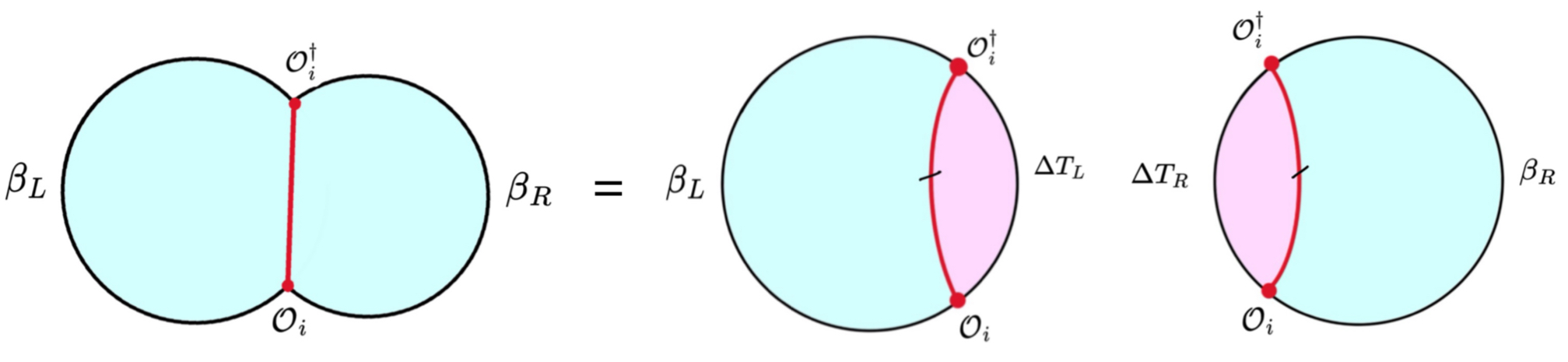}
            \caption{The saddlepoints for the shell norm path integral $\overline{\langle i|i\rangle}$ are constructed by gluing together two disks. In particular the shell homology regions (purple) on each disk are discarded and the resulting geometries glued together along the shell worldvolume.}
            \label{fig:shell_norm}
\end{figure}

\begin{figure}
    \centering
    \includegraphics[width=1\linewidth]{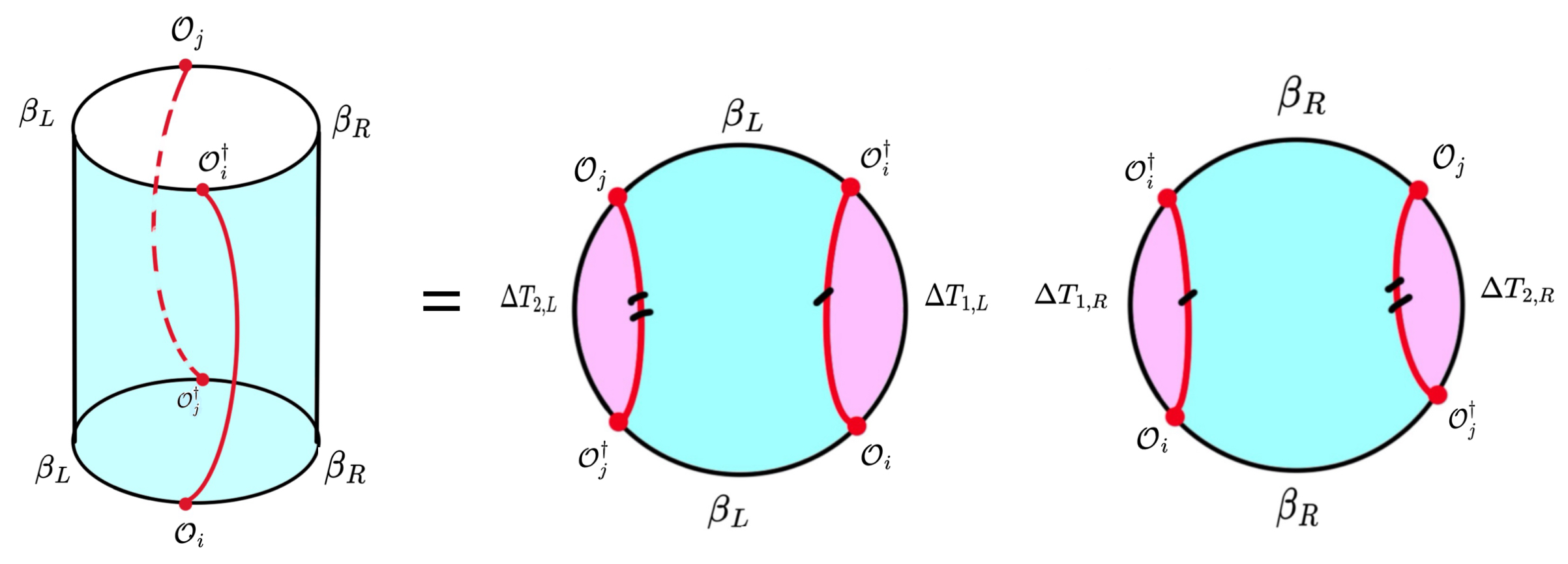}
    \caption{Construction of shell wormhole saddles $Z_2$. We glue two disks together by discarding the regions indicated in purple and  attaching the remainder along the shell worldvolumes (red, black dashes indicate the glued shells) while obeying the Israel junction conditions.  Saddlepoints, i.e., solutions to the equation of motion, exist with this topology \cite{Balasubramanian:2022gmo}.}
    \label{fig:shell_wormhole}
\end{figure}

We will want to compute the Gram matrix $G_{ij}\equiv \langle i|j\rangle$ of overlaps between the shell states. We will say that the path integrals calculating these quantities have \textit{shell boundaries} which include operator insertions $\mathcal{O}_{j}$ and  $\mathcal{O}^{\dagger}_{i}$ inserted on the asymptotic boundary separated by asymptotic times  $\beta_{L}$ and $\beta_{R}$ respectively (Fig.~\ref{fig:shell_bdry}).  The overlap $\overline{\langle i|j \rangle}$ is evaluated by the gravitational path integral over geometries satisfying this boundary condition (Fig.~\ref{fig:shell_norm}).  Following~\cite{Balasubramanian:2022gmo},  shell states of different mass can be rendered orthogonal in terms of the overlaps computed in Fig.~\ref{fig:shell_norm} by taking their  inertial mass differences to be arbitrarily large, as it takes $|m_{i}-m_{j}|$ bulk interactions in Planck units to match such shells in the bulk, leading to 
\begin{equation}
\overline{\langle i|j \rangle} =\delta_{ij} Z_{1} \,.
\label{eq:defZ1}
\end{equation} 
However, higher topology wormhole contributions to the path integral  are stabilized by the shell matter and modify the overlap to

\begin{equation}
\overline{|\langle i|j \rangle|^2}=\overline{\langle i|j \rangle\langle j|i \rangle} = Z_{2} + \delta_{ij}Z_{1}^2 \, ,
\end{equation}
where $Z_{2}$ is the wormhole saddlepoint contribution (Fig.~\ref{fig:shell_wormhole}).

\subsubsection{Shell wormholes}\label{sec:shellwh}

\begin{figure}[h]
    \centering
    \includegraphics[width=0.4\linewidth]{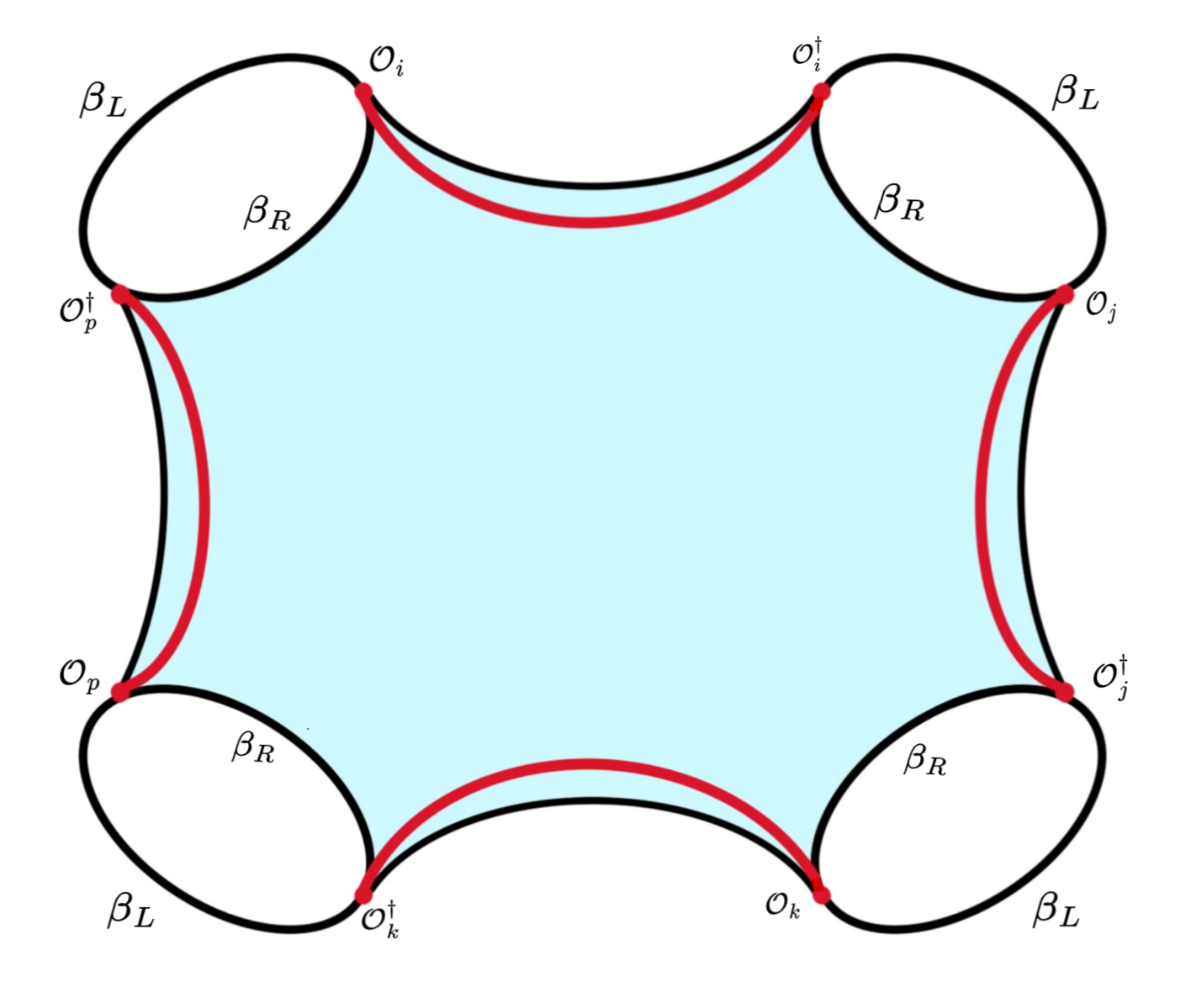}
    \caption{Fully connected wormhole geometry for $\overline{(G^n)_{ii}}$ consisting of $n$ shell boundaries connected by a single bulk, pictured here for $n=4$.}
    \label{fig:pinwheel}
\end{figure}

The shell states are useful because the associated wormholes and their action can be explicitly computed, at least in the large shell mass limit. The wormholes are constructed by identifying shell worldvolumes across two disks subject to the Israel junction conditions as indicated in Figs.~\ref{fig:shell_norm} and \ref{fig:shell_wormhole}. The saddlepoint geometry on either side of the shell can be anything that locally solves the equations of motion; hence by symmetry these are portions of any of the saddlepoints contributing to the Z-partition function in (\ref{eq:Zpartitionfunction}).
Generalizing this construction, we construct wormhole saddlepoints contributing to the amplitude $\overline{(G^n)_{ii}}=\overline{\braket{i|j_1}\braket{j_1|j_2} \cdots \braket{j_n|i}}$  (no sum on $i,j_k$) by connecting $n$ asymptotic boundaries (arising from each factor in $(G^n)_{ii}$) with a single multiboundary wormhole. The $k$-th shell is able to propagate through the bulk of the wormhole from $\mathcal{O}_{k}$ to $\mathcal{O}^{\dagger}_{k}$ on different asymptotic boundaries, and hence the index structure of $\overline{(G^n)_{ii}}$ manifests in the bulk as the wormhole in Fig.~\ref{fig:pinwheel}.  We will call this wormhole geometry the \textit{pinwheel}. There are also disconnected saddlepoints connecting different subsets of the boundaries together.  These look like collections of smaller pinwheels.

\begin{figure}[h]
\begin{subfigure}{\linewidth}
    \centering
    \includegraphics[width=0.5\linewidth]{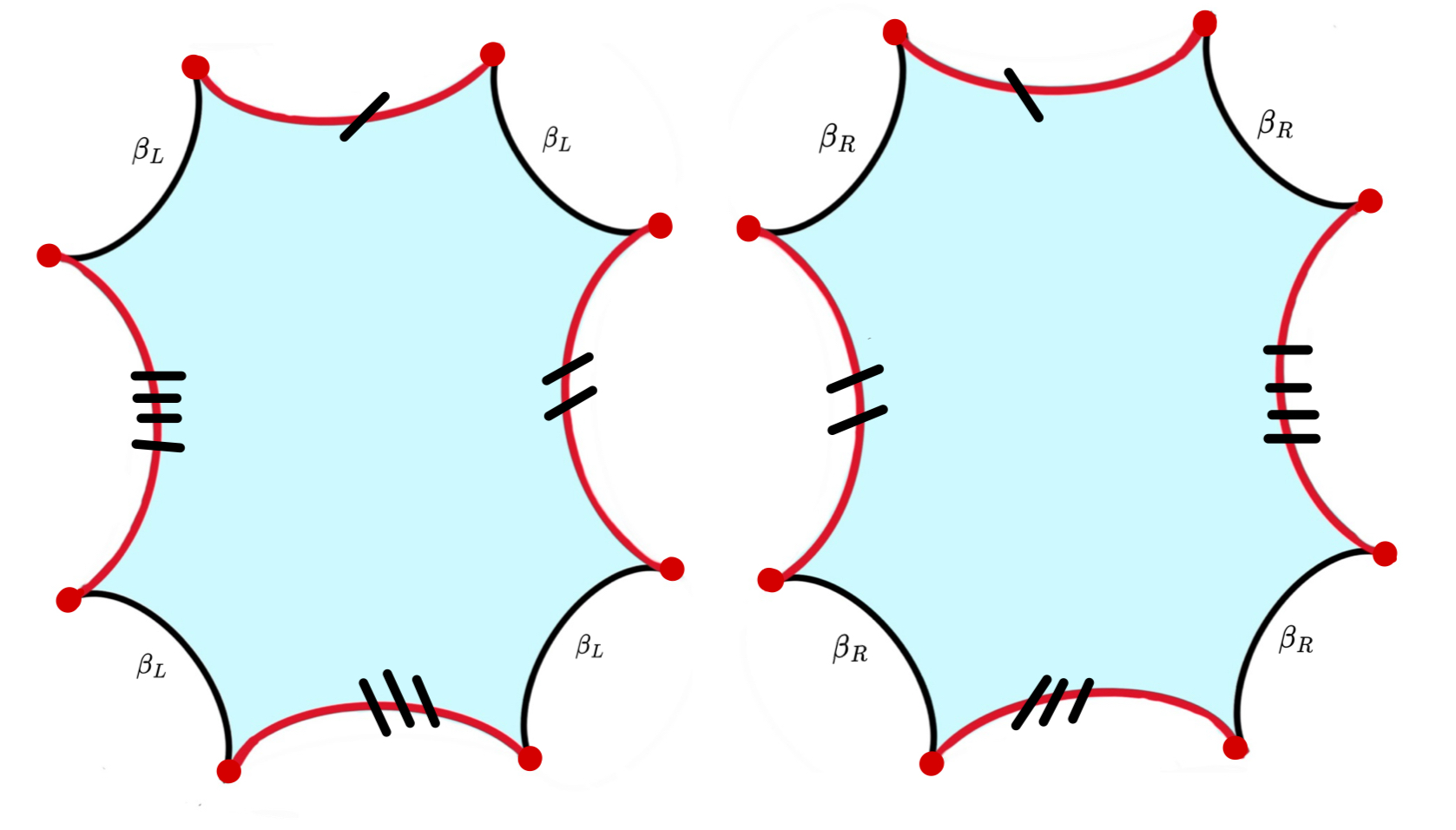}
    \caption{}
    \label{fig:pinwheelsheet}
\end{subfigure}
\begin{subfigure}{\linewidth}
\centering
    \includegraphics[width=0.6\linewidth]{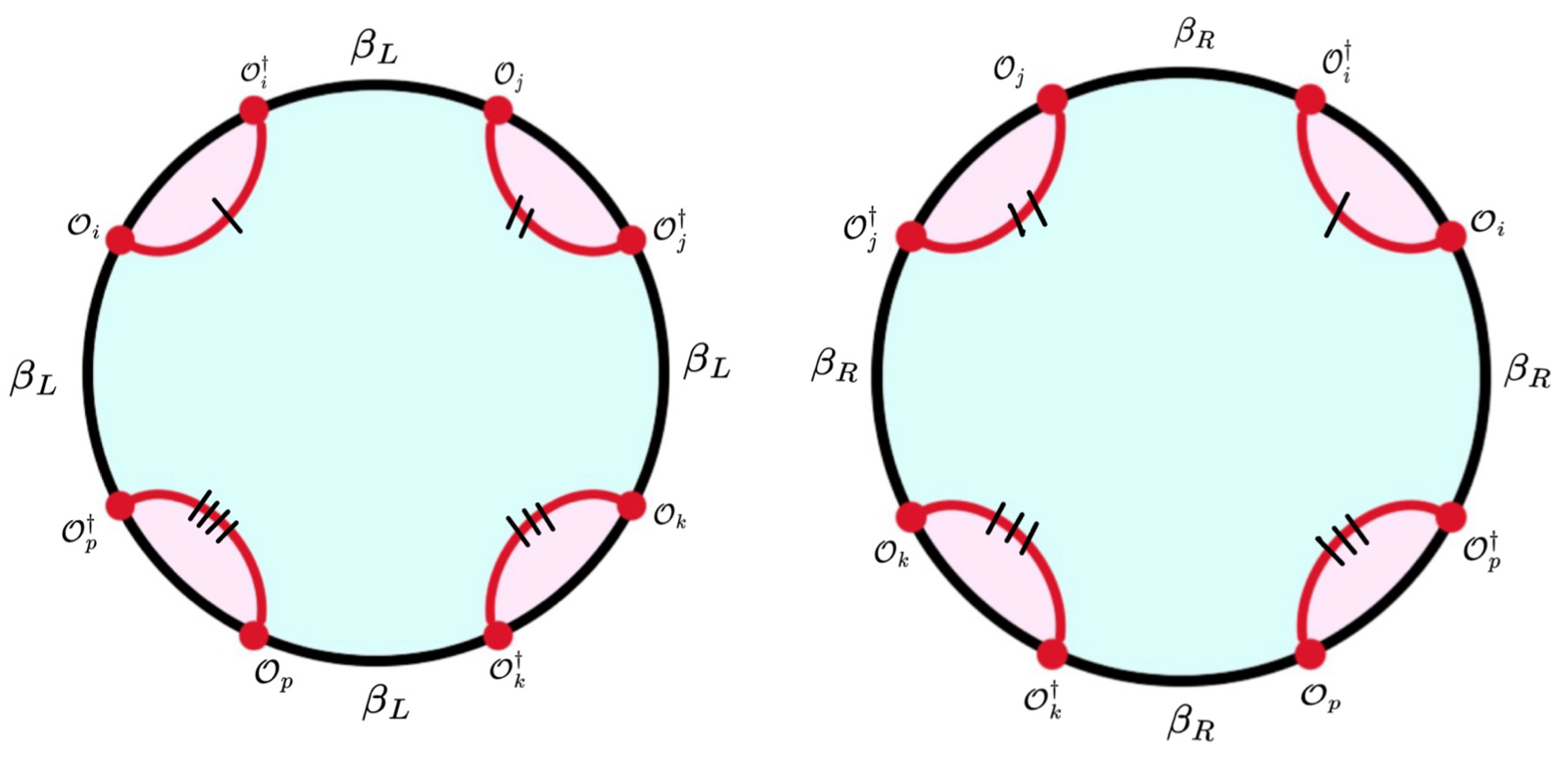}
    \caption{}
    \label{fig:pinwheeldisks}
    \end{subfigure}
    \caption{Schematic of the construction of the pinwheel wormhole. ({\bf a}) Sheet diagrams for the pinwheel obtained by cutting Fig.~\ref{fig:pinwheel} along the shell worldvolumes.  ({\bf b}) Each of the pinwheel sheets can be regarded a portion of a disk. The pinwheel saddle constructed by gluing to disks together along the corresponding shell worldvolumes by the junction conditions.}
\end{figure}

To construct a pinwheel saddlepoint imagine that 
we cut Fig.~\ref{fig:pinwheel} open along the shell worldvolumes producing the so-called $L$- and  $R$- \textit{sheet diagrams} in Fig.~\ref{fig:pinwheelsheet}. We can produce these sheets by removing the pink shell homology regions from the $L$ and $R$ disks in Fig.~\ref{fig:pinwheeldisks}.    
The $L$ ($R$) disk boundary contains $n$ shell operator insertions  $\mathcal{O}_{1,2, \cdots n}, \mathcal{O}^{\dagger}_{1,2, \cdots n}$ with  $\mathcal{O}_{i}$ and $\mathcal{O}^{\dagger}_{i}$ separated by a time $\Delta T_{L,i}$ ($\Delta T_{R,i}$) and $\mathcal{O}^{\dagger}_{i},\mathcal{O}_{i+1} $  separated by a time $\beta_L$ ($\beta_R$). Constructively, then, we start with the disks in  Fig.~\ref{fig:pinwheeldisks}, remove the pink shell homology regions, and then use the Israel junction conditions \cite{Israel:1966rt}  to glue along the indicated shell worldvolumes. The propagation times $\Delta T_i$ are determined dynamically by the junction conditions to obtain an on-shell wormhole geometry.  This procedure can be applied to any saddle of the Z partition function defined above, and hence there are in general multiple saddles. 

As shown in \cite{Balasubramanian:2022gmo,Balasubramanian:2022lnw,Antonini:2023hdh}  the bulk turning points of the shells approach the asymptotic boundary  of the disk in the large shell mass ($m_i \to \infty$) limit and hence the purple shell homology regions pinch off. In this limit, $\Delta T_i \to 0$, and each shell contributes simply contributes a factor $Z_{m_i}\sim e^{-2(d-1)log(\mathrm{G_N} m_i)}$ that is a universal function of its mass.   Furthermore, this behavior is independent of the disk saddle geometries that are being glued along the shells \cite{Balasubramanian:2022gmo,Balasubramanian:2022lnw,Antonini:2023hdh}. Hence, in the saddle approximation the fully connected contribution to the n-boundary path integral decomposes into two Z-partition functions and a shell contribution: $\overline{(G^n)_{ii}}|_{connected} = \overline{Z}(n\beta_L)\overline{Z}(n\beta_R )\Pi_{i=1}^n Z_{m_i}$. 

In the $m_i \to \infty$ limit the diagram in Fig.~\ref{fig:shell_norm} computing the norm $\braket{i|i}$ also simplifies -- the propagation time shell propagator shrinks to zero, so that the diagram decomposes into two disks of circumference $\beta_{L,R}$ attached via a pointlike shell trajectory so that Eq.~\ref{eq:defZ1} for the norm becomes $\braket{i|j} = \delta_{ij} Z_1 = \delta_{ij} \overline{Z}(\beta_L) \overline{Z}(\beta_R)Z_{m_i}$ within the saddlepoint approximation.  We can also see this by setting $n=1$ in the above expression for $\overline{(G^n)_{ii}}|_{connected}$. If we then consider the normalized set of states $\ket{k}/\sqrt{Z_1}$, we get

\beq \label{eq:pinwheel}
\overline{(G^n)_{ii}}|_{connected} = \frac{\overline{Z}(n\beta_L)\overline{Z}(n\beta_R)}{\overline{Z}(\beta_L)^n \overline{Z}(\beta_R)^n},
\eeq
within the saddlepoint approximation, where again the overbars indicate coarse-grained averages computed by the path integral.  Note that this equality is not restricted to the leading saddlepoint; each Z partition sum on the right side is a sum over all saddlepoints, accounting for all the saddles of $\overline{(G^n)_{ii}}|_{connected}$.

\paragraph{On the large-mass limit}
Note that the ``universality'' of heavy shells is a property of a single shell worldvolume
on the background geometry --- its turning point approaches the boundary, the homology
region pinches off, and it contributes the factor $Z_{m_i}$, cancelled by normalisation. It
does not erase the data distinguishing the states. Two pieces survive: the states keep their
distinct masses, with disk-level orthogonality $\overline{\langle i|j\rangle}=\delta_{ij}Z_1$,
matching unequal masses requiring $|m_i-m_j|$ Planckian interactions; and the moments
$\overline{\mathrm{Tr}\,G^n}$ retain nontrivial dependence on $n$ through
$\overline{Z}(n\beta_{L,R})$ even at infinite mass. The latter, not the per-shell factor,
fixes the spectrum of $G$. The universal factor does not affect the rank, because
normalisation acts by congruence $G\to D^{-1/2}GD^{-1/2}$ with $D$ positive and invertible,
which preserves rank; the rank is fixed by the moments. Equivalently, in the language of
Sec.~\ref{sec:QTFTPI} the path integral computes the moments of the overlap distribution:
the heavy-shell limit makes the mean tractable, while the variance
$\overline{\langle i|j\rangle\langle j|i\rangle}=Z_2$ and higher moments --- also computed
in this limit --- carry the Hilbert-space structure.

\subsubsection{Shell geometry} \label{sec:shellgeom}
\begin{figure}[h]
    \centering
    \begin{subfigure}[c]{0.45\linewidth}
        \centering
        \includegraphics[width=\linewidth]{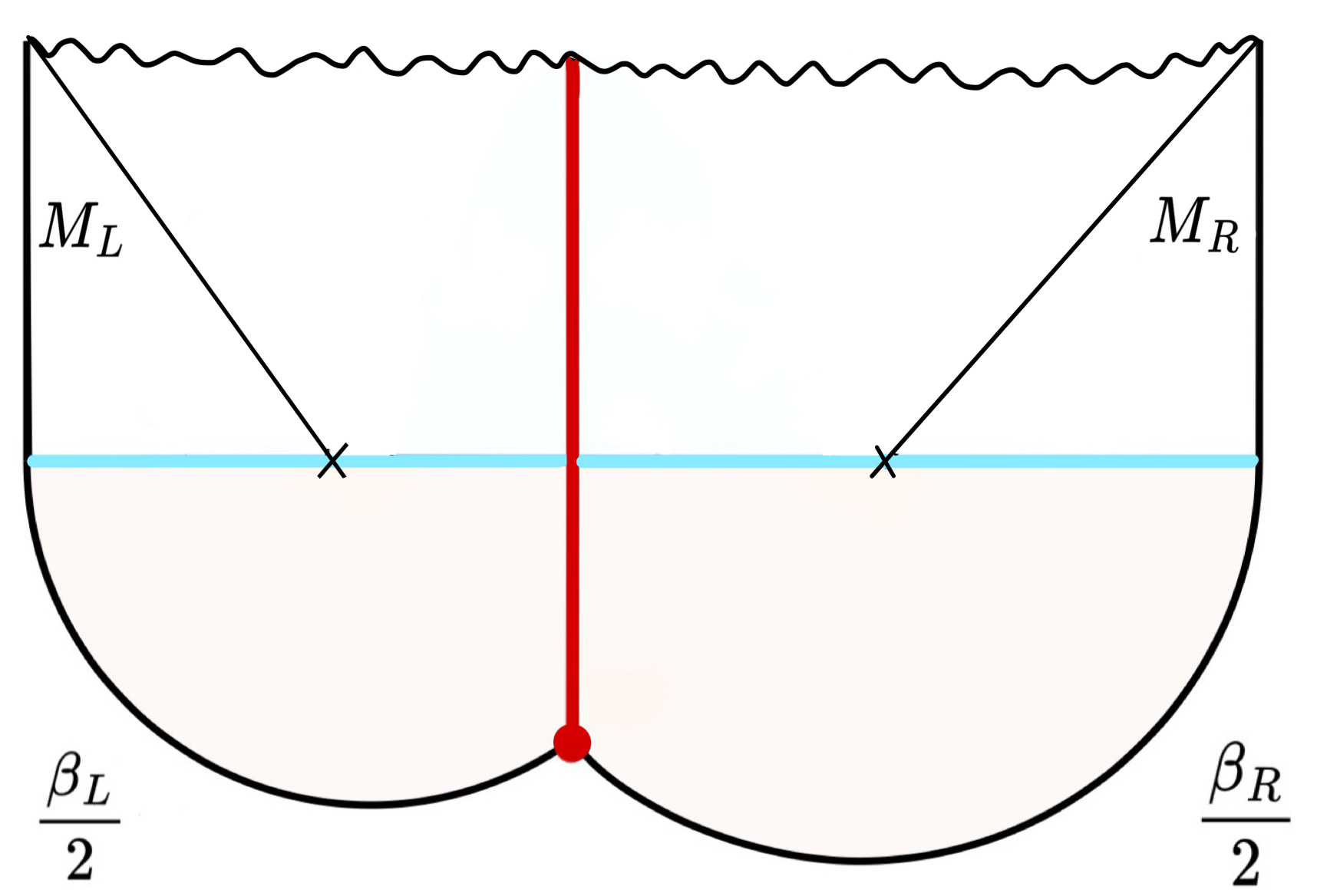}
        \caption{}
        \label{type1}
    \end{subfigure}
    \hfill
    \begin{subfigure}[c]{0.45\linewidth}
        \centering
        \includegraphics[width=\linewidth]{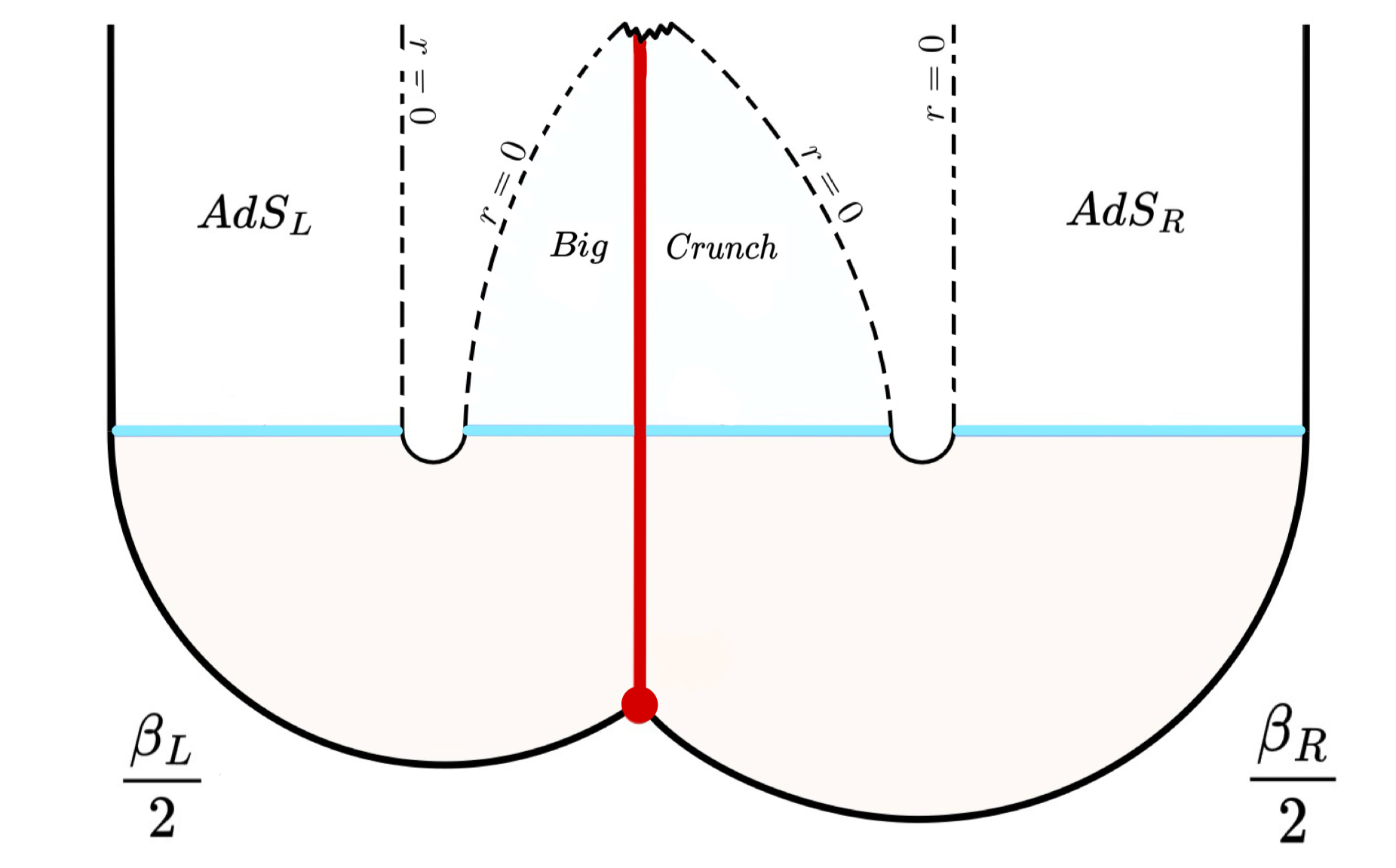}
        \caption{}
        \label{type2}
    \end{subfigure}
    \hfill
    \begin{subfigure}[c]{0.45\linewidth}
        \centering
        \includegraphics[width=\linewidth]{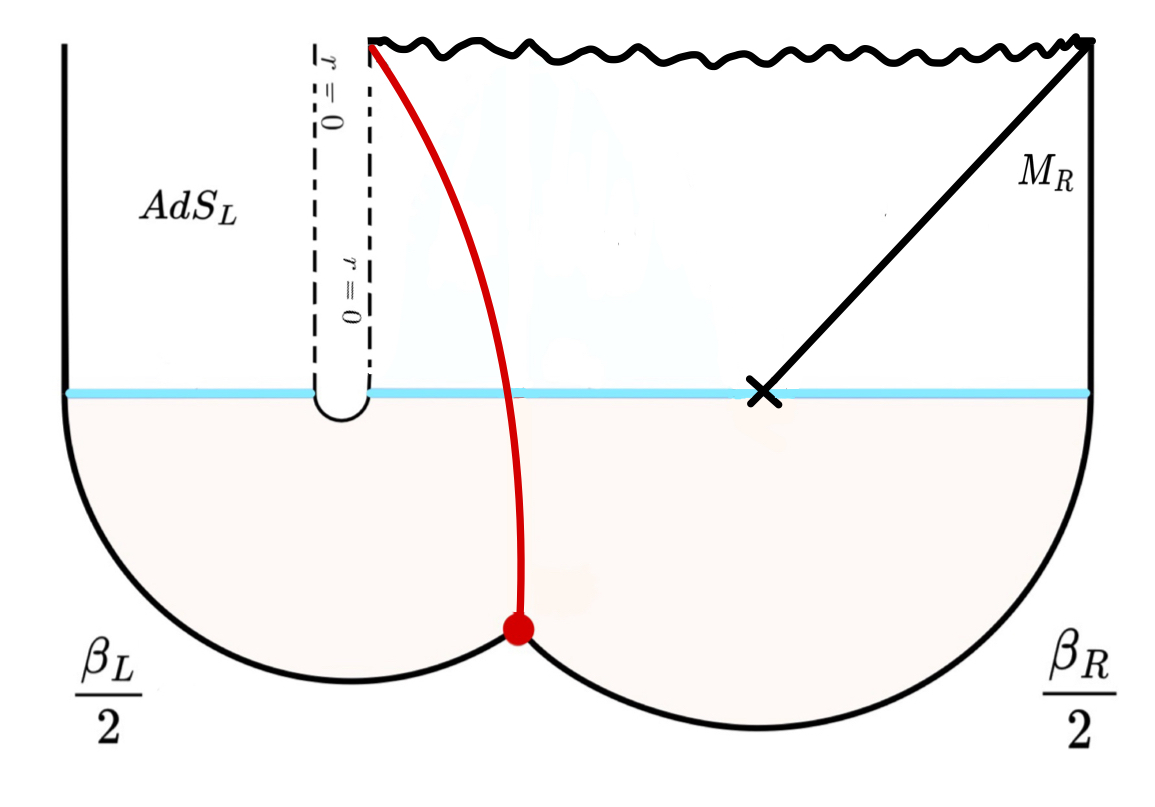}
        \caption{}
        \label{type3}
     \end{subfigure}
\caption{Diagrams showing the analytic continuation of the type 1-3 shell states to  Lorentzian signature.  ({\bf a}) Type 1 shell state with $\beta_L < \beta_{HP}$ and $\beta_R < \beta_{HP}$ corresponding to a two-sided black hole. ({\bf b}) Type 2 shell state with $\beta_L > \beta_{HP}$ and $\beta_R>\beta_{HP}$ consisting of two copies of thermal AdS with the addition of a compact Big-Crunch AdS cosmology. ({\bf c}) Type 3 shell state with $\beta_L > \beta_{HP}$ and $\beta_R < \beta_{HP}$ corresponding to a disconnected geometry with an $R$ single sided BH and an $L$ thermal AdS factor. }
\end{figure}

Focusing now on the asymptotically AdS case for concreteness, we can associate a geometry to a shell state by considering the leading gravitational saddlepoint contributing to the norm $\overline{\langle i|i\rangle}=\overline{Z}(\beta_L) \overline{Z}(\beta_R)Z_{m_i}$ (Fig.~\ref{fig:shell_norm}).  As shown above, in the large shell mass limit this is determined by the leading saddlepoint for the Z partition functions $\overline{Z}(\beta_L),\overline{Z}(\beta_R )$.  The leading Z-saddle depends on whether the inverse effective temperature, i.e. Euclidean time period $\beta_{L,R}$, is above or below the Hawking-Page transition \cite{Hawking:1982dh} at $\beta_{HP}$. For example if $\beta_{L} >\beta_{HP}$ the $L$ disk saddle is thermal AdS, whereas for $\beta_{L} < \beta_{HP}$ it is the Euclidean AdS black hole (similarly for $R$). The time reflection symmetric bulk slice of these saddles can  be analytic continued to Lorentzian signature to associate a Lorentzian geometry to the state. As $\beta_L, \beta_R$ can be tuned independently there are 4 regimes to consider:

\begin{enumerate}
    \item $\beta_{L}, \beta_{R} < \beta_{HP}$ : The leading  saddles are two-sided Schwarschild-AdS black holes with a long wormhole interior (Fig.~\ref{type1}). We will refer to these  as \textit{type 1} shell states and the space spanned by these states as $\mathcal{H}_{1} \subseteq \mathcal{H}_{Shell}$. These are the states originally considered in \cite{Balasubramanian:2022gmo}. 
   
    \item $\beta_{L}, \beta_{R} > \beta_{HP}$: The leading saddles  are two copies of thermal AdS with the addition of a compact Big-Crunch AdS cosmology (Fig.~\ref{type2}). This construction was studied in detail in \cite{Antonini:2023hdh}. We will refer to these  as \textit{type 2} shell states and the space spanned by them as $\mathcal{H}_{2} \subseteq \mathcal{H}_{Shell}$. 
    
    \item $\beta_{L}  < \beta_{HP}$ and  $\beta_{R} > \beta_{HP}$  or vice versa : The leading saddles correspond to an $L$ ($R$) single-sided AdS Black hole and a disconnected $R$ ($L$) copy of thermal AdS (Fig.~\ref{type3}). We refer to these  as \textit{type 3} shell states and the space spanned by them states as $\mathcal{H}_{3} \subseteq \mathcal{H}_{Shell}$. Such states were studied in \cite{Balasubramanian:2022lnw}, but they did not include the disconnected thermal AdS.

\end{enumerate} 

\subsection{Inserting a complete set of shell states} \label{sec:Zid}

We will now argue that the shell states span $\mathcal{H}_{LR}$ at least in a coarse-grained sense.  There is already some evidence for this from the argument of \cite{Balasubramanian:2022gmo,Balasubramanian:2022lnw,Climent:2024trz}  that any sufficiently large collection of shell states, projected into a microcanonical band where the dominant saddlepoint is an eternal black hole, spans a space of dimension  $e^{\mathbf{S}(E_L) + \mathbf{S}(E_R)}$ where  where $\mathbf{S}(E)$ is the Beckenstein-Hawking entropy of a horizon associated to energy $E$. However, this approach to  completeness  of the shell states requires an independent argument that the dimension of the full Hilbert space in this regime is $e^{\mathbf{S}(E_L) + \mathbf{S}(E_R)}$. There are also technical complexities involving the microcanonical projection. To develop the tools for the full argument in Sec.~\ref{sec:scspan} and Sec.~\ref{sec:tool3}, we start with an alternative argument that the shell Hilbert space $\mathcal{H}_{shell}$  at fixed preparation temperature spans $\mathcal{H}_{LR}$ by showing that the projector onto $\mathcal{H}_{shell}$ (\ref{eq:id}) acts as the identity on $\mathcal{H}_{LR}$:
\beq \label{eq:project1}
\mathds{1}_{\mathcal{H}_{LR}} \stackrel{?}{=} \Pi_{\mathcal{H}_{shell}} \equiv G^{-1}_{ij} |i\rangle\langle j|.
\eeq
The gravitational path integral  for $\overline{Z(\beta)}$, which has a boundary manifold of topology $\mathbb{S}_{\beta}\times \mathcal{X}^{d-1}$, has a standard interpretation following Gibbons and Hawking as the thermal partition function of quantum gravity. If this is indeed the case, the state $|\beta\rangle \in \mathcal{H}_{LR}$ obtained by cutting this path integral in half (recall that $Z(\beta)=\langle \beta|\beta\rangle$) must have support on all the energy eigenstates of the theory. Indeed, in the AdS/CFT setting the holographic dictionary equates $|\beta\rangle$ to the thermofield-double (TFD) state defined on two copies of the CFT: $|TFD(\beta)\rangle =\sum_{n} e^{\frac{-\beta E_{n}}{2}} |n\rangle _{L}|n\rangle_R$. We can test 
the validity of (\ref{eq:project1}) by checking whether the following equality is true:
\beq \label{eq:Zid}
\overline{Z(\beta)}=\overline{\langle \beta|\beta\rangle}=\overline{\langle 
\beta|\mathds{1}_{\mathcal{H}_{LR}}|\beta\rangle}\stackrel{?}{=} \overline{\langle 
\beta|\Pi_{\mathcal{H}_{shell}}|\beta\rangle} =\overline{G^{-1}_{ij}\langle \beta|i\rangle\langle 
j|\beta\rangle}\equiv\lim_{n\to -1} \overline{G^{n}_{ij}\langle \beta|i\rangle\langle 
j|\beta\rangle} \, .
\eeq

\subsubsection{Resolvent Resummation} \label{sec:resolvant}
\begin{figure}[h]
    \centering
    \includegraphics[width=0.3\linewidth]{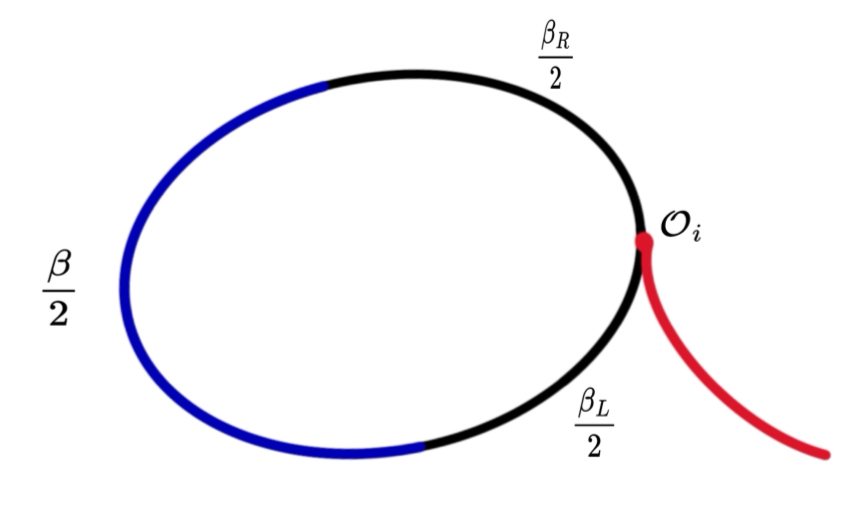}
    \caption{Tadpole diagram representing the asymptotic boundary associated to the overlap $\langle \beta|i\rangle$ of length $\frac{\beta + \beta_L+ \beta_R }{2}$ with a single $\mathcal{O}_i$ shell operator insertion indicated in red.}
    \label{fig:beta1}
\end{figure}

To try to establish (\ref{eq:Zid}) we will use the gravitational path integral to evaluate $\overline{G^{n}_{ij}\langle \beta|i\rangle\langle j|\beta\rangle}$. The overlap  $\langle \beta|i\rangle$ in (\ref{eq:Zid}) corresponds to a periodic asymptotic boundary of length $\frac{\beta_L+\beta_R+\beta }{2}$ with a shell insertion $\mathcal{O}_i$ (Fig.~\ref{fig:beta1}). Here $\beta_{L,R}$ here are the $L,R$ shell preparation temperatures as in Sec.~\ref{sec:shellstates}.  We will refer diagrams like Fig.~\ref{fig:beta1} as \textit{tadpoles}. The $\langle \beta|i\rangle\langle j|\beta\rangle$ factor in (\ref{eq:Zid}) is thus described by two tadpoles, and the powers of the Gram matrix $G_{ij}$ correspond to products of shell overlaps $\langle i|j\rangle$ each associated a boundary condition of the  form depicted in
Fig.~\ref{fig:shell_bdry}.  The saddlepoints of the path integral with these boundary conditions include all ways of connecting the two shells from the tadpole boundaries to the $n$ shell boundaries coming from powers of the Gram matrix.  Suppose  that we include $\kappa$ shell states in our candidate basis.  A shell sourced by  $\mathcal{O}_{i}$ can only propagate through the bulk and annihilate with $\mathcal{O}^{\dagger}_{j}$ if $i=j$. Therefore less than fully-connected geometries have to break the shell-index loops in the boundary condition, and will be relatively suppressed by factors of $1/\kappa$. 

For example, consider  $Tr{G^2}= \sum_{i,j} \langle i | j \rangle\langle j | i \rangle = \sum_{i,j} Z_2 + Z_1^2 \delta_{ij} = \kappa^2 Z_2 + \kappa Z_1 ^2$. As the disconnected contribution requires $j=i$ it must break the $j$ index loop and is suppressed by a factor $1/\kappa$, whereas higher topology contributions, like $Z_2$, are suppressed by factors of $e^{-1/G_{N}}$.  We are going to consider a ``planar'' limit in which $\kappa,e^{1/G_{N}}$ are both large while $\kappa \, e^{-1/G_{N}}\sim \mathcal{O}(1)$, discussed in \cite{Penington:2019kki,Balasubramanian:2022gmo}. In this limit both planar partially-connected and planar fully-connected geometries contribute at leading order in  $1/\kappa$ and $e^{-1/\mathrm{G_N}}$. By planar we mean a pattern of connectivity between the asymptotic boundaries that involves no crossings when depicted on a 2D plane (see for example Sec.~5.3 in \cite{Boruch:2024kvv}). Non-planar wormhole geometries require breaking at least two index loops, and for every such geometry there is a planar geometry with the same pattern of connectivity between boundaries. The non-planar geometries are thus subleading and we will exclude them from the sum over saddles in the planar limit.

\begin{figure}[h]
\begin{subfigure}[b]{0.45\linewidth}
    \centering
    \includegraphics[width=1.2\linewidth]{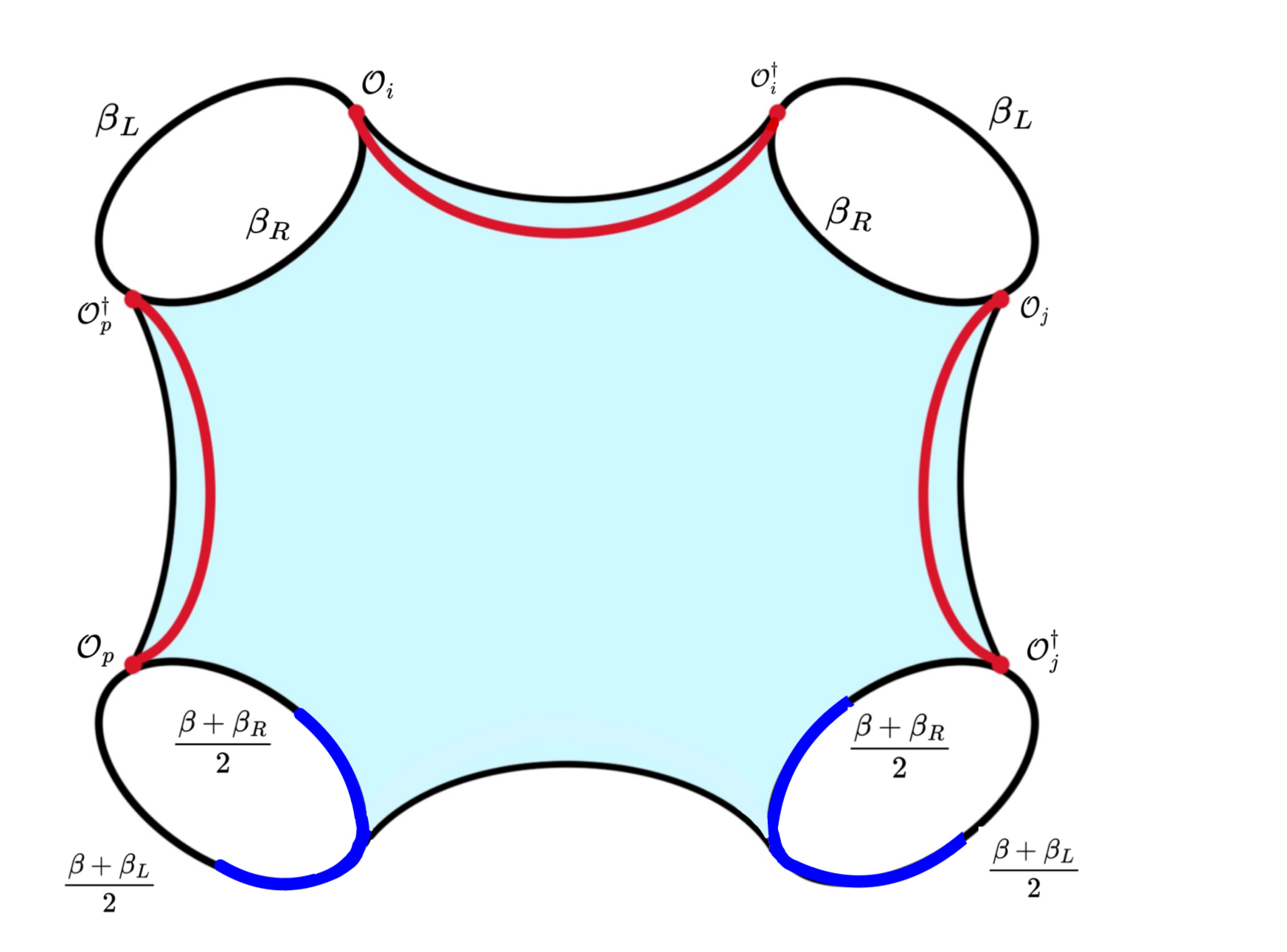}
    \caption{}
    \label{fig:foldedWH}
  \end{subfigure}
  \hfill
\begin{subfigure}[b]{0.45\linewidth}
    \centering
    \includegraphics[width=\linewidth]{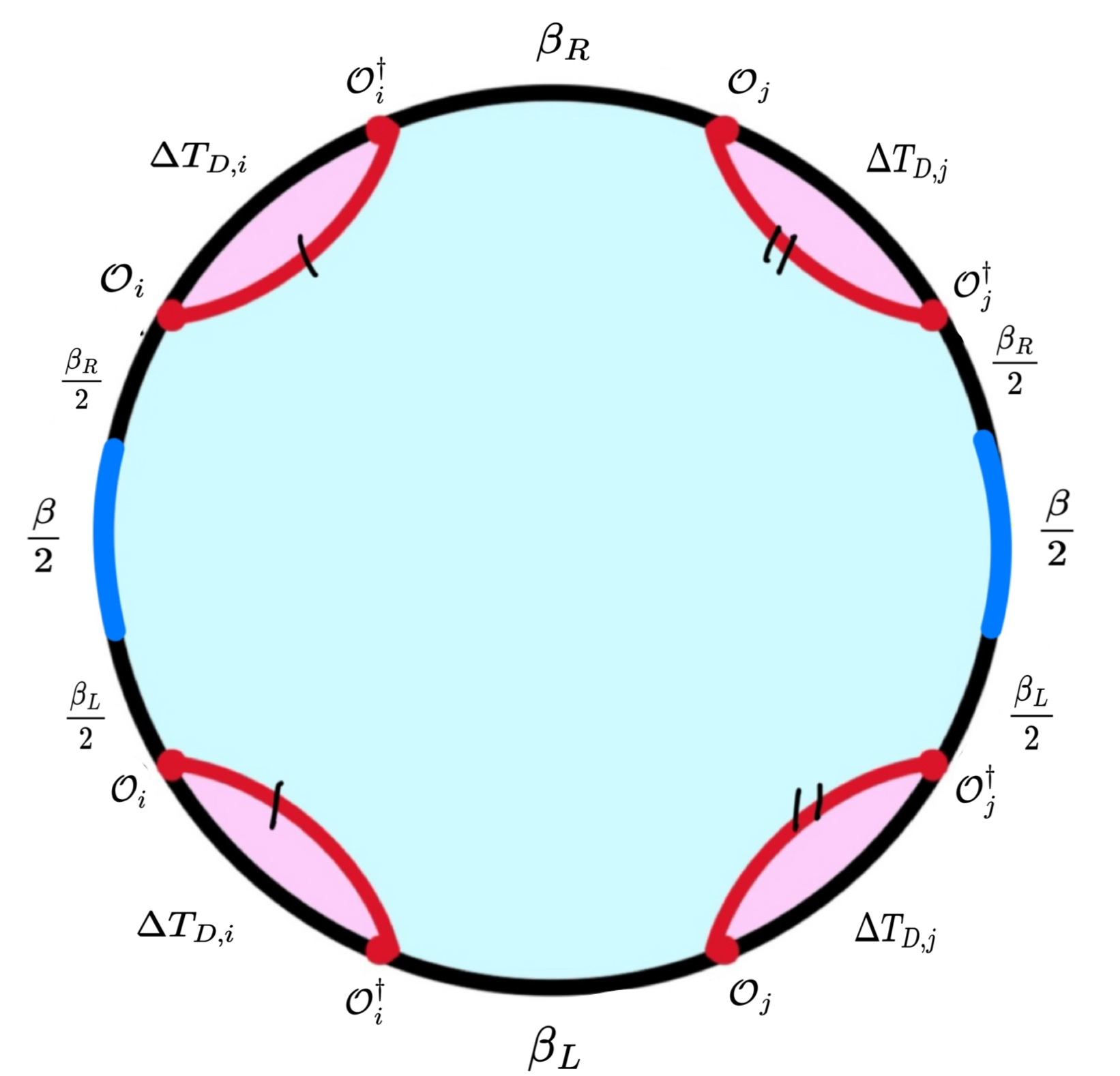}
    \caption{}
    \label{fig:Folded_worm_constr}
    \end{subfigure}
    \caption{({\bf a}) Folded wormhole geometry $\tilde{Z}_{n+2}$,  consisting of $n$ shell boundaries and two tadpole boundaries connected by a single bulk geometry, pictured here for $n=2$. ({\bf b}) Construction of the folded wormhole by gluing portions of the same disk together. To construct the saddle for $\tilde{Z}_{n+2}$, we consider a disk with $2(n+1)$ shell insertions, cut out the pink shell homology regions, and identify the corresponding shells. This procedure is pictured here for $n=1$.}
\end{figure}

In the planar limit the sum over saddles can be performed by considering the fully-connected  saddle geometries connecting two of the tadpoles to $n$ shell boundaries (Fig.~\ref{fig:foldedWH}), which we refer to as a \textit{folded wormhole}. We denote the sum of all folded wormhole saddles as $\tilde{Z}_{n+2}$. These \textit{folded wormholes} are distinct from the pinwheels  explained above and considered in \cite{Balasubramanian:2022gmo,Balasubramanian:2022lnw,Climent:2024trz,Balasubramanian:2024yxk}. Rather, these wormholes are generalizations of the $n=0$ case discussed in Sec.~3 of \cite{Sasieta:2022ksu}.

\paragraph{Folded Wormhole construction} \label{sec:foldedWH}
The  folded wormhole saddles are constructed by identifying shell wordvolumes across a single disk using the junction conditions.  Hence they correspond to saddles of the Z-partition function that are folded in on themselves. To see this we imagine cutting open the folded wormhole in Fig.~\ref{fig:foldedWH} along the shell worldvolumes. This results in a single sheet, which can be obtained by discarding portions of a disk saddle.  In particular, consider the pattern of shell operators insertions on the Euclidean disk boundary in Fig.~\ref{fig:Folded_worm_constr}. The saddle geometries satisfying this boundary condition consists of shells that propagate through bulk and are re-absorbed at the asymptotic boundary after a boundary time  $\Delta T_{D,i}$. A given folded wormhole saddle is then obtained by cutting off the disk regions homologous to the shell worldvolumes (pink in Fig.~\ref{fig:Folded_worm_constr}) and gluing the remaining portion of the disk saddle together along the shell worldvolumes using the Israel junction conditions. In particular, the $\mathcal{O}_i$ shell is identified with the other $\mathcal{O}_i$ shell etc. The shell propagation times $\Delta T_D$ are dynamically determined by the junction conditions such that the resulting wormhole geometry satisfies the equations of motion. In the large mass limit the bulk turning point of the shells approaches the asymptotic boundary, the homology regions pinch off, the propagation times $\Delta T$ all tend to zero and each shell contributes universally to the action. Hence in this limit the total asymptotic boundary length of the pre-gluing disk becomes $(n+1)(\beta_{R}+\beta_{L}) + \beta$. The resulting $\tilde{Z}_{n+2}$ wormhole saddles are therefore products of the Z partition function saddles $\overline{Z}((n+1)(\beta_{R}+\beta_{L}) + \beta)$ and universal shell contributions $\Pi_{i=1}^{n+1}Z_{m_i}$. The sum over fully connected wormhole saddles therefore takes the simple form $
\tilde{Z}_{n+2} = \overline{Z}((n+1)(\beta_{R}+\beta_{L})+\beta)\Pi_{i=1}^{n+1}Z_{m_i}$.  After normalizing the shell states this becomes independent of the shell masses:
\beq \label{eq:foldedaction}
\tilde{Z}_{n+2} = \frac{\overline{Z}((n+1)(\beta_{R}+\beta_{L})+\beta)}{\overline{Z}(\beta_L)^{n+1}\overline{Z}(\beta_R)^{n+1}}.
\eeq
This equality is not restricted to the leading saddlepoint; each $Z$ partition sum on the right side is a sum over saddlepoints. The overline once again indicates that the result is coarse-grained because its is computed by the gravitational path integral.  The left hand side should also therefore be understood in this coarse-grained sense.

\begin{figure}[H]
    \centering
    \includegraphics[width=0.5\linewidth]{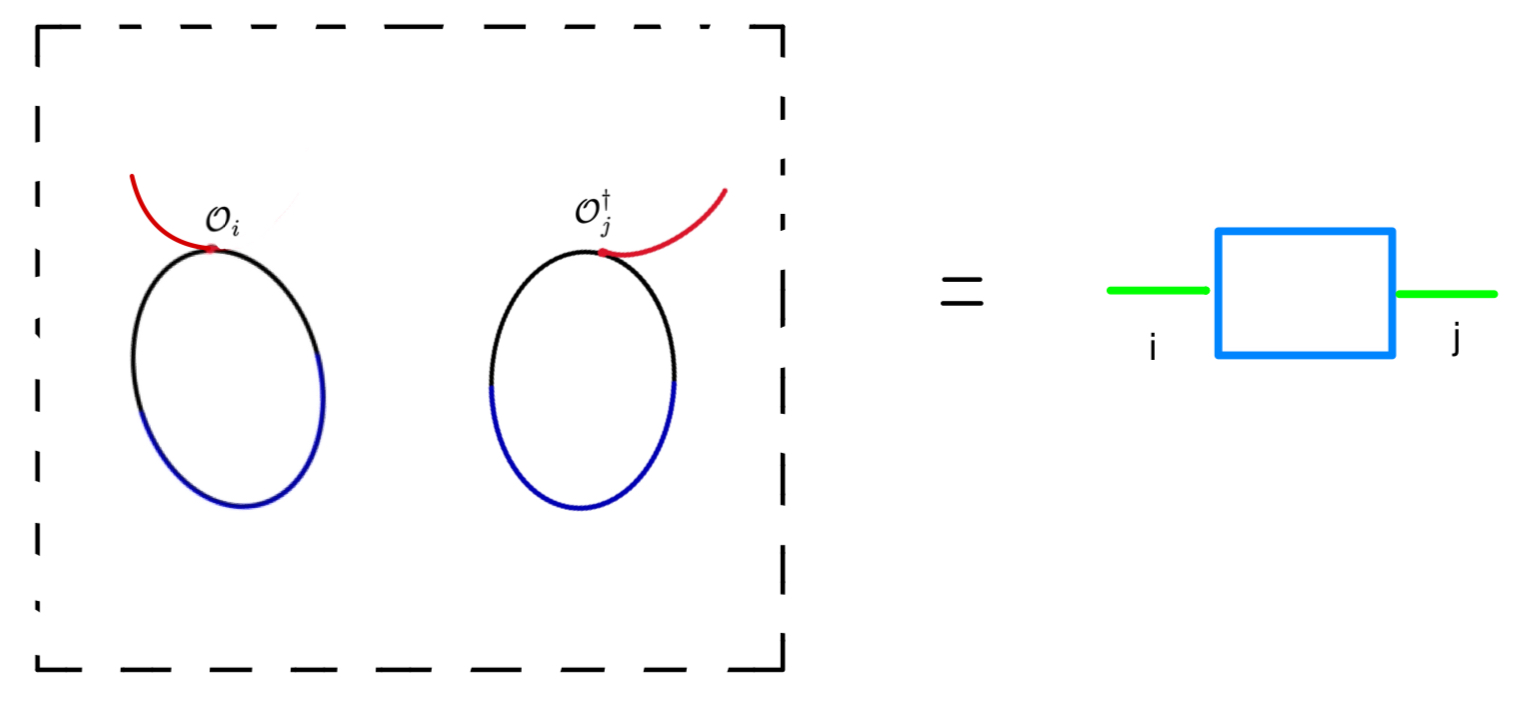}
    \caption{A graphical representation of the two tadpole boundaries $\langle \beta|i\rangle\langle j|\beta\rangle$ in (\ref{eq:Z_resolv}).}
    \label{fig:TFD_bdr_box}
\end{figure}

\begin{figure}[H]
    \centering
    \includegraphics[width=1.1\linewidth]{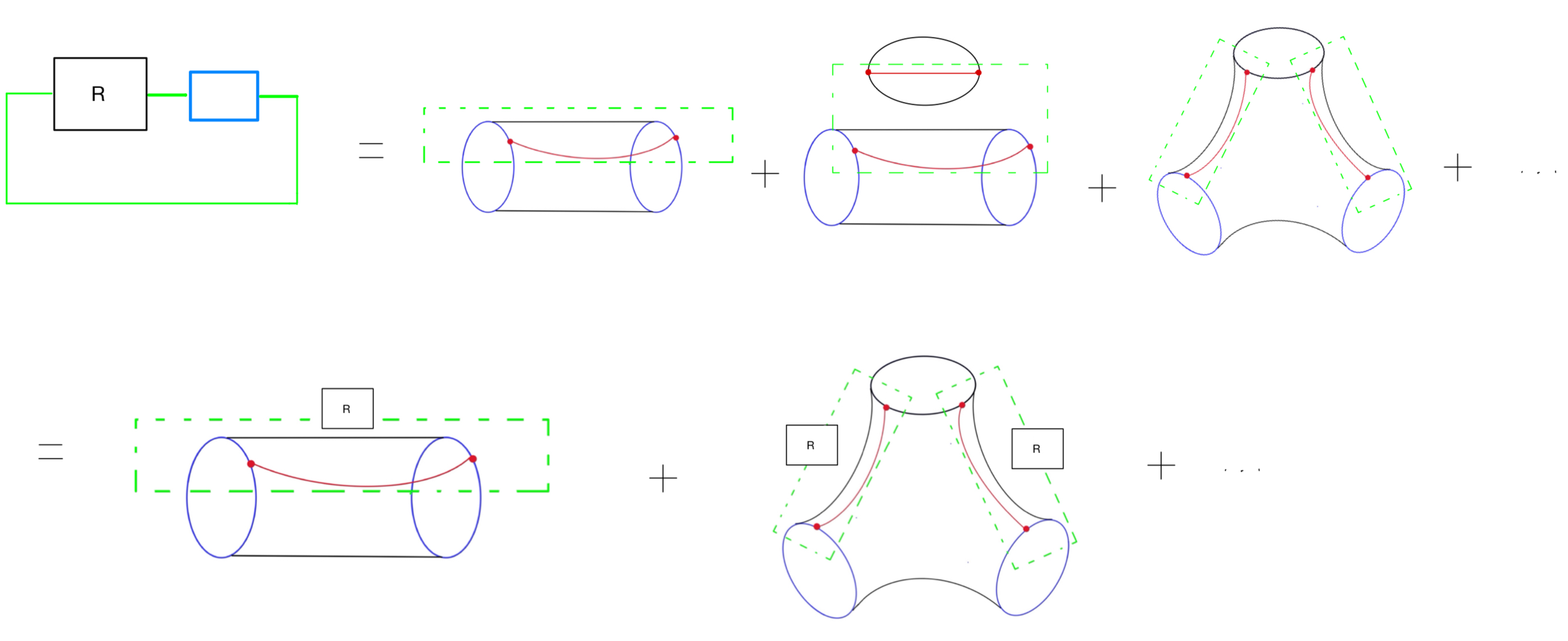}
    \caption{Schematic of the Swinger-Dyson re-summation for (\ref{eq:Z_resolv}). The sum over saddles is re-organized in terms of which shell boundaries are connected to the two tadpole shells. Insertions of $R$ correspond to the trace of the shell state resolvant.}
    \label{fig:Z_resolv_sum}
\end{figure}

\paragraph{Resummation}In the planar limit we can use a resolvent resummation method \cite{Penington:2019kki,Balasubramanian:2022gmo,Balasubramanian:2024yxk,Boruch:2024kvv} to express the sum over all planar saddles for $\overline{G^{n}_{ij}\langle \beta|i\rangle\langle 
j|\beta\rangle}$ 
in terms of just the fully-connected wormhole saddles via a Swinger-Dyson equation. In particular, we define the resolvent matrix 
\beq
R_{ij}(\lambda)= \frac{\delta_{ij}}{\lambda} +\sum_{n=1}^{\infty}\frac{\overline{G^{n}_{ij}}}{\lambda^{n+1}}
\eeq
and re-express the required path integral (\ref{eq:Zid}) as 
\beq \label{eq:Z_resolv}
\overline{G^{n}_{ij}\langle \beta|i\rangle\langle j|\beta\rangle} = \oint_{c_0} \frac{d\lambda}{2\pi i} \lambda^n \overline{R_{ij}(\lambda)\langle \beta|i\rangle\langle j|\beta\rangle}.
\eeq
See \cite{Boruch:2024kvv} for details on the $c_0$ integration contour. As we discussed above, the shell boundaries have two shells each that must be matched up by grouping together the tadpole boundaries. Diagrammatically, we can write the tadpoles as  in Fig.~\ref{fig:TFD_bdr_box}, leading to the schematic resummation in Fig.~\ref{fig:Z_resolv_sum}. 
Thus, in the planar limit, the combinatorics of the resolvent sum reduces to 
\beq \label{resum}
\overline{R_{ij}(\lambda)\langle \beta|i\rangle\langle j|\beta\rangle}= \sum_{n=0}^{\infty} \kappa^{n+1}R^{n+1}\tilde{Z}_{n+2} =\sum_{n=0}^{\infty}\kappa^{n+1}R^{n+1}\frac{\overline{Z}((n+1)(\beta_{R}+\beta_{L}) +\beta)}{\overline{Z}(\beta_L)^{n+1}\overline{Z}(\beta_R)^{n+1}} \,
\eeq
where $R$ is the trace of the resolvent matrix. This simple resummation is possible because the path integral for the wormhole saddles (\ref{eq:foldedaction}) are independent of the masses of the shells in the index loop.

\subsubsection{Microcanonical Projection}\label{sec:microcanon}
The sum (\ref{resum}) cannot be performed analytically due to the $\overline{Z}((n+1)\beta_L + (n+1)\beta_R +\beta)$ factor.
 Even within the leading saddle approximation, at large enough $n$ the asymptotic disk length will be above $\beta_{HP}$, causing the dominant disk saddle to shift from the  large black hole to thermal AdS.  We can evade these difficulties by considering instead the insertion of the projector onto the \textit{microcanonical} shell Hilbert space $\mathcal{H}_{shell,E_L,E_R}$ at fixed $L,R$ energy windows of with $\Delta E_{L,R}$ centered around $E_L,E_R$. As discussed in \cite{Balasubramanian:2024yxk}, microcanonical shell states are obtained by Laplace transform of the canonical shell states.\footnote{This follows by noting the Hamiltonian generates boundary time evolution and assuming the factorisation (\ref{eq:facprob}).} In particular, we can project the two-boundary shell states into $L,R$ microcanonical windows $[E_{L,R}-\frac{\Delta E_{L,R}}{2},E_{L,R}+\frac{\Delta E_{L,R}}{2}] $ by the double Laplace transform:
\beq
\ket{i,\beta_{L,R},E_{L,R}}= \int_{E_L-\frac{\Delta E_L}{2}}^{E_L+\frac{\Delta E_L}{2}} \int_{E_R-\frac{\Delta E_R}{2}}^{E_R+\frac{\Delta E_R}{2}} d\tilde{E}^{L,R}  \, e^{-\frac{\beta_{L,R} }{2}\tilde{E}^{L,R}  } \int d\tilde{\beta}_{L,R}  e^{\tilde{E}^{L,R}  \frac{\tilde{\beta}_{L,R} }{2}} \ket{i,\tilde{\beta}_{L,R}}.
\eeq
Performing the microcanonical projection of (\ref{resum}) in this way we get
\beq \label{eq:mircoproject}
\overline{\langle \beta|\Pi_{\mathcal{H}_{shell,E}}|\beta\rangle} = \oint_{c_0} \frac{d\lambda}{2\pi i} \lambda^n\sum_{n=0}^{\infty}\kappa^{n+1}R^{n+1}\frac{\overline{Z}((n+1)(\beta_L + \beta_R) +\beta)_{E_L=E_R=E} }{(\overline{Z}(\beta_L)_{E_L=E})^{n+1}(\overline{Z}(\beta_R)_{E_R=E})^{n+1}},
\eeq
where we have defined $\overline{Z}(\beta_L)_{E_L}\equiv e^{-E_L\beta_L+\mathbf{S}(E_L)}$ (similarly for $R$), and $\overline{Z}((n+1)(\beta_L + \beta_R) +\beta)_{E_L,E_R} \equiv \delta_{E_L,E_R}e^{-((n+1)(\beta_L + \beta_R)E} e^{-\beta E} e^{\mathbf{S}(E)}$. The $E_L=E_R=E$ constraint in the numerator arises from the fact that the thermo-field double state $|\beta\rangle$ is diagonal in the energy basis, and for ease we have taken $\Delta{E_L}=\Delta{E_R}$. Explicit calculation shows that for energies above a certain threshold the black hole saddle dominates so that  $\mathbf{S}(E)$ is the Bekenstein-Hawking entropy at energy $E$  with small corrections from the  subleading saddles. Below  this threshold $\mathbf{S}(E)=1$ because the thermal saddle dominates. Plugging this into (\ref{eq:mircoproject})  yields: 
\beq
\sum_{n=0}^{\infty}\kappa^{n+1}R^{n+1}\frac{\overline{Z}((n+1)(\beta_L + \beta_R) +\beta)_{E_L=E_R=E} }{(\overline{Z}(\beta_L)_{E_L=E})^{n+1}(\overline{Z}(\beta_R)_{E_R=E})^{n+1}} =e^{-\beta E}e^{\mathbf{S}(E)}\sum_{n=0}^{\infty}\left(\kappa R e^{-2\mathbf{S}(E)}\right)^{n+1} \\
= e^{-\beta E}e^{\mathbf{S}(E)} \frac{\kappa R e^{-2\mathbf{S}(E)}}{1-\kappa R e^{-2\mathbf{S}(E)}}.
\eeq
Putting everything together we obtain
\beq 
\overline{\langle \beta|\Pi_{\mathcal{H}_{shell,E}}|\beta\rangle}= e^{-\beta E}e^{\mathbf{S}(E)} \lim_{n\to -1} \oint_{c_0} \frac{d\lambda}{2\pi i}  \lambda^n  \frac{\kappa R e^{-2\mathbf{S}(E)}}{1-\kappa R e^{-2\mathbf{S}(E)}}.
\eeq
The analytic structure of the shell state resolvent was discussed in detail in  \cite{Balasubramanian:2024yxk}. The upshot is that once $\kappa \geq  e^{\mathbf{S}(E)}$ the resolvent develops a pole at $\lambda=0$ and  we can deform the contour $c_0 \rightarrow c_2$ which encircles $\lambda=0$ clockwise, obtaining: 
\beq
\lim_{n\to -1} \oint_{c_2} \frac{d\lambda}{2\pi i}  \lambda^n   \frac{\kappa R e^{-2\mathbf{S}(E)}}{1-\kappa R e^{-2\mathbf{S}(E)}} = 1 ,
\eeq
and hence, in the saddlepoint approximation:
\beq\label{eq:microZid}
\overline{\langle \beta|\Pi_{\mathcal{H}_{shell,E}}|\beta\rangle} =e^{-
\beta E}e^{\mathbf{S}(E)}= \overline{Z(\beta)}|_{E} \equiv \overline{\langle 
\beta|\beta\rangle}|_{E}.
\eeq

Note that if $\kappa < e^{\mathbf{S}(E)}$, evaluating $\overline{\langle \beta | \Pi_{\mathcal{H}{\text{shell}}} | \beta \rangle}$ requires a detailed analysis of the functional form of the resolvent $R(\lambda)$. Only when $\kappa \geq e^{\mathbf{S}(E)}$ can the projected overlap be computed straightforwardly, yielding equation (\ref{eq:microZid}). We emphasize that the insertion of $G^{n}_{ij} | i \rangle \langle j |$ and the subsequent analytic continuation $n \to -1$ are therefore nontrivial operations, in contrast to a simpler insertion such as $e^{-\frac{1}{a}\hat{\mathcal{O}}}$ followed by the limit $a \to \infty$.

We have found that either the family of would-be TFD states has support only on shell states, or in the microcanonical ensemble any set of $\kappa \geq e^{\mathbf{S}(E)}$ shell states resolves the identity, i.e., (\ref{eq:project1}) is true, and hence spans the complete microcanonical Hilbert space. The latter also demonstrates that the dimension of this Hilbert space is $e^{\mathbf{S}(E)}$, within the leading saddlepoint approximation, consistently with the Bekenstein-Hawking entropy formula. This result implies that the state counting performed in \cite{Balasubramanian:2022lnw,Balasubramanian:2022gmo,Climent:2024trz,Balasubramanian:2024rek} indeed measures the dimension of the full (semiclassical) microcanonical Hilbert space, rather than that of some restricted “shell superselection sector".

\section{Tool 2: Infinite overcompleteness and topological simplification}
\label{sec:tool2}

The result (\ref{eq:microZid}), computed in the saddlepoint approximation of the gravitational path integral, is a coarse-grained statement. In particular, equation (\ref{eq:microZid}) shows that the projector onto the shell states acts as the identity \textit{on average}. In this section, we lift this result to a fine-grained statement that holds in each realization of the ensemble computed by the path integral, while remaining agnostic about the precise nature of this ensemble. We obtained equation (\ref{eq:microZid}) by resumming the planar topologies and performing a microcanonical projection to enable the evaluation of this sum. When extending to fine-grained equalities, however, the topological resummations become intractable. Building on \cite{Balasubramanian:2024yxk,Boruch:2024kvv}, we demonstrate that this difficulty can be overcome by resolving the identity in the limit of an infinitely overcomplete basis, in which only fully connected topologies contribute. We argue that this limit arises naturally in the canonical ensemble and removes the need for a microcanonical projection when evaluating expressions such as (\ref{eq:Zid}). We demonstrate the utility of these methods by re-deriving key results from \cite{Balasubramanian:2022gmo,Balasubramanian:2022lnw,Climent:2024trz,Balasubramanian:2024yxk} directly in the canonical ensemble and without  requiring topological resummation. We then use these methods to provide a more general argument that the shell states span at a fine grained level and proivde some explicit examples. First, we continue the example from the previous section and show the fine-grained equality:

\beq \label{eq:finegrained}
Z(\beta)_{E}=\langle \beta|\Pi_{\mathcal{H}_{shell,E}}|\beta\rangle _{E} \, .
\eeq
As discussed in Sec.~\ref{sec:QTFTPI}, a coarse-grained gravitational path integral equality $\overline{\mathcal{A}}=\overline{\mathcal{B}}$ can be lifted to the fine-grained theory by showing $\overline{(\mathcal{A}-\mathcal{B})^2}=0$ also, from which $\mathcal{A}=\mathcal{B}$ follows. Hence, we want to show
\beq \label{eq:sq}
\overline{\left(Z(\beta)_E- \langle \beta|\Pi_{\mathcal{H}_{shell,E}}|\beta\rangle \right)^2} = 0,
\eeq
where again each term in (\ref{eq:sq}) is given by the Laplace transform of the corresponding canonical term. For example, the term $\overline{Z(\beta)_EZ(\beta)_E}$ is given by the Laplace transform of $\overline{Z(\beta)Z(\beta)}$, which in the saddle point approximation is simply $\overline{Z(\beta)}\times \overline{Z(\beta)}$, and hence  $\overline{Z(\beta)_EZ(\beta)_E}=\overline{Z(\beta)}_E\times \overline{Z(\beta)}_E$. \footnote{For the factorization of $\overline{Z(\beta)} \times \overline{Z(\beta)}$ considered here, we focus on real and positive $\beta$, so the microcanonical saddle points obtained for $\beta \to iT$ in \cite{Saad:2018bqo} do not contribute. In this work, we study shell states inserted on a boundary of topology $\mathbb{S}^{1}_{\beta} \times \mathbb{S}^{d-1}$, with particular emphasis on the case $d = 3$. However, it was shown in \cite{cotler:2020lxj} (see also \cite{Cotler:2022rud,Cotler:2021cqa}) that in the $\mathbb{S}^{1}_{\beta} \times \mathbb{S}^{2}$ and $\mathbb{S}^{1}_{\beta} \times \mathbb{S}^{4}$ cases, the path integral for $\overline{Z(\beta) Z(\beta)}$ admits so-called constrained instanton contributions. These connected wormhole geometries correspond to saddle points subject to the constraint of either fixed length or that the two ADM energies of the boundaries are fixed to be equal, $E_{1} = E_{2} = E$. Although we work within the saddle-point approximation here, it would be interesting to investigate whether these constrained instanton contributions can also be recovered in our framework. This could be achieved by deriving the constrained junction conditions and subsequently gluing the constrained wormholes into a connected contribution to, for example, $\overline{Z(\beta)_{E} \times \langle \beta | \Pi{\mathcal{H}_{\text{shell},E}} | \beta \rangle}$ in the manner depicted in Fig.~\ref{fig:finegrainedwh}.}

Similarly $\overline{Z(\beta)\langle \beta|\Pi_{\mathcal{H}_{shell}}|\beta\rangle}=\overline{Z(\beta)} \times \overline{\langle \beta|\Pi_{\mathcal{H}_{shell}}|\beta\rangle}$ , as there are no wormholes saddles connecting the $Z(\beta)$ boundary to those of $\langle \beta|\Pi_{\mathcal{H}_{shell}}|\beta\rangle$ 
because the $Z(\beta)$ boundary condition contains no shell matter insertions. Hence by the Laplace transform, and using (\ref{eq:microZid}),  $\overline{Z(\beta)_E\langle \beta|\Pi_{\mathcal{H}_{shell,E}}|\beta\rangle}=\overline{Z(\beta)}_E\times \overline{Z(\beta)}_E$.

Thus showing (\ref{eq:sq}) in the saddle approximation only requires showing $\overline{ \langle \beta|\Pi_{\mathcal{H}_{shell}}|\beta\rangle)^2} =\left(\overline{Z(\beta)}\right)^2$. This is  nontrivial as the sum over saddle geometries for 
\beq \label{eq:sqprojectedbeta}
\overline{(\langle \beta|\Pi_{\mathcal{H}_{shell}}|\beta\rangle)^2} = \lim_{n,m \to -1} \overline{G^{n}_{ij} \langle \beta|i\rangle\langle j|\beta\rangle G^{m}_{kl}\langle \beta|k\rangle\langle l|\beta\rangle} 
\eeq
contains planar geometries connecting boundaries from the two $\langle\beta|\Pi_{\mathcal{H}_{shell}}|\beta\rangle$ insertions together in the bulk.  Such planar geometries break the two maximal index loops $G^{n}_{ij} \langle \beta|i\rangle\langle j|\beta\rangle $ and $G^{m}_{kl}\langle \beta|k\rangle\langle l|\beta\rangle$ into multiple non-maximal loops, making a Swinger-Dyson re-summation intractable.

An analogous problem was dealt with in \cite{Balasubramanian:2024yxk,Boruch:2024kvv} by considering the $\kappa \to \infty,\frac{\kappa}{e^{1/G_{N}}} \to \infty $ limit.  The calculations considered in this work involve gravitational path integrals with some index loop $\overline{G^{n}_{ij}M_{ji} \cdots}$ where $M_{ji}$ is a quantity with two shell indices. This sum over saddle geometries includes fully-connected, partially-connected and disconnected contributions. Since the shell sourced by $\mathcal{O}_i$ can only propagate in the bulk if absorbed by $\mathcal{O}^{\dagger}_i$, for every index loop implied by the boundary conditions the saddlepoint geometry must have a connected component that joins the associated bra and ket.  Any saddlepoint that breaks the index loop will be suppressed by $1/\kappa$.   Hence in the strict $\kappa \to \infty,\frac{\kappa}{e^{1/G_{N}}} \to \infty $ limit only wormhole geometries that support  maximal index loops  contribute.    We will refer to this as the $\kappa \to \infty$ or large $\kappa$ limit.  Note that this does not mean that only the fully connected geometry contributes in this limit.  If there are multiple maximal loops, each loop just has to appear in a connected component of the saddlepoint geometry.


Putting everything together, in the large $\kappa$ limit the two maximal index loops in (\ref{eq:sqprojectedbeta}) must be unbroken, leading to $\overline{(\langle \beta|\Pi_{\mathcal{H}_{shell,E}}|\beta\rangle)^2}= \left(\overline{\langle \beta|\Pi_{\mathcal{H}_{shell,E}}|\beta\rangle}\right)^2 =\left(\overline{Z(\beta)}_{E}\right)^2$ within the saddlepoint approximation. This establishes  (\ref{eq:finegrained}). 
Note that even in the models that offer a great deal of analytic control like JT gravity, the large $\kappa$ limit was needed to extend coarse-grained statements to fine-grained ones, see \cite{Boruch:2024kvv}.

\paragraph{Interpretation of the large $\kappa$ limit}
Since $\kappa$ is the size of the candidate basis, not a parameter of the underlying
theory, we delineate what depends on it. The physical quantities --- the resolution of
the identity, the Hilbert space dimension, the factorisation below --- carry no
$\kappa$-dependence; $\overline{Z(\beta)}$ makes no reference to $\kappa$ at all. The limit
only simplifies the sum over topologies needed to compute them. This works differently in
the two ensembles. Microcanonically $\mathcal{H}_{LR}|_E$ is finite-dimensional with
$\dim\mathcal{H}_{LR}|_E=e^{\mathbf{S}(E)}$, and any $\kappa\geq e^{\mathbf{S}(E)}$ already
resolves the identity: the resolvent pole of Sec.~\ref{sec:microcanon} appears exactly at
$\kappa=e^{\mathbf{S}(E)}$, and beyond it added states are linearly dependent, leaving
$\Pi_{\mathcal{H}_{shell}}=\mathds{1}_{\mathcal{H}_{LR}|_E}$ unchanged. There is thus a sharp
threshold above which the result is strictly $\kappa$-independent. Canonically
$\mathcal{H}_{LR}$ is infinite-dimensional and admits no finite spanning set, so
$\Pi_{\mathcal{H}_{shell}}$ is a finite-rank projector that depends on $\kappa$ at any finite
value --- $\overline{\langle\beta|\Pi_{\mathcal{H}_{shell}}|\beta\rangle}$ is strictly below
$\overline{\langle\beta|\beta\rangle}$ and approaches it only as $\kappa\to\infty$. Here
$\kappa\to\infty$ is not an optional strengthening but the natural basis size.

What survives the limit is not a single geometry but a single {connectivity class},
the folded wormhole, itself a sum over all $Z$-partition saddles (large black hole, thermal
AdS, small black hole). Nor are wormholes switched off: the limit suppresses the
{less}-connected topologies. In the example
$\overline{\mathrm{Tr}\,G^2}=\kappa^2 Z_2+\kappa Z_1^2$ of Sec.~\ref{sec:resolvant} the
wormhole $Z_2$ is enhanced by $\kappa^2$ while the disconnected $Z_1^2$ carries the relative
$1/\kappa$; the rule is combinatorial, each broken maximal index loop costing a power of
$1/\kappa$. Taking $\kappa\to\infty$ is merely the economical route to the
$\kappa$-independent answer: above threshold the resolvent resummation of
Sec.~\ref{sec:resolvant} retains and sums the partially-connected topologies and yields the
same content. In this respect the limit is unlike the large-$N$ limit, where $1/N$ is a
physical coupling organising a new expansion; here it is a representation choice, as in
resolving the identity with an overcomplete set of coherent states.

\subsection{Canonical path integrals} \label{sec:canon}
While the microcanonical example above illustrates  the main methods, the Laplace transform prescription for obtaining microcanonical states requires some assumptions about the path integral and its meaning.  Additionally, we are interested in obtaining results about the entire Hilbert space, not just a microcanonical band, and would prefer to avoid subtleties in sewing these bands together to build the complete Hilbert space. Thus for the remainder of this paper we consider canonical path integrals and aim to derive canonical fine-grained equality:
\beq \label{eq:finegrainZinsert}
Z(\beta)=\langle 
\beta|\Pi_{\mathcal{H}_{shell}}|\beta\rangle.
\eeq
For the reasons outlined above we therefore consider $ \overline{G^{n}_{ij}\langle \beta|i\rangle\langle j|\beta\rangle}$ in the $\kappa\to \infty$ limit. 
There is only one index loop here, and thus only fully connected saddles contribute as $\kappa\to \infty$. These geometries are precisely the $\tilde{Z}_{n+2}$ folded wormholes of Sec.~\ref{sec:foldedWH}. We therefore directly obtain:
\beq
\lim_{n\to -1} \overline{G^{n}_{ij}\langle \beta|i\rangle\langle j|\beta\rangle} = \lim_{n\to -1}\kappa^{n+1} \tilde{Z}_{n+2} \, .
\label{eq:result1-4.1}
\eeq
Using (\ref{eq:foldedaction}) this gives
\beq \label{eq:idinZ}
=\lim_{n\to -1}\kappa^{n+1} \frac{\overline{Z}((n+1)(\beta_{R}+\beta_{L})+\beta)}{\overline{Z}(\beta_L)^{n+1}\overline{Z}(\beta_R)^{n+1}}=\overline{Z(\beta)}.
\eeq
Hence the path integral equality (\ref{eq:Zid}) holds in the sum over saddlepoints.

We should be careful with the $\kappa \rightarrow \infty$ limit beause of the the $\kappa^{n+1}$ factors in Eqs.~\ref{eq:result1-4.1} and \ref{eq:idinZ}. Formally we will consider the theory to have some very high energy cut off, rendering the dimension of $\mathcal{H}_{LR}$ and therefore  $\kappa, \frac{\kappa}{e^{1/G_{N}}}$ very large but finite. After taking the $n \to -1$ limit  the cutoff can be taken to infinity upon which the $\mathcal{O}(1/\kappa)$ terms vanish. This order of limits makes (\ref{eq:idinZ}) well defined. The extension to the fine-grained statement via $\overline{(Z(\beta)- \langle \beta|\Pi_{\mathcal{H}_{shell}}|\beta\rangle)^2}=0$ now  follows trivially, showing the fine-grained relation (\ref{eq:finegrainZinsert}), in the path integral approximation where we sum overall saddles. This suggests the equality of the full Hilbert spaces $\mathcal{H}_{LR}=\mathcal{H}_{shell}$. 

Previous work, as discussed in Sec.~\ref{sec:resolvant}, attempted to perform resolvent resummations like (\ref{resum}) with the canonical states at large but finite $\kappa$. This resulted in infinite sums that could not be computed analytically.  Therefore the authors worked instead with the microcanoncal projections.  From our perspective, the canonical shell states at finite $\kappa$ cannot span $\mathcal{H}_{LR}$; they instead some complicated hard to probe sub-space, that becomes tractable only after a microcanonical projection.   In the in $\kappa \to \infty$ limit this subspace grows to span the entire Hilbert space $\mathcal{H}_{LR}$, with the added advantage that the sum over topologies simplifies drastically.

\paragraph{Note on instabilities}
In the large shell mass limit the wormhole saddles are proportional those of the Z partition function $\overline{Z(\beta)}$. In the AdS case  the Z-partition function has 3 saddles: the large black hole, thermal AdS and the small black hole. Dominance between the large BH and thermal AdS saddle is determined by the Hawking-Page transition while the small BH is always subleading. Recently, \cite{Barbon:2025bbh} discussed the inabilities of the small BH due to negative modes contributing to the one-loop determinant around this saddle. Such instabilities can challenge the interpretation of the shell-wormhole saddles computing coarse-grained overlaps of a positive-definite inner product. However, in the canonical ensemble the small black holes is never the dominant saddle, and after taking all the saddles into account $\overline{Z(\beta)}$ is still positive-definite. Furthermore \cite{Barbon:2025bbh} showed that in a careful treatment of the microcanonical ensemble these issues can also be resolved. Hence, we will not consider such instabilities in this work. 

\subsubsection{Semiclassical factorisation of the trace in the large $\kappa$ limit} \label{sec:facSC}
To further stress the simplifying power of the large $\kappa$ limit, we economically derive the factorisation of the thermal partition function of $\mathcal{H}_{shell}$:
\beq \label{eq:factorisation}
Tr_{\mathcal{H}_{shell}}(e^{-\tilde{\beta}_LH_{L}}e^{-\tilde{\beta}_RH_{R}}) = Z(\beta_1)\times Z(\beta_2)
\eeq
shown in \cite{Balasubramanian:2024yxk}, without a microcanonical projection and resolvent resummation. As discussed in Sec.~\ref{hamiltonian}, the action of the operators $e^{-\tilde{\beta}_{L,R}H_{L,R}}$ on the states in $\mathcal{H}_{LR}$ is to append boundary cylinders of length $\tilde{\beta}_{L,R}$ respectively to the $L,R$ cuts defining state. Defining $k_L = e^{-\tilde{\beta}_LH_{L}}$ and $k_R = e^{-\tilde{\beta}_RH_{R}}$ and using (\ref{eq:trace}) we obtain
 
\beq \label{eq:shelltr}
\overline{Tr_{\mathcal{H}_{Shell}}(k_Lk_R)}=\overline{G^{-1}_{ij}\langle i|k_Lk_R|j\rangle}\equiv\lim_{n \to -1} \overline{G^{n}_{ij}\langle j|k_Lk_R|i\rangle}.
\eeq
 
\begin{figure}[h]
     \centering
     \includegraphics[width=0.4\linewidth]{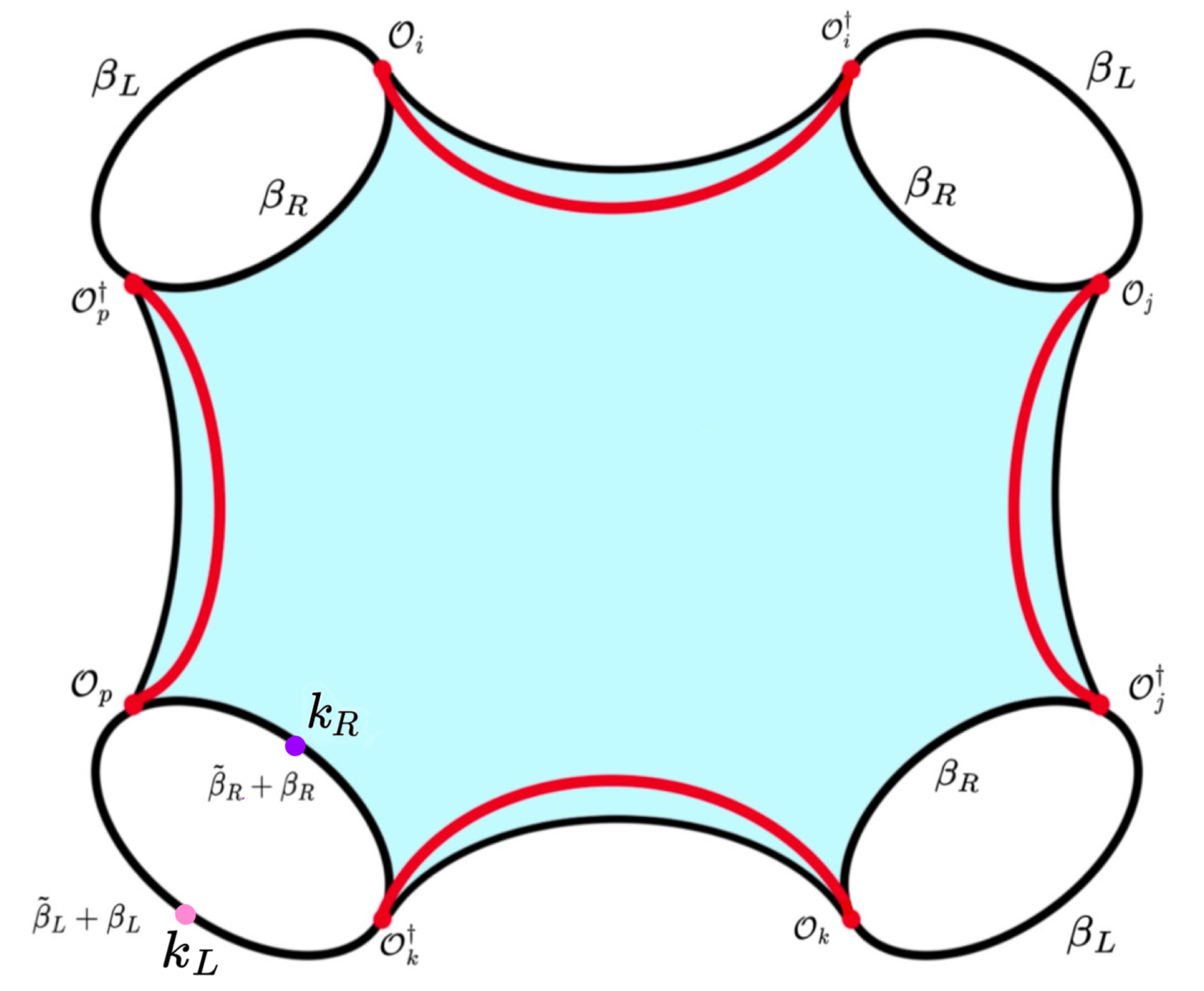}
     \caption{Fully-connected trace pinwheel geometry consisting of $n+1$ asymptotic shell boundaries, one of which is of $L,R$ size $\beta_L+\tilde{\beta}_L$ and $\beta_R+\tilde{\beta}_R$ respectively, pictured for $n=3$.}
     \label{fig:tracewheel}
 \end{figure}

We again consider the $\kappa \to \infty$ limit in which only fully-connected wormhole geometries contribute to $\overline{G^{n}_{ij}\langle i|k_Lk_R|j\rangle}$. We will call the geometry at a given $n$ the \textit{trace pinwheel} $Z_{n+1}(k_L,k_R)$ with $n+1$ shell boundaries, depicted in Fig.~\ref{fig:tracewheel}:
   \beq \label{fullconnect}
 \lim_{n \to -1} \overline{G^{n}_{ij}\langle j|k_Lk_R|i\rangle} =  \lim_{n \to -1}\kappa ^{n+1}Z_{n+1}(k_L,k_R).
 \eeq

\begin{figure}[h]
  \begin{subfigure}{\linewidth}
        \centering
        \includegraphics[width=0.7\linewidth]{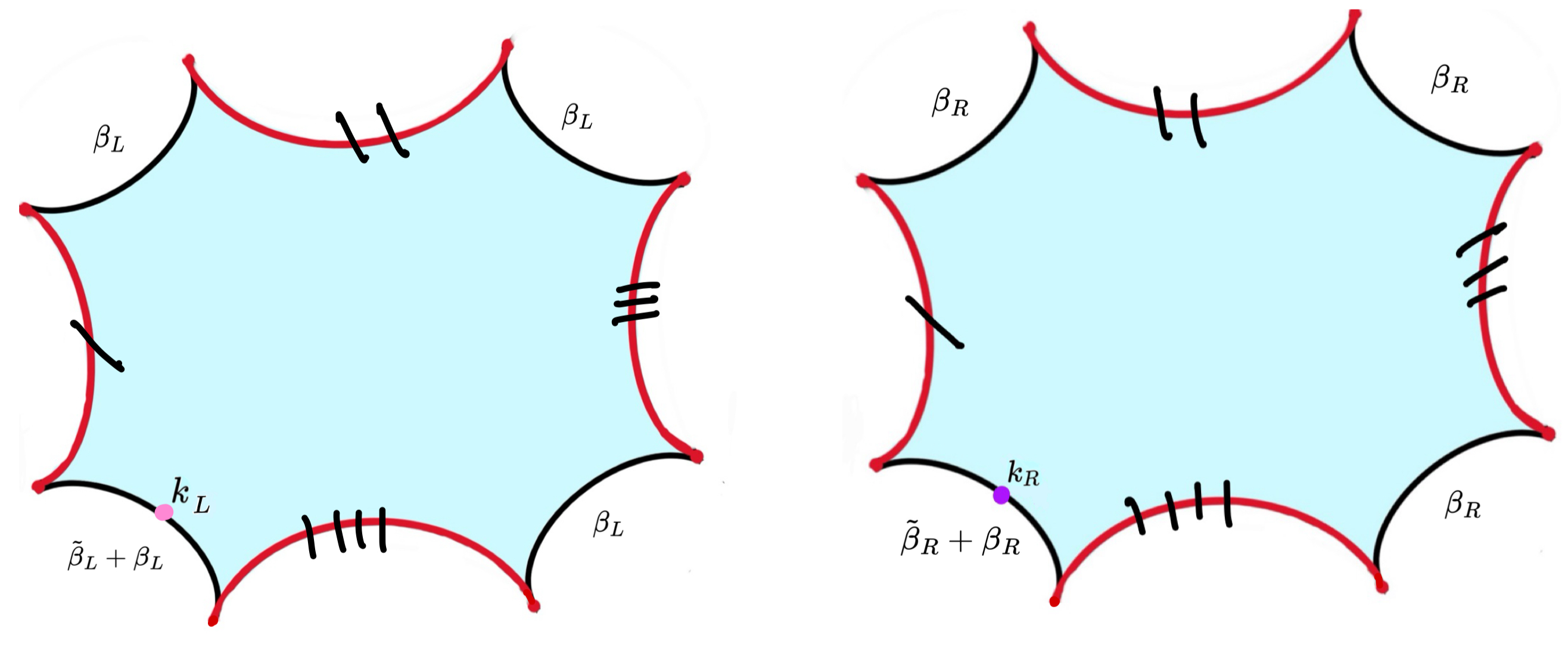}
         \caption{}
         \label{fig:Tr_pinhweel_sheets}
 \end{subfigure}
      \centering
        \begin{subfigure}{\linewidth}
            \centering
             \includegraphics[width=0.7\linewidth]{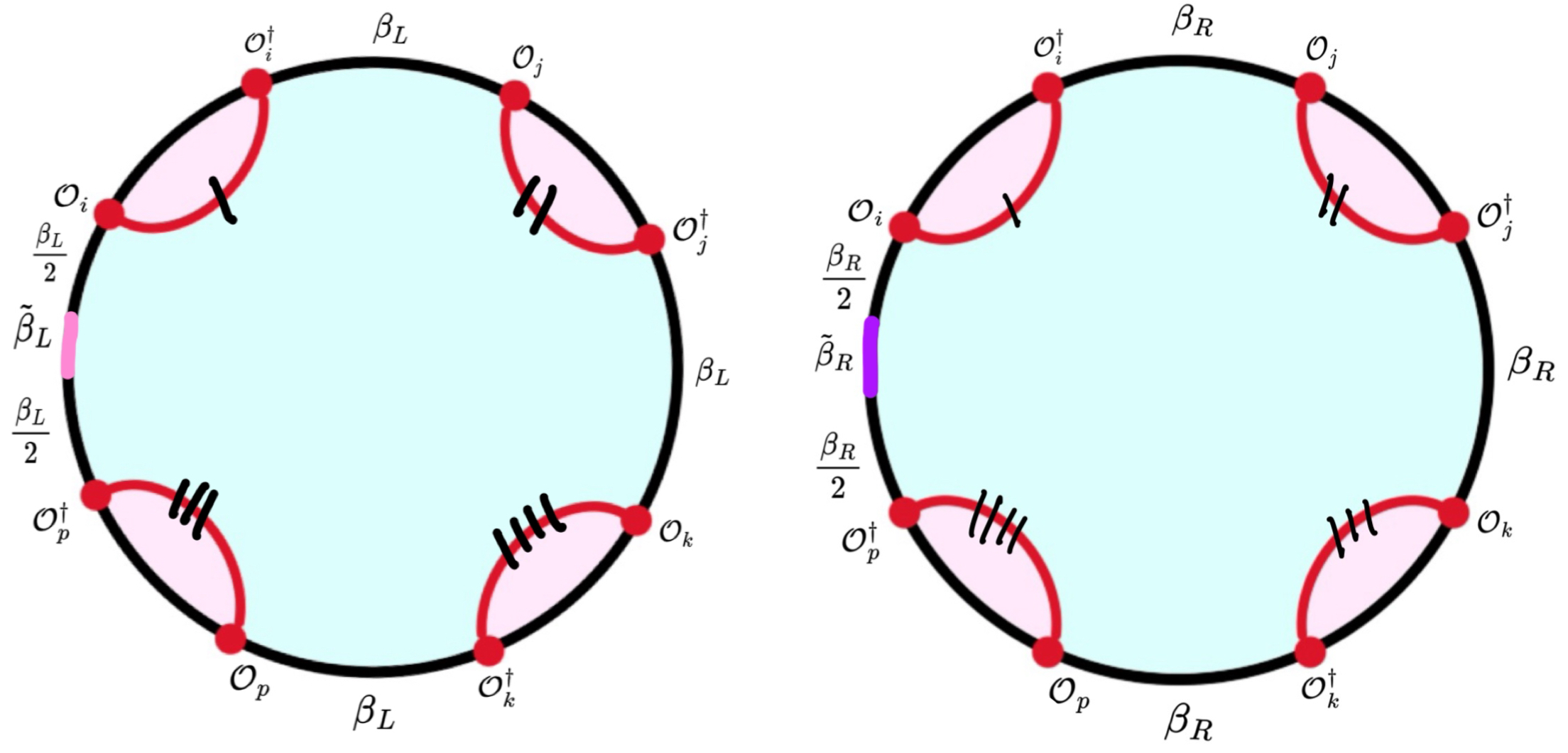}
            \caption{}
            \label{fig:Tr_wheel_disk}
    \end{subfigure}
     
 \caption{({\bf a}) Sheet diagrams of the trace pinwheel in Fig.~\ref{fig:tracewheel} obtained by cutting along the shell world volumes. ({\bf b}) Construction of the pinwheel from the disk by cutting out the purple homology regions and gluing together the remainder by identifying the shell world volumes, leaving behind the sheets of Fig.~\ref{fig:Tr_pinhweel_sheets} }
 \label{fig:Tr_wheel_disk }
 \end{figure}

The construction of the trace pinwheel saddles is nearly identical to the $n+1$ pinwheel saddles discussed in Sec.~\ref{sec:shellwh} apart from the boundary insertion $\langle j|k_Lk_R|i\rangle$ consisting of $\mathcal{O}^{\dagger}_{j}$ and $\mathcal{O}_{i}$ insertions separated by $L,R$ times $\beta_L+\tilde{\beta}_L$  and $\beta_R+\tilde{\beta}_R$ respectively. Hence on the  $L$ sheet diagram  $\mathcal{O}^{\dagger}_{n+1},\mathcal{O}_{1} $  are separated by a boundary time $\beta_L+\tilde{\beta}_L$ and similarly for the $R$ sheet (see Fig.~\ref{fig:Tr_pinhweel_sheets}). The wormhole saddles are constructed by gluing the two disks in Fig.~\ref{fig:Tr_wheel_disk} via the junction conditions in a manner identical to that in Sec.~\ref{sec:shellwh}. In the large shell mass limit we obtain:

\beq
Z_{n+1}(k_L,k_R)= \frac{\overline{Z}((n+1)\beta_L + \tilde{\beta}_L)\overline{Z}((n+1)\beta_R + \tilde{\beta}_R)}{\overline{Z}(\beta_L)^{n+1}\overline{Z}(\beta_R)^{n+1}},
\eeq
  
and hence

\beq \label{eq:SCTrfac}
\overline{Tr_{\mathcal{H}_{Shell}}(e^{-H_{L}\tilde{\beta}_L}e^{-H_{R}\tilde{\beta}_R})}= \lim_{n \to -1} \overline{G^{n}_{ij}\langle j|k_Lk_R|i\rangle} = \overline{Z}(\tilde{\beta}_L) \times \overline{ Z}(\tilde{\beta}_R).
\eeq

Repeating this argument to show $\overline{\left(Tr_{\mathcal{H}_{Shell}}(e^{-H_{L}\tilde{\beta}_L}e^{-H_{R}\tilde{\beta}_R})-Z(\tilde{\beta}_L) \times  Z(\tilde{\beta}_R)\right)^2}=0$ is now straightforward because there are no saddles that interpolate between $Z$ and $Tr$ boundaries as $Z$ contains no shell insertions.  This establishes the fine-grained result (\ref{eq:factorisation}) in the sum over saddles.
The simplicity of this calculation as compared to the finite $\kappa$ microcanonical methods of \cite{Balasubramanian:2024yxk} will be crucial for extending the result away from the semiclassical limit in Sec.~\ref{sec:Trzz}. 

\paragraph{State counting} \label{statecount}
In general terms, the micro-canonical density of states of a theory is obtained by the Laplace transform of the thermal partition function. The left hand side of (\ref{eq:SCTrfac}) is by definition the thermal partition function of the shell Hilbert space. Adopting the results of Sec.~\ref{sec:microcanon} for the Laplace transform of the Z-partition functions on the RHS of (\ref{eq:SCTrfac}), we find   $dim(\mathcal{H}_{Shell,E})|=e^{\mathbf{S}_{L}(E_L)}e^{\mathbf{S}_{R}(E_R)}$.
Therefore, as a corollary of (\ref{eq:SCTrfac}) we can count the dimension of $\mathcal{H}_{shell,E}$ after taking $\kappa \to \infty$ without having to deal with the subtleties of micro-canonically projecting the shell states themselves. Furthermore, this derivation can be performed without taking $m_s \to \infty$ limit, which  played an important role in the state-counting arguments of \cite{Balasubramanian:2022gmo,Balasubramanian:2022lnw,Climent:2024trz}. Away from this limit shell propagation times $\Delta T_{D,i}>0$ are  determined by the junction conditions.  However the details turn out to be unimportant:
\beq 
\overline{Tr_{\mathcal{H}_{shell}}(e^{-H_{L}\tilde{\beta}_L}e^{-H_{R}\tilde{\beta}_R})}= \lim_{n \to -1} \kappa^{n+1} \frac{(n+1)\beta_L+ \tilde{\beta}_L}{(n+1)\beta_L + \tilde{\beta}_L+ \sum_{i=1}^{n+1}\Delta T_{i,L}} \overline{Z}((n+1)\beta_L + \tilde{\beta}_L+ \sum_{i=1}^{n+1}\Delta T_{i,L}) \times
\\
\frac{(n+1)\beta_R+ \tilde{\beta}_R}{(n+1)\beta_R + \tilde{\beta}_R+\sum_{i=1}^{n+1}\Delta T_{i,R}} \overline{Z}((n+1)\beta_R + \tilde{\beta}_R+\sum_{i=1}^{n+1}\Delta T_{i,R}) \times \Pi^{n+1}_{i} Z_{shell,i} = \overline{Z}_{L}(\tilde{\beta}_L)\overline{Z}_{R}(\tilde{\beta}_R)
\eeq
where now the shell contributions $Z_{shell,i}$ no longer cancel at general $n$ by normalizing the shell states because for finite shell masses they will depend on the trajectory followed by the shell in the different saddlepoint geometries.  Formally, however, in the $n\to-1$ limit the shell contributions are  absent, giving the product of left and right partition sums on the right side. We will return in later sections to examine the sleight of hand of naively taking $n\to -1$.

\subsection{A semiclassically complete basis} \label{sec:scspan}
In Sec.~\ref{sec:microcanon} and Sec.~\ref{sec:canon} above we have shown that the microcanonical and canonical versions of the state $\ket{\beta}$ are in the span of a sufficiently large set of shell states \textit{for any value of} $\beta$. By noting that the Hamiltonian of the gravity theory generates boundary time evolution and assuming the factorisation (\ref{eq:facprob}) , it follows that the states $\ket{\beta}$ are in fact the TFD states of the gravity theory $\ket{\beta}=\sum_{n} e^{\frac{-\beta E_{n}}{2}} |E_n\rangle _{L}|E_n\rangle_R$ .\footnote{ We show this factorisation in \cite{Balasubramanian:2025zey}}  Hence we learn that the shell states span the diagonal subspace $\ket{n}_L \ket{n} _R$. This is somewhat surprising, as by the same logic the shell states correspond to PETS states ($|\mathcal{O}\rangle = \sum_{n,m}e^{-(\beta_{L}E_{n} +\beta_{R}E_{m})}\mathcal{O}_{mn}|E_n\rangle_L |E_m\rangle_R$) which have nonzero support on many off diagonal elements too. While this supports the claim that the shell states form a basis, we would like to strengthen this argument by showing other large classes of states are also in this span. To do so, we outline the structure of the argument in this section, and then work out a few examples in Sec.~\ref{sec:examples}.

The equality $\mathcal{H}_{shell}=\mathcal{H}_{LR}$ can be derived by showing  
 \beq \label{eq:projectid}
 \langle \Psi|\Psi\rangle=\langle \Psi|\Pi_{\mathcal{H}_{Shell}}|\Psi\rangle  
 \eeq
for $\forall \, |\Psi\rangle \in \mathcal{H}_{LR}$. By definition all states in $\mathcal{H}_{LR}$ are states made by cutting the path integral or superpositions thereof, see Sec.~\ref{sec:QTFTPI}. The former can be written in the notation $|\Psi\rangle= |e^{-\frac{\tilde{\beta}_LH_L}{2}}\mathcal{O}_{\Psi}e^{\frac{-\tilde{\beta}_R H_R}{2}}\rangle$ where the operator $\mathcal{O}_{\Psi}$ may be non-local, and we define $\langle \Psi|\Psi\rangle=Z_{\mathcal{O}_{\Psi}}(\tilde{\beta}_L + \tilde{\beta}_R)$. To derive the fine-grained statement (\ref{eq:projectid}) via the gravitational path integral we require that
\beq \label{eq:idinsert}
\overline{\langle \Psi|\Pi_{\mathcal{H}_{Shell}}|\Psi\rangle}=\overline{\langle \Psi|\Psi\rangle}
\eeq
and
\beq \label{eq:idsqr}
\overline{(\langle \Psi|\Pi_{\mathcal{H}_{Shell}}|\Psi\rangle-\langle \Psi|\Psi\rangle)^2}=0.
\eeq
If (\ref{eq:projectid}) holds for all states $|\Psi\rangle \in \mathcal{H}_{LR}$, it follows from the Cauchy-Schwarz identity that $\mathcal{H}_{Shell}=\mathcal{H}_{LR}$.  
 
In this section we argue that (\ref{eq:projectid}) holds (in the $\kappa\to\infty$ limit)  for the subset of states  $|\Psi\rangle \in \mathcal{H}_{LR}$  for which the path integral $\overline{\langle \Psi|\Psi\rangle}$ has at least some saddlepoint. By extension (\ref{eq:projectid})  then also holds for arbitrary superpositions of these states. If one supposes that there exists \textit{some} basis set for $\mathcal{H}_{LR}$ consisting of states whose norms are dominated by a saddlepoint, then our argument shows $\mathcal{H}_{shell}=\mathcal{H}_{LR}$.  We are not seeking to prove this assumption here, but give some heuristic arguments in Sec.~\ref{sec:completebasis}.
The shell states discussed here are analogous to coherent states, which generically provide a non-orthonormal, over-complete basis for the Hilbert space. 

\begin{figure}[h]
    \centering
    \includegraphics[width=0.4\linewidth]{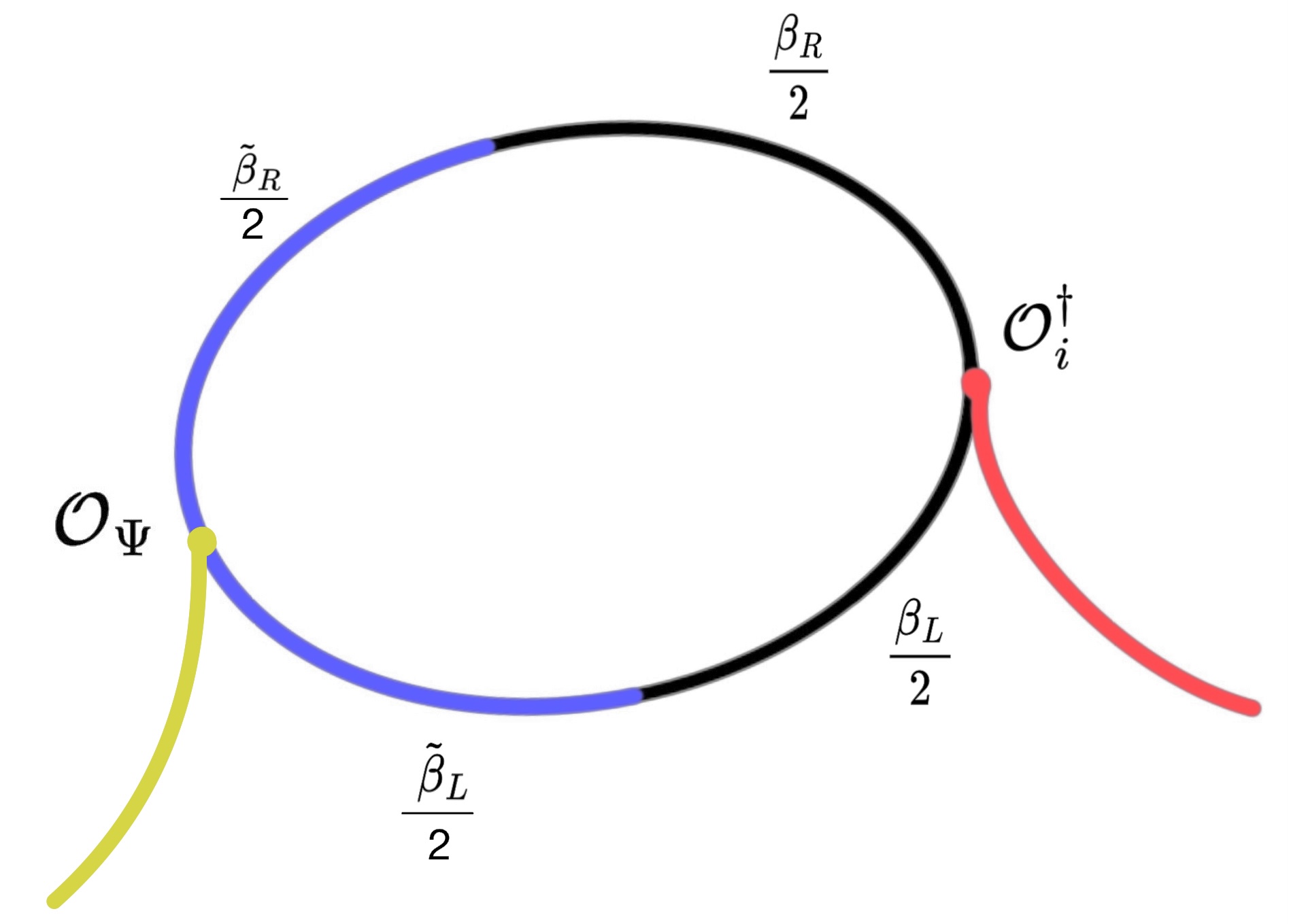}
    \caption{So-called \textit{operator boundary} condition for $\langle i|\Psi\rangle$, which consist of the $\mathcal{O}_{\Psi}$ and $\mathcal{O}^{\dagger}_{i}$ shell insertion separated by $L,R$ times  $\frac{\beta_{L,R} +\tilde{\beta}_{L,R}}{2}$  respectively. The stress-energy from the $\mathcal{O}_{\Psi}$ operator is depicted in the bulk as a dotted yellow line.}
    \label{fig:Obdry}
\end{figure}

\begin{figure}[h]
    \centering
    \includegraphics[width=1\linewidth]{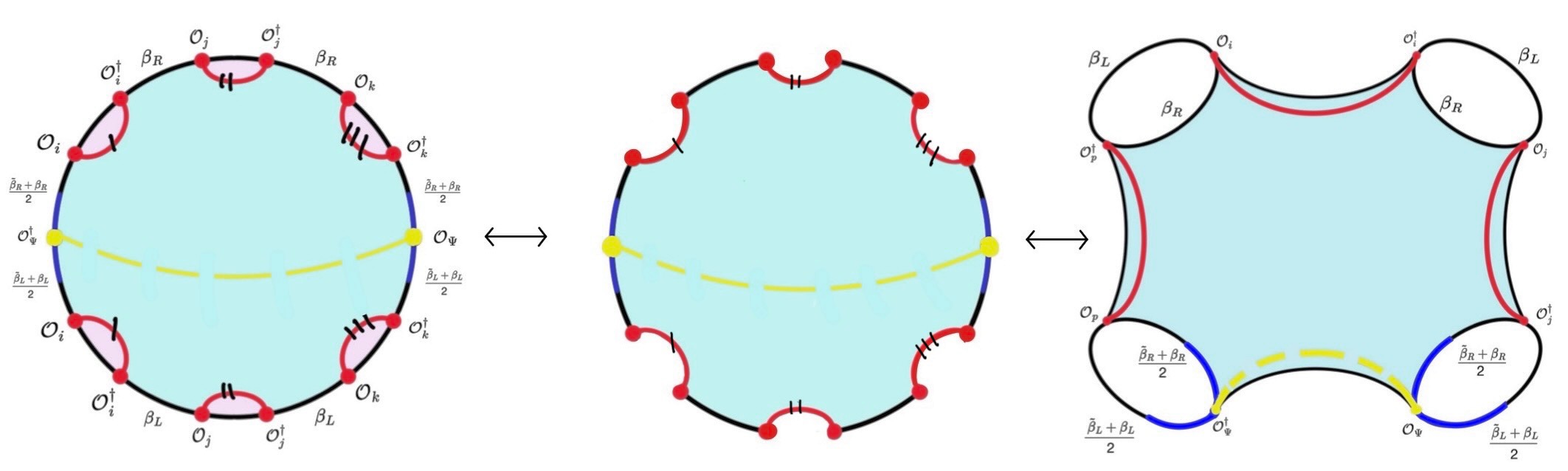}
    \caption{Disk-sheet-wormhole: the operator pinwheel saddles for (\ref{eq:1}) are constructed by considering the disk saddles, discarding the shell homology regions and gluing the resulting sheet along the corresponding shell worldvolumes into the wormhole, depicted here for $n=2$.}
    \label{fig:cut_operator_wheel}
\end{figure}

\paragraph{Coarse-grained span}
The path integral (\ref{eq:idinsert}) is computed by the analytic continuation
\beq \label{eq:1}
\overline{\langle \Psi|\Pi_{\mathcal{H}_{Shell}}|\Psi\rangle}= \lim_{n \to -1 } \overline{G^{n}_{ij}\langle \Psi|i\rangle\langle j|\Psi\rangle}. 
\eeq In the $\kappa \to \infty$ limit, only fully-connected saddles contribute to the right hand side because of the the single index loop in $G^{n}_{ij}\langle \Psi|i\rangle\langle j|\Psi\rangle$.   These geometries connect $n$ shell boundaries (Fig.~\ref{fig:shell_bdry}) to two \textit{operator boundaries} (see Fig.~\ref{fig:Obdry}) in a single bulk.
There could be multiple saddles of this kind.  We will show later that all of these saddles give contributions that are independent of the particular shell running around the loop. Collectively denoting these saddles as $\hat{Z}_{n+2}[\Psi]$ we therefore obtain:
\beq \label{eq:idinnorm}
\overline{G^{n}_{ij}\langle \Psi|i\rangle\langle j|\Psi\rangle} =\kappa^{n+1} \hat{Z}_{n+2}[\Psi].
\eeq
The $\hat{Z}_{n+2}[\Psi]$ wormholes are conceptually similar to the $\tilde{Z}_{n+2}$ folded wormhole of Sec.~\ref{sec:Zid} where the special case $|\Psi\rangle=|\beta\rangle$ was considered. To see how these saddles are constructed we imagine we cutting open the $\hat{Z}_{n+2}[\Psi]$ wormholes along the shell worldvolumes as in Fig.~\ref{fig:cut_operator_wheel}, producing a sheet diagram. These sheet geometries are each a cut-out of the saddle geometries for an asymptotic boundary circle  with some operator insertions. In particular, these ``disk" saddles involve asymptotic boundary insertions of the $\mathcal{O}_{\Psi},\mathcal{O}^{\dagger}_{\Psi}$ operators and $2(n+1)$ pairs of shell operators separated by preparation times $\Delta T_i$ on the asymptotic boundary, see the left hand side of Fig.~\ref{fig:cut_operator_wheel}. We will call the  asymptotic boundary condition  of this disk $\Sigma_n$. These disk saddles will be some function of the propagation times $\Delta T_i$. This disk is then glued into the $\hat{Z}_{n+2}[\Psi]$ wormhole via the junction conditions according to the shell identification pattern in Fig.~\ref{fig:cut_operator_wheel}, which dynamically determine the $\Delta T_i$.
The key point, as discussed in earlier sections, is that in the large shell mass limit the shell homology regions and propagation times on the disk shrink to zero, and hence the sheet grows to cover the entire disk. In this limit the only shell mass dependence is in the universal factor $\Pi_{i=1}^{n+1}Z_{m_i}$, which will be canceled by the shell state normalisation factors. As the shell action decouples, the contribution from  the bulk geometry is simply that of the disk computing the norm of a time evolved  version of the $|\Psi\rangle$ state: $\overline{Z}_{\mathcal{O}_{\Psi}}((n+1)(\beta_L + \beta_R) +\tilde{\beta}_L + \tilde{\beta}_R)$.\footnote{Here again $\overline{Z}_{\mathcal{O}_{\Psi}}((n+1)(\beta_L + \beta_R) +\tilde{\beta}_L + \tilde{\beta}_R)$ collectively denotes all saddles to $\overline{\langle \Psi | \Psi\rangle}((n+1)(\beta_L + \beta_R) +\tilde{\beta}_L + \tilde{\beta}_R)$ and for each of these saddles individually the above construction follows.} The upshot is thus that given a saddle for the norm of time-evolved version of the state $\ket{\Psi}$ given by $\overline{Z}_{\mathcal{O}_{\Psi}}((n+1)(\beta_L + \beta_R) +\tilde{\beta}_L + \tilde{\beta}_R)$, the saddles for (\ref{eq:idinnorm}) can automatically be constructed in the large shell mass limit, because of the universal pinching-off behavior of the shell insertions. Normalizing the shell states and putting everything together we obtain 
\beq
\hat{Z}_{n+2}(\Psi)= \frac{\overline{Z}_{\mathcal{O}_{\Psi}}((n+1)(\beta_L + \beta_R) +\tilde{\beta}_L + \tilde{\beta}_R)}{\overline{Z}(\beta_L)^{n+1}\overline{Z}(\beta_R)^{n+1}},
\eeq
and hence
\beq \label{eq:3}
\overline{\langle \Psi|\Pi_{\mathcal{H}_{Shell}}|\Psi\rangle}= \lim_{n \to -1 } \kappa^{n+1} \frac{\overline{Z}_{\mathcal{O}_{\Psi}}((n+1)(\beta_L + \beta_R) +\tilde{\beta}_L + \tilde{\beta}_R)}{\overline{Z}(\beta_L)^{n+1}\overline{Z}(\beta_R)^{n+1}} = \overline{Z}_{\mathcal{O}_{\Psi}}(\tilde{\beta}_L + \tilde{\beta}_R) \equiv \overline{\langle \Psi|\Psi\rangle}
\eeq
showing the coarse-grained equality (\ref{eq:projectid}) as an equality between the sum over saddlepoints of the LHS and RHS. 

\paragraph{Extension to fine-grained theory}
We can extend this to the fine-grained theory by showing (\ref{eq:idsqr}). For each term in the expansion of the square on the left side of (\ref{eq:idsqr}), for example $\overline{\langle \Psi|\Pi_{\mathcal{H}_{Shell}}|\Psi\rangle\langle \Psi|\Psi\rangle}$, there is a disconnected contribution which can be computed by simply multiplying the amplitudes discussed above for each factor.  There can also be  contributions from wormholes connecting  the boundaries associated to one factor with boundaries associated to the other.

\begin{figure}[h]
\centering
\begin{subfigure}[b]{0.25\linewidth}
        \centering
        \includegraphics[width=0.9\linewidth]{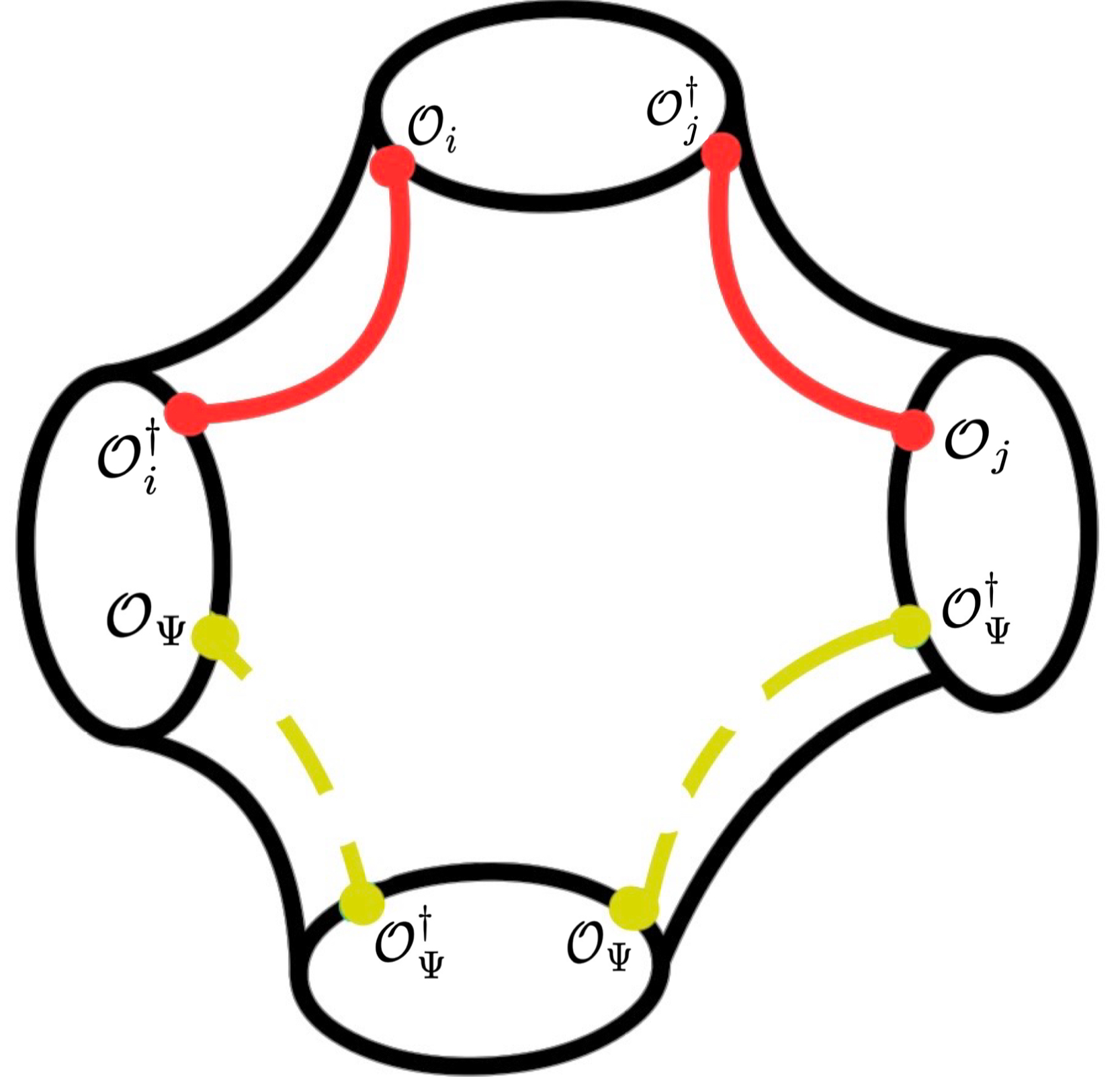}
         \caption{}
         \label{fig:sqWHa}
\end{subfigure}
\hfill
\begin{subfigure}[b]{0.25\linewidth}
            \centering
             \includegraphics[width=0.8\linewidth]{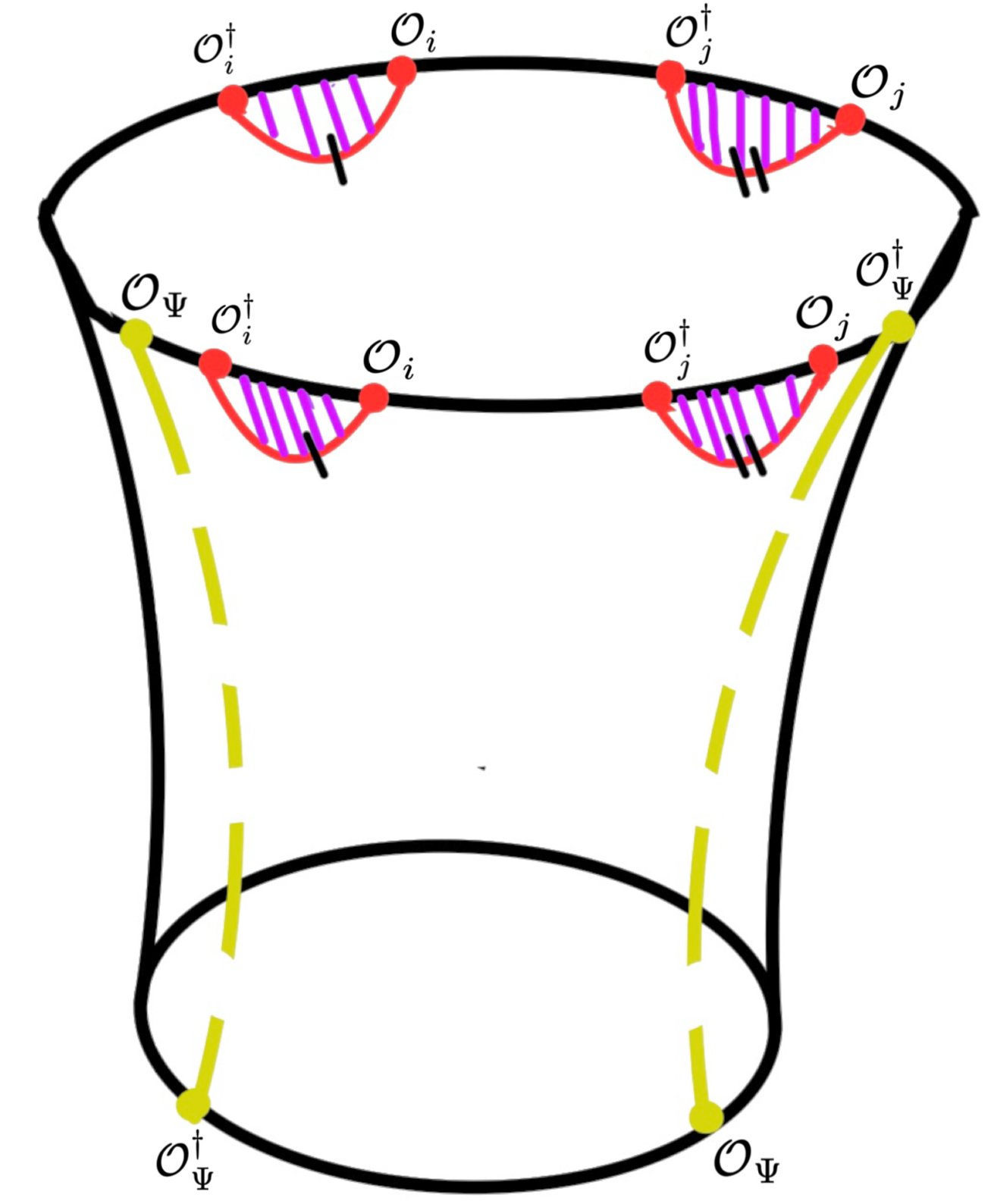}
            \caption{ }
           \label{fig:sqWHb}
\end{subfigure}
\hfill
\begin{subfigure}[b]{0.25\linewidth}
            \centering
             \includegraphics[width=0.6\linewidth]{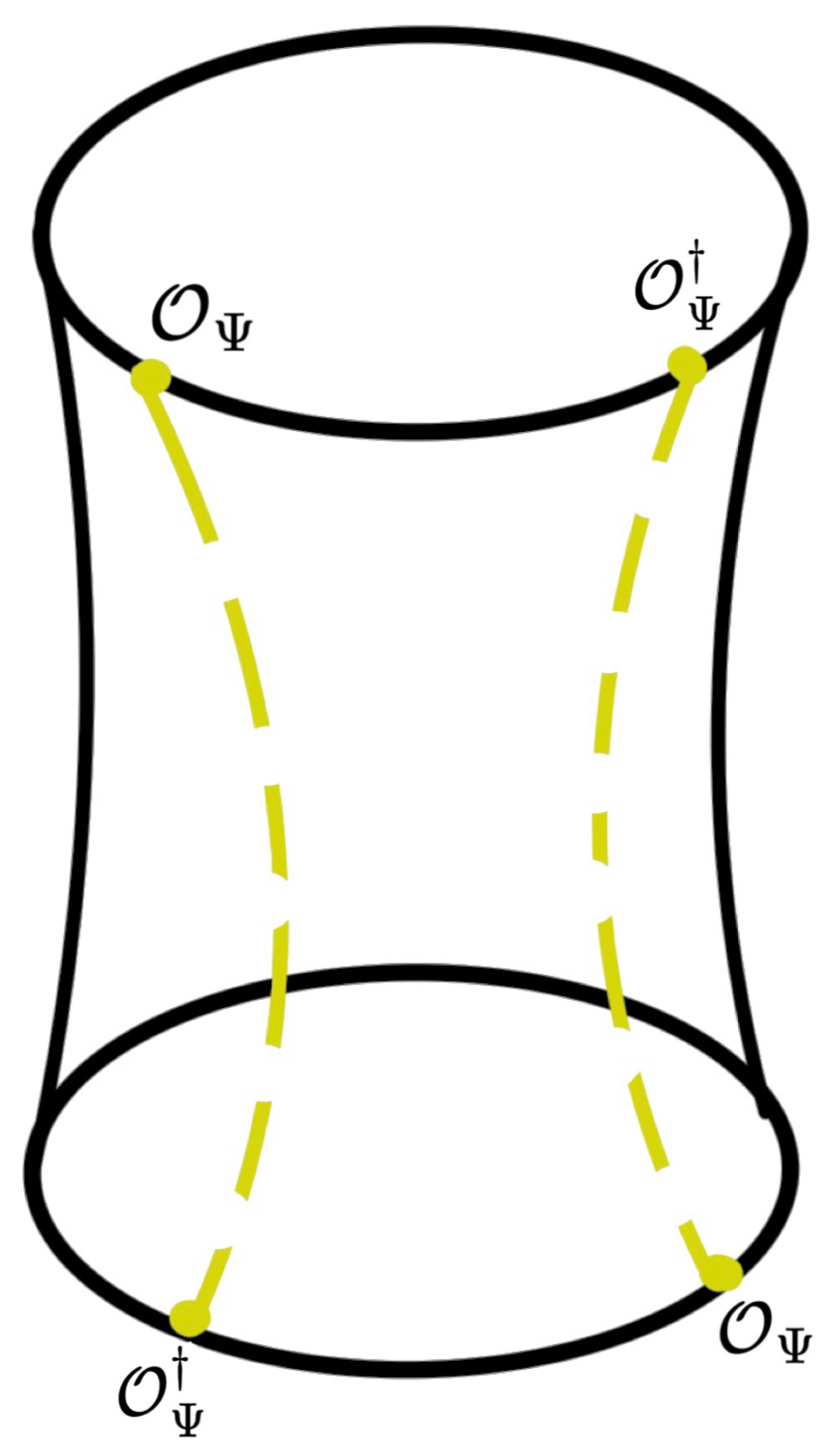}
            \caption{ }
           \label{fig:sqWHlim}
\end{subfigure}
    
\caption{({\bf a}) Fully connected contributions to $\overline{G^{n}_{ij}\langle \Psi|i\rangle \langle j|\Psi\rangle\langle \Psi|\Psi\rangle}$ for $n=1$. ({\bf b}) Construction of these wormholes by gluing shells. ({\bf c}) As $n \to -1$ these limits to the connected contributions to $\overline{\langle \Psi|\Psi\rangle^2}$. The coloring scheme in this figure is consistent with the discussion above; red denotes shell worldvolume and chartreuse denotes matter sources by $\mathcal{O}_\Psi$.}
\label{fig:finegrainedwh}
\end{figure}

For example, consider the fully connected contributions to
\beq \label{eq:2}
\overline{\langle \Psi|\Pi_{\mathcal{H}_{Shell}}|\Psi\rangle\langle \Psi|\Psi\rangle}= \lim_{n \to -1}\overline{G^{n}_{ij}\langle \Psi|i\rangle \langle j|\Psi\rangle\langle \Psi|\Psi\rangle}
\eeq
depicted in Fig.~\ref{fig:sqWHa}. At a given $n$, these saddle geometries can be constructed by first considering a different wormhole: the cylinder connecting one operator boundary $\langle \Psi|\Psi\rangle$ to the $\Sigma_n$ boundary condition discussed above and depicted on the left hand side of Fig.~\ref{fig:cut_operator_wheel}.  On this geometry, the shells propagate into the cylinder wormhole bulk and are re-absorbed at the boundary. The saddle is constructed by discarding the shell homology regions and gluing the geometry along the corresponding shell worldvolumes via the junction conditions (Fig.~\ref{fig:sqWHb}). In the large shell mass limit the shell homology regions pinch off, for the same reasons as described in previous sections, contributing a factor $\Pi_{i=1}^{n+1}Z_{m_i}$. 

The remaining wormhole bulk action is therefore computed by the path integral of the topology shown in Fig.~\ref{fig:sqWHlim} connecting a boundary of the same length  $\Sigma_n$ but \textit{without} any shell insertions to the $\langle \Psi|\Psi\rangle$ boundary. In the large shell mass limit, this no-shell $\Sigma_n$ boundary is of length $(n+1) \beta_R + \tilde{\beta}_R +(n+1) \beta_L + \tilde{\beta}_L$, as the shell propagation times $\Delta T_i \to 0$. This is nothing more than an elongated version of the  $\langle \Psi|\Psi\rangle$ boundary condition; we will call the path integral  with these boundary conditions $Z_{\Psi \tilde\Psi}\left((n+1)\beta_L,(n+1)\beta_R\right)$. For each saddle to $Z_{\Psi \tilde\Psi}$ there is therefore a corresponding saddle to (\ref{eq:2}). Upon normalizing the shell states to have unit norm, their contributions cancel in the same manner as described above, and we obtain
\beq \label{eq:4}
\overline{\langle \Psi|\Pi_{\mathcal{H}_{Shell}}|\Psi\rangle\langle \Psi|\Psi\rangle}= \lim_{n \to -1} \kappa^{n+1}\times \frac{Z_{\Psi \tilde\Psi}\left((n+1)\beta_L,(n+1)\beta_R\right)}{\overline{Z}(\beta_L)^{n+1}\overline{Z}(\beta_R)^{n+1}}= Z_{\Psi \tilde\Psi}(0,0)\equiv\overline{\langle \Psi|\Psi\rangle^2}.
\eeq
 A similar argument applies to the fully connected wormhole relevant for $\overline{\left(\langle \Psi|\Pi_{\mathcal{H}_{Shell}}|\Psi\rangle\right)^2}$
and hence in the $n \to -1$ limit each term in the expansion of the square on the left hand side of (\ref{eq:idsqr}) limits to $\overline{\langle \Psi|\Psi\rangle^2}$, proving the equality.



\paragraph{Conclusion} Because we have shown both (\ref{eq:idinsert}) and (\ref{eq:idsqr}),  fine-grained equality
\beq \label{Hshellspan}
\mathcal{H}_{LR}=\mathcal{H}_{shell}
\eeq
follows, in the sum over saddelpoints approximation to the path integral.  Therefore the shell states span the entire Hilbert space if we take $\mathcal{H}_{shell}$ to consist of $\kappa \to \infty$ shell states. As the derivations of (\ref{eq:idinsert}) and (\ref{eq:idsqr}) were both independent of the shell preparation temperature, the fine-grained result (\ref{Hshellspan}) holds for \textit{any} choice of preparation temperature. Also note that the above arguments extend  easily to $\langle \Psi|\Phi\rangle =\langle \Psi|\Pi_{\mathcal{H}_{Shell}}|\Phi\rangle + \mathcal{O}(e^{G^{0}_N})$ for any $|\Psi \rangle, |\Phi \rangle \in \mathcal{H}_{LR}$.

After the completion of this work the authors learned of the works \cite{Hsin:2020mfa,Boruch:2023trc} in which a similar argument is used in JT gravity to show any microcanonical gravity state can be reconstructed from a certain basis of $e^{\mathbf{S}}$ states. In light of this, our work can be seen as an explicit higher dimensional realization, and extension the fine-grained level, of this argument.

\subsection{Some more examples} \label{sec:examples}
At the beginning of this section, we demonstrated the above argument explicitly for the case in which the state under consideration is the TFD state $|\Psi\rangle = |\beta\rangle$. Two additional simple examples can be worked out straightforwardly:

\paragraph{Shell state} 
Let $|\Psi\rangle$ be some shell state created with shell operator $\mathcal{O}_S$ and preparation temperatures $\tilde{\beta}_L,\tilde{\beta}_R$. For brevity we consider the heavy shell mass limit $m_s\to \infty$, although this is not required, and fix $m_s$ to be distinct from the masses of the shell states defining the basis. Furthermore, we could instead choose $\mathcal{O}_S$ to carry some global charge, whilst our basis does not. In this case $m_s$ may equal one of the masses in the basis. As discussed in Sec.~\ref{sec:shellstates}, the norm is given by $\overline{\braket{\Psi|\Psi}}=\overline{Z}(\tilde{\beta}_L)\times\overline{Z}(\tilde{\beta}_R)\times Z_{m_s} $, with $Z_{m_s} $ the shell mass dependent term. The boundary conditions defining $\overline{G^{n}_{ij}\langle \Psi|i\rangle\langle j|\Psi\rangle}$ is now simply that of the pinwheel discussed in (Fig.~ \ref{fig:pinwheel}) with $n+2$ boundaries. A simple adaptation of (\ref{eq:pinwheel})  yields:
\beq
\overline{G^{n}_{ij}\langle \Psi|i\rangle\langle j|\Psi\rangle}= Z_{m_s} \times \frac{\overline{Z}((n+1)\beta_L+\tilde{\beta}_L)\overline{Z}((n+1) \beta_R)+\tilde{\beta}_R)}{\overline{Z}(\beta_L)^{n+1} \overline{Z}(\beta_R)^{n+1}}
\eeq
and hence $\overline{\langle \Psi|\Pi_{\mathcal{H}_{Shell}}|\Psi\rangle}= \lim_{n \to -1 } \overline{G^{n}_{ij}\langle \Psi|i\rangle\langle j|\Psi\rangle}=\overline{Z}(\tilde{\beta}_L)\times\overline{Z}(\tilde{\beta}_R)\times Z_{m_s}=\overline{\braket{\Psi|\Psi}}$. The cross terms required for the fine grained extension can be dealt with analogously. Consider for example (\ref{eq:2}). The connected contribution to $\overline{G^{n}_{ij}\langle \Psi|i\rangle \langle j|\Psi\rangle\langle \Psi|\Psi\rangle}$ is given by a pinwheel with $n+3$ boundaries:
\beq
\overline{G^{n}_{ij}\langle \Psi|i\rangle \langle j|\Psi\rangle\langle \Psi|\Psi\rangle}=  Z_{m_s}^2 \frac{\overline{Z}((n+1)\beta_L+2\tilde{\beta}_L)\overline{Z}((n+1) \beta_R)+2\tilde{\beta}_R)}{\overline{Z}(\beta_L)^{n+1} \overline{Z}(\beta_R)^{n+1}}.
\eeq
Taking $n \to -1$ we obtain: 

\beq
\overline{\langle \Psi|\Pi_{\mathcal{H}_{Shell}}|\Psi\rangle\langle \Psi|\Psi\rangle}=  Z_{m_s}^2 \overline{Z}(2\tilde{\beta}_L)\overline{Z}(2\tilde{\beta}_R),
\eeq
which can be recognized as the cylinder topology connected contribution to $\overline{\braket{\Psi|\Psi}\braket{\Psi|\Psi}}$ (see Fig.~\ref{fig:shell_wormhole}). Again, the other cross term can be dealt with similarly. We conclude that at a fine-grained level 
$\langle \Psi|\Psi\rangle =\langle \Psi|\Pi_{\mathcal{H}_{Shell}}|\Psi\rangle$. In the case where $\mathcal{O}_S$ is charged under some global symmetry, the above argument shows that this symmetry is broken by the non-perturbative wormhole contributions, analogous to \cite{Hsin:2020mfa}.

\paragraph{Light operator}
We now consider the case where the mass $m_{\Psi}$ of the operator $\mathcal{O}_{\Psi}$  is heavy enough for its two-point function to be accurately described by the geodesic approximation, yet light enough to avoid significant back-reaction on the geometry. The norm $\overline{\braket{\Psi|\Psi}}$ then takes the schematic form:
\beq \label{eq:case1}
\overline{\braket{\Psi|\Psi}}= \sum_X \overline{Z}_X(\tilde{\beta}_L+\tilde{\beta}_R) \times ( \sum_i e^{-m_{\Psi}L^i_X (\tilde{\beta}_L,\tilde{\beta}_R) }), 
\eeq
where $\{ X\}$ label the different saddles to the Z-partition function $\overline{Z}(\tilde{\beta}_L+\tilde{\beta}_R) $ and $L^i_X $ denotes the length of the $i$-th bulk geodesic connecting the boundary insertion points of the operator on the background geometry $X$. These insertion points are separated on the boundary by $\tilde{\beta}_L,\tilde{\beta}_R$ the $L,R $ way around. The index $i$ here denotes the sum over windings. Crucially, the geodesic lengths $L^i_X $ are simple functions of the boundary separations $\tilde{\beta}_L,\tilde{\beta}_R$. See, for example, Sec.~4 of \cite{Antonini:2023hdh} and references therein for details. As $\mathcal{O}_{\Psi}$ is not heavy enough to significancy back-react, the fully connected contribution to $\overline{G^{n}_{ij}\langle \Psi|i\rangle\langle j|\Psi\rangle}$ is given by $\Psi$-matter geodesics propagating on the folded wormhole background geometry constructed in Sec.~\ref{sec:foldedWH} above. Recall that this wormhole was obtained by gluing a single disk onto itself. Hence the $\Psi$-matter geodesics are again simply geodesics of the Z-partition saddles with a longer boundary:

\beq\label{eq:case2}
\overline{G^{n}_{ij}\langle \Psi|i\rangle\langle j|\Psi\rangle}= \sum_X \overline{Z}_X((n+1)\beta_L+\tilde{\beta}_L+(n+1)\beta_R+\tilde{\beta}_R) \times ( \sum_i e^{-m_{\Psi}L^i_X ((n+1)\beta_L+\tilde{\beta}_L,(n+1)\beta_R+\tilde{\beta}_R )}). 
\eeq
In both (\ref{eq:case1}) and (\ref{eq:case2}) we sum over all winding, such the geodesics summed over are in one-to-one correspondence.  It then follows that the $n \to -1 $ limit of (\ref{eq:case2}) recovers (\ref{eq:case1}). Note that if we had kept only the leading terms, subtleties about the relative sizes of $\beta_L,\beta_R,\tilde{\beta}_L,\tilde{\beta}_R$ arize, as for example there might be some positive integer $n$ at which the $i=0$ and $i=-1$ windings degenerate for (\ref{eq:case2}) but not for (\ref{eq:case1}). 

The fine-grained extension is trivial, as the $\mathcal{O}_{\Psi}$ insertions are not heavy enough to stablize new connected wormhole contributions to the cross terms such as $\overline{\langle \Psi|\Pi_{\mathcal{H}_{Shell}}|\Psi\rangle\langle \Psi|\Psi\rangle}$ (see Fig.~\ref{fig:finegrainedwh}), and hence each of the cross terms factorises.\footnote{See  footnote 1 for a potential subtlety.} We therefore conclude  that  $\langle \Psi|\Psi\rangle =\langle \Psi|\Pi_{\mathcal{H}_{Shell}}|\Psi\rangle$ at the fine grained level, and hence that the shell states span these states also.

\section{Tool 3: Surgery for relational equivalence}
\label{sec:tool3}
As discussed in Sec.~\ref{sec:QTFTPI}, the gravitational path integral appears to compute coarse-grained averages over an underlying fine-grained theory. The path integral  is difficult to compute explicitly, and so we resorted above to the saddlepoint approximation to derive fine-grained equalities like (\ref{eq:factorisation}) and (\ref{eq:projectid}). However, the latter approach at most shows fine-grained equality in the \textit{approximate} effective distribution defined by the saddlepoint sum. In principle, the ``true" distribution is computed by the full path integral over all geometries and topologies satisfying the boundary conditions (although this sum might not be strictly well defined). The additional contributions include perturbative corrections around a given saddlepoint, but perhaps also topologies that do not admit any saddlepoints. Most of these contributions are highly suppressed; for example  non-smooth geometries will be suppressed by their  infinite action arising from  derivative terms in the Lagrangian.
Nevertheless, the additional contributions to the path integral could violate fine-grained equalities derived in the saddlepoint approximation. For example, consider  the trace factorization (\ref{eq:factorisation}) in Sec.~\ref{sec:facSC}. Any finite action contribution to the trace path integral with a handle connecting the two sides of the pinwheel as in Fig.~\ref{fig:nonfactorcont} will potentially violate factorization.

Below, we propose a method for establishing equality between fine-grained observables in the full path integral.  The idea is that to show that two quantities are equal we do not have to explicitly calculate the gravitational path integral for either.  We simply need to show that every topology and  geometry that contributes to one also contributes equally to the other.  We will demonstrate how to use a cut-and-paste procedure on the relevant geometries to argue for such equality.   To this end, we will strategically insert the shell state resolution of the identity in the $\kappa \to \infty$ limit that we developed in previous sections.  In this limit, the sum over topologies simplifies if we construct our basis out of parametrically heavy shells. This is because, as explained  above, in this circumstance only fully connected contributions to the gravitational path integral survive the $\kappa \to \infty$ limit.

\subsection{A complete basis} \label{sec:completebasis}
First, we want to extend the fine-grained equality (\ref{Hshellspan}) between $\mathcal{H}_{LR}$ and $\mathcal{H}_{shell}$ beyond the saddlepoint approximation.  It is straightforward to include perturbative corrections around the semiclassical saddles if we take the shells to be parametrically heavy.  As we discussed, in that case only fully-connected contributions survive the $\kappa \to \infty$ limit because all diagrams that break index loops in (\ref{eq:1}) are suppressed in powers of $\kappa$.  What is more, as we explained, when the shells become heavier the turning points of their trajectories approach the asymptotic boundary, and the effect of the shells is simply to glue disk geometries into wormholes.  The shells themselves then contribute a universal factor to the path integral, and by the arguments of Sec.~\ref{sec:scspan} the same diagrams contribute to the demonstration of the equalities in (\ref{eq:3}) and (\ref{eq:4}) and hence also
(\ref{Hshellspan}).  Earlier we studied these in the sum over saddlepoints, but by construction the perturbative corrections around the saddlepoints will also agree.  

In fact, for any topology that supports a saddlepoint, the sum over geometries associated to each topology will match. To see this, consider the contribution of any such geometry.  If we fix the geometry consistently with the boundary conditions, in the heavy shell limit the sum over shell trajectories will again localize to the vicinity of the fixed spacetime boundary. So the interior gravitational action will be the same, while the shells contribute a universal piece from their trajectory near the fixed boundary.

We can also attempt to include topologies that do not support a classical saddle point, if such contributions exist. Below we outline a speculative schematic procedure for how this might be done. As a cautionary note, we do not claim that the path-integral manipulations presented here are strictly well defined, particularly in dimensions $d\geq 4$. Our argument also relies on several heuristic assumptions about the integration measure and the analytic continuation of certain contributions. Although this presentation is non-rigorous, we expect the schematic to capture the essential features of a careful implementation of these ideas in lower-dimensional gravity models, where the relevant analytic and geometric structures are better controlled.

To do so we have to show the relevant configurations associated to any of these topologies can we represented as shells propagating on disks with interior topological decorations.  We will argue for this below. Then, by the same reasoning as above, provided we take the heavy shell limit first, then for any fixed geometry the shell path integral will universally localize near the boundary, and the $\kappa \to \infty$ limit will restrict the sum to connected geometries.  Then the equalities (\ref{eq:3}) and (\ref{eq:4}) and  (\ref{Hshellspan}) follow for the complete gravitational path integral.  This argument relies on the heavy shell and $\kappa \to \infty$ limits, which lead to no loss of generality because we are always free to make the basis as overcomplete as we want ($\kappa \to \infty$), and the inertial mass of the shells can be arbitrarily chosen in the basis construction.

We will attempt to make the above reasoning somewhat more formal. To do so we re-arrange the $\kappa \to \infty$ path integral (\ref{eq:1}) defined as the $n \to -1$ limit of $\overline{G^{n}_{ij}\langle \Psi|i\rangle\langle j|\Psi\rangle}$ at a fixed positive integer $n$, 
\beq \label{eq:spanpathint}
\zeta[G^{n}_{ij}\langle \Psi|i\rangle\langle j|\Psi\rangle] = \int_{g, \phi \sim G^{n}_{ij}\langle \Psi|i\rangle\langle j|\Psi\rangle} \mathcal{D}g \mathcal{D}\phi\mathcal \, \,  e^{-I_{tot}[g]}
\eeq
into  parts that vanish once $n \to -1$, and a remainder that limits to $\overline{\langle \Psi|\Psi\rangle}$.  We are using notation of Sec.~\ref{sec:QTFTPI}, where $\zeta[\cdots]$ denotes the gravity path integral subject to some boundary conditions and $\phi$ denotes the shell matter fields. 

\begin{figure}[h]
    \begin{subfigure}{\linewidth}
    \centering
    \includegraphics[width=0.4\linewidth]{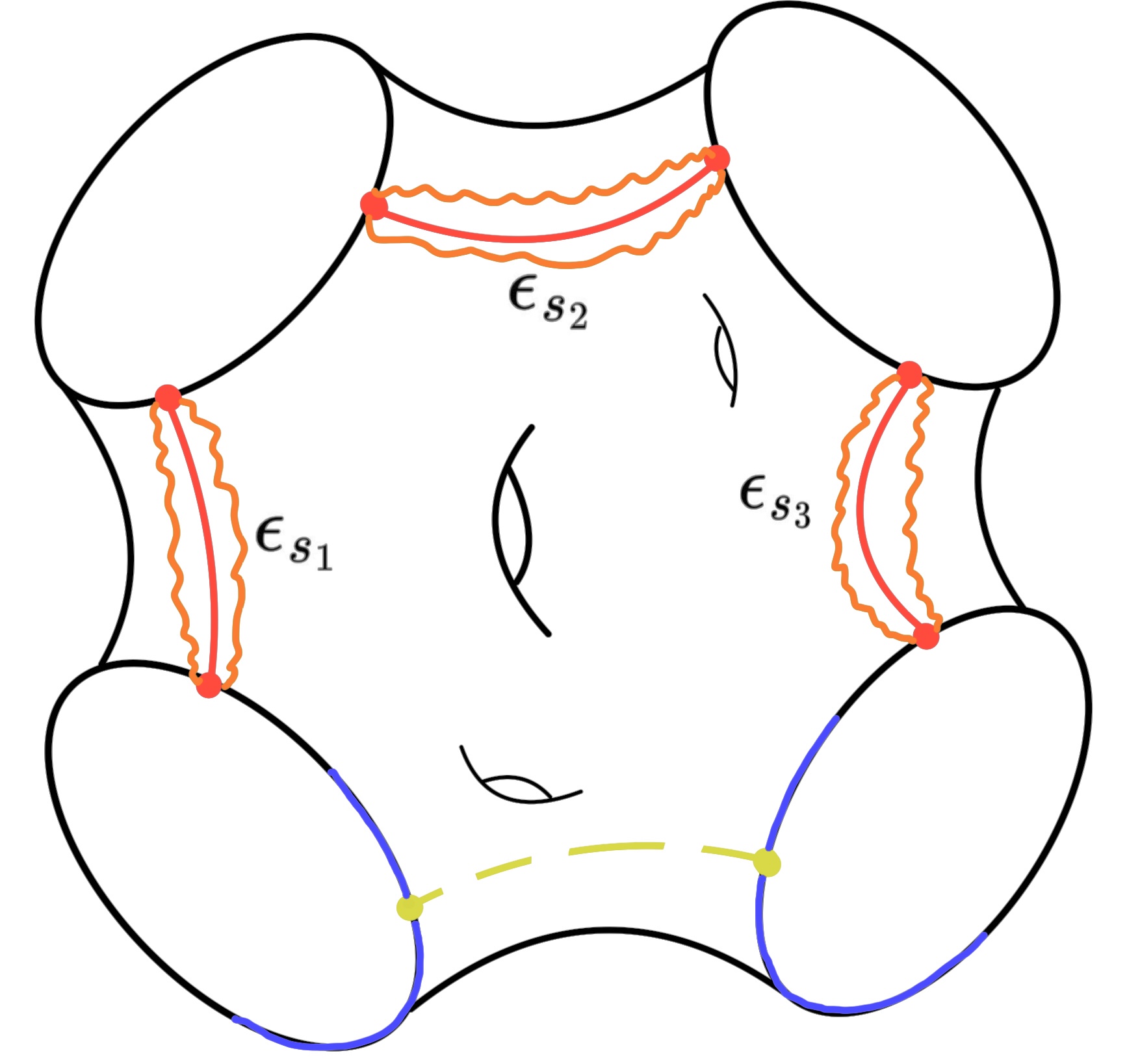}
    \caption{}
    \label{fig:cutout}
    \end{subfigure}
    \begin{subfigure}[b]{0.45\linewidth}
    \centering
    \includegraphics[width=0.8\linewidth]{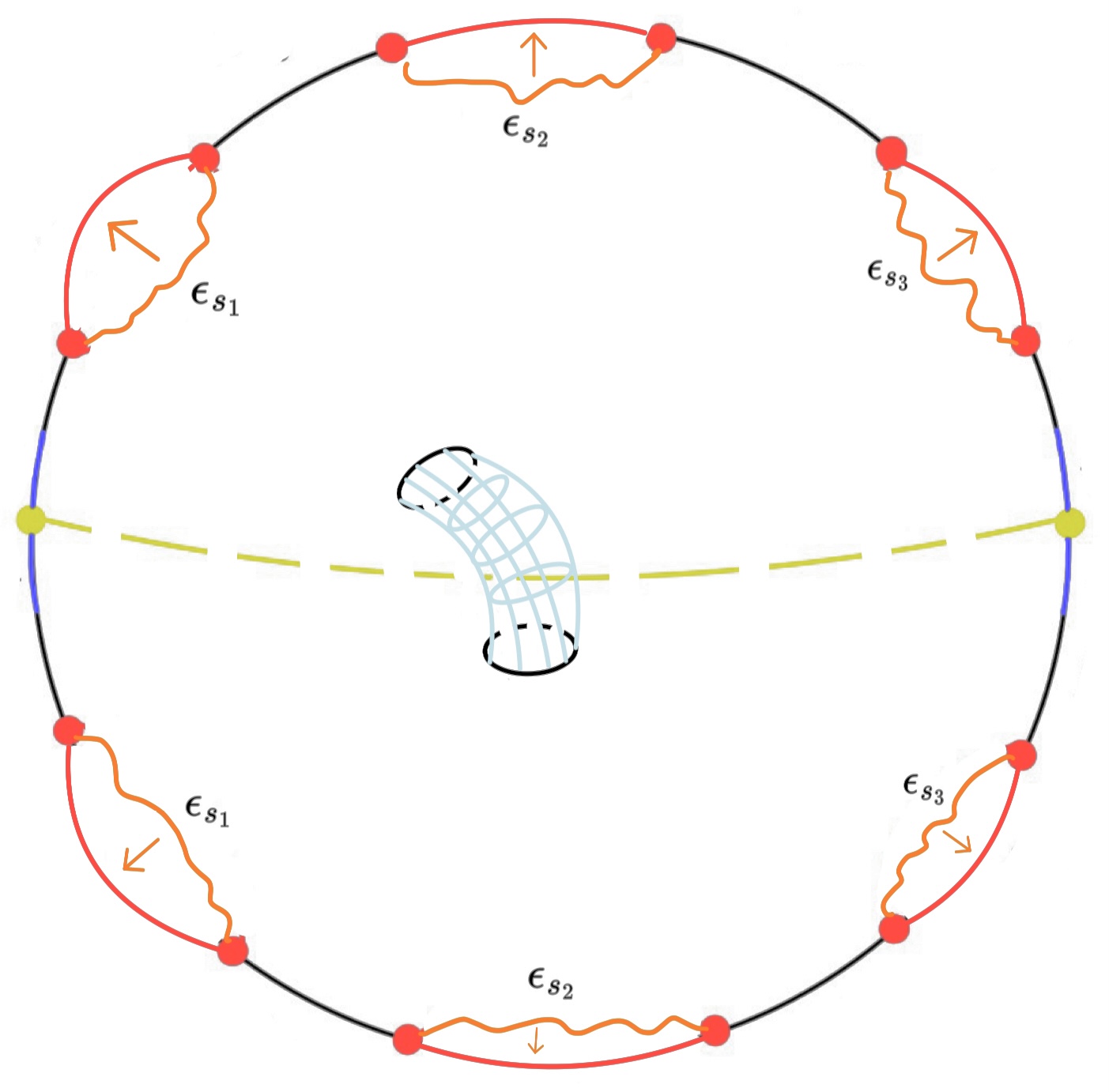}
    \caption{}
    \label{fig:sheet_embed}
    \end{subfigure}
    \hfill
    \begin{subfigure}[b]{0.45\linewidth}
    \centering
    \includegraphics[width=0.8\linewidth]{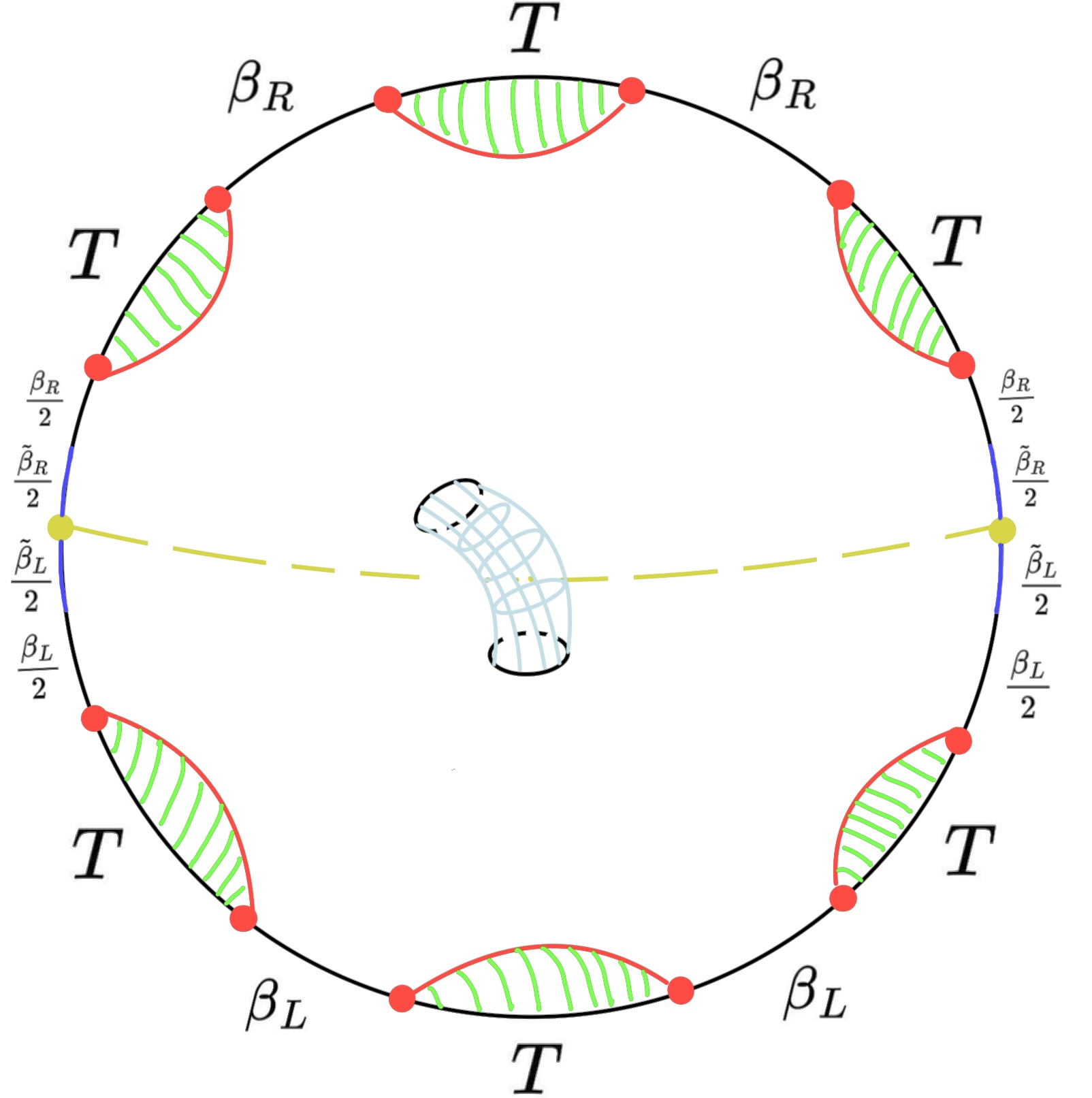}
    \caption{}
    \label{fig:allOsheet_embed}
    \end{subfigure}
    \caption{ ({\bf a}) Cartoon of the $\epsilon_{s_i}$ cut outs around the shell worldvolume for an off-shell contribution to (\ref{eq:spanpathint}) for depicted for $n=3$. ({\bf b}) Sheet diagram for an non-saddle contribution, where the cutouts $\epsilon_{s_i}$ have been shrunk to be arbitrarily close to the shell worldvolume $\mathcal{W}_i$. ({\bf c}) Embedding of this sheet onto the disk by filling in homology regions.}
\end{figure}

We start by organizing the full sum over connected geometries by topology.  We assume the path integral (\ref{eq:spanpathint}) can be performed by 
first performing performing the matter path integral on each geometry $g$ contributing at a fixed topology, and subsequently integrating over these geometries. For a given metric, we can then consider the semiclassical limit of the matter path integral such that the shell trajectories  in  the heavy shell  limit  localize to geodesics on that background.  So for each geometry $\mathcal{M}$ associated to a given topology we can cut out a  small region around the $i$-th shell worldvolume of size $\epsilon_{s_i}$.   There are $n+1$ such regions, one for each shell, see Fig.~\ref{fig:cutout}. We denote the part of the geometry excluding these regions  as $\mathcal{M}_{bulk}$, which contains all handles of $\mathcal{M}$. As the Einstein-Hilbert action consists of local and boundary terms the total action for a given geometry is the sum of the action of these individual patches: 
\beq
I_{tot}[\mathcal{M}]= \sum_{i=1}^{n+1} I_{\epsilon_{s_i}}+ I_{\mathcal{M}_{bulk}} \, .
\eeq

Next, on a given geometry $g$  the contribution of the $i$-th shell stress-energy to the action is given by the co-dimension 1 integral $ \int_{\mathcal{W}_i[g]} \sigma_i$ of the shell matter density ($\sigma_i$) over the world-volume ($\mathcal{W}_i[g]$).  Then shrinking the cut-out regions $\epsilon_{s_i}$ to include just the shell world-volume we write
 \beq
 \sum_{i=1}^{(n+1)} I_{\epsilon_{s_i}}=\sum_{i=1}^{(n+1)} \int_{\mathcal{W}_i} \sigma_i
 \eeq
and hence 
\beq \label{eq:5}
\zeta[G^{n}_{ij}\langle \Psi|i\rangle\langle j|\Psi\rangle] = \int_{g \sim G^{n}_{ij}\langle \Psi|i\rangle\langle j|\Psi\rangle} \mathcal{D}g\mathcal \, \, e^{-I_{bulk}} \Pi_{i=1}^{n+1} e^{-\int_{\mathcal{W}_i[g]} \sigma_i}.
\eeq

For each topology, we cut a given geometry $g$ along the shell world-volumes to produce a sheet diagram, see Fig.~\ref{fig:sheet_embed}. To make contact with the desired result, the boundary sections of the bulk sheets can be completed onto an asymptotic circle of length $(n+1)(\beta_L + \beta_R) +\tilde{\beta}_L + \tilde{\beta}_R +\sum_{i=1}^{n+1} T$ by adding in ``homology" regions $g_{\mathcal{Y}_i}$ between an asymptotic boundary section of length $T$ and the shell cutout $\epsilon_{s_i}$ on the sheet bulk, see Fig.~\ref{fig:allOsheet_embed}. We can sum over all possible completions of the sheet 
geometry $g$ by summing over all possible homology region geometries $g_{\mathcal{Y}_i}$ 
ending on the worldvolume $\mathcal{W}_i$: $ \int\mathcal{D}g_{\mathcal{Y}_i} \, e^{-
I_{g_{\mathcal{Y}_i}}}\equiv Z_{\mathcal{Y}_i}[g] $. Of course, these terms were not present 
in the original path integral, but can rewrite (\ref{eq:5}) in a suggestive manner by inserting $1=\frac{\Pi_{i=1}^{n+1}Z_{\mathcal{Y}_i}[g]}{\Pi_{i=1}^{n+1}Z_{\mathcal{Y}_i}[g]}$:
 \beq
\int \mathcal{D}g\, \, e^{-I_{{bulk}}}\Pi_{i=1}^{n+1} e^{-\int_{\mathcal{W}_i[g]} \sigma_i} = \int \mathcal{D}g\mathcal \, \, \frac{\Pi_{i=1}^{n+1}Z_{\mathcal{Y}_i}[g]}{\Pi_{i=1}^{n+1}Z_{\mathcal{Y}_i}[g]} e^{-I_{{bulk}}} \Pi_{i=1}^{n+1} e^{-\int_{\mathcal{W}[g]_i} \sigma_i} \\ = \int \mathcal{D}g \, \, e^{-I_{bulk}} \Pi_{i=1}^{n+1} \int \mathcal{D}g_{\mathcal{Y},i} e^{-I_{g_{\mathcal{Y},i}}}
\left(\Pi_{i=1}^{n+1}\frac{e^{-\int_{\mathcal{W}_i[g]} \sigma_i}}{Z_{\mathcal{Y}_i}[g]}\right).
\eeq

This is useful because summing over all possible completions of the $g$ sheet regions into a geometry with a circular asymptotic boundary of length $(n+1)(\beta_L + \beta_R) +\tilde{\beta}_L + \tilde{\beta}_R +\sum_{i=1}^{n+1} T$ and doing this for all possible geometries $g$ is the same as summing over all geometries with this circular asymptotic boundary condition, which we denote $g_{\circ}$:
\beq \label{eq:final}
\int \mathcal{D}g\mathcal \, \, e^{-I_{{bulk}}} \Pi_{i=1}^{n+1} \int \mathcal{D}g_{\mathcal{Y},i} e^{-I_{g_{\mathcal{Y},i}}}
\left(\Pi_{i=1}^{n+1}\frac{e^{-\int_{\mathcal{W}_i[g]} \sigma_i}}{Z_{\mathcal{Y}_i}[g]}\right) = \int \mathcal{D}g_{\circ} \, \, e^{-I_{{bulk}}[g_{\circ}]} \left(\Pi_{i=1}^{n+1}\frac{e^{-\int_{\mathcal{W}_i[g]} \sigma_i}}{Z_{\mathcal{Y}_i}.[g]}\right)
\eeq
 In the $n \to -1$ limit the factor in parenthesis in (\ref{eq:final}) is absent, and the asymptotic boundary condition for the sum over geometries $g_{\circ}$ is that of a periodic boundary of length $\tilde{\beta}_L,\tilde{\beta}_R $ separating $\mathcal{O}_{\Psi}, \mathcal{O}^{\dagger}_{\Psi}$. The latter is precisely the boundary condition for the norm of $|\Psi\rangle$ and hence
\beq
\lim_{n \to -1} \zeta[G^{n}_{ij}\langle \Psi|i\rangle\langle j|\Psi\rangle]  = \overline{\langle \Psi | \Psi \rangle} \, .
\eeq

Schematically, this approach establishes full equality between the path integrals,
$\overline{\langle \Psi | \Pi_{\mathcal{H}_{\text{shell}}} | \Psi \rangle} = \overline{\langle \Psi | \Psi \rangle}$.
Although many of the intermediate steps require a more rigorous justification, we expect this argument captures the essential qualitative features of a more complete treatment. Similarly, the argument could be extended to show $\overline{(\langle \Psi|\Pi_{\mathcal{H}_{shell}}|\Psi\rangle-\langle \Psi|\Psi\rangle)^2}=0$ by including all handles connecting $\langle \Psi|\Pi_{shell}|\Psi\rangle$ with $\langle \Psi|\Pi_{shell}|\Psi\rangle$ (or $\langle \Psi|\Psi\rangle$ in the cross term) into the $I_{bulk}$ region and repeating the above argument. This concludes our argument that the  $\kappa \to \infty$ set of shell states at any preparation temperature spans the fine-grained Hilbert space $\mathcal{H}_{LR}$ in the full gravitational path integral.

\subsection{Application: full factorisation of the thermal partition function } \label{sec:Trzz} 
As the shell states span the fine-grained Hilbert space $\mathcal{H}_{LR}$ in the full gravitational path integral, we can use insertions of the shell resolution of the identity to simplify the task of showing full equality between gravitational path integrals. We demonstrate the main ideas by means of an example, and summarize the systematic method in Sec.~\ref{sec:alg}; readers can skip directly to that section if they wish.

We aim to show the fine-grained two-sided thermal partition function factorises:
\beq \label{eq:tracefac}
Tr_{\mathcal{H}_{LR}}(e^{-\beta_1H_{L}}e^{-\beta_2H_{R}}) = Z(\beta_1)\times Z(\beta_2),
\eeq
based on full path integral equalities, without explicitly evaluating either side of (\ref{eq:tracefac}).

In Sec.~\ref{sec:Zid} we showed that the thermal partition function on $\mathcal{H}_{shell}$ factorizes at the fine-grained level in the sum over saddlepoints, see (\ref{eq:factorisation}).  This followed because the  pinwheel saddles contributing to the trace path integral are constructed by gluing an $L$- and $R$-disk together along shells via the junction conditions. In the large shell mass limit these disks decouple and the saddle action is simply the sum of the $L,R$ disk actions and universal shell action terms. As we have shown, $\mathcal{H}_{LR} = \mathcal{H}_{shell}$, the result (\ref{eq:factorisation}) extrapolates to (\ref{eq:tracefac}) in the sum over saddlepoints.  
\begin{figure} 
    \centering
    \includegraphics[width=0.3\linewidth]{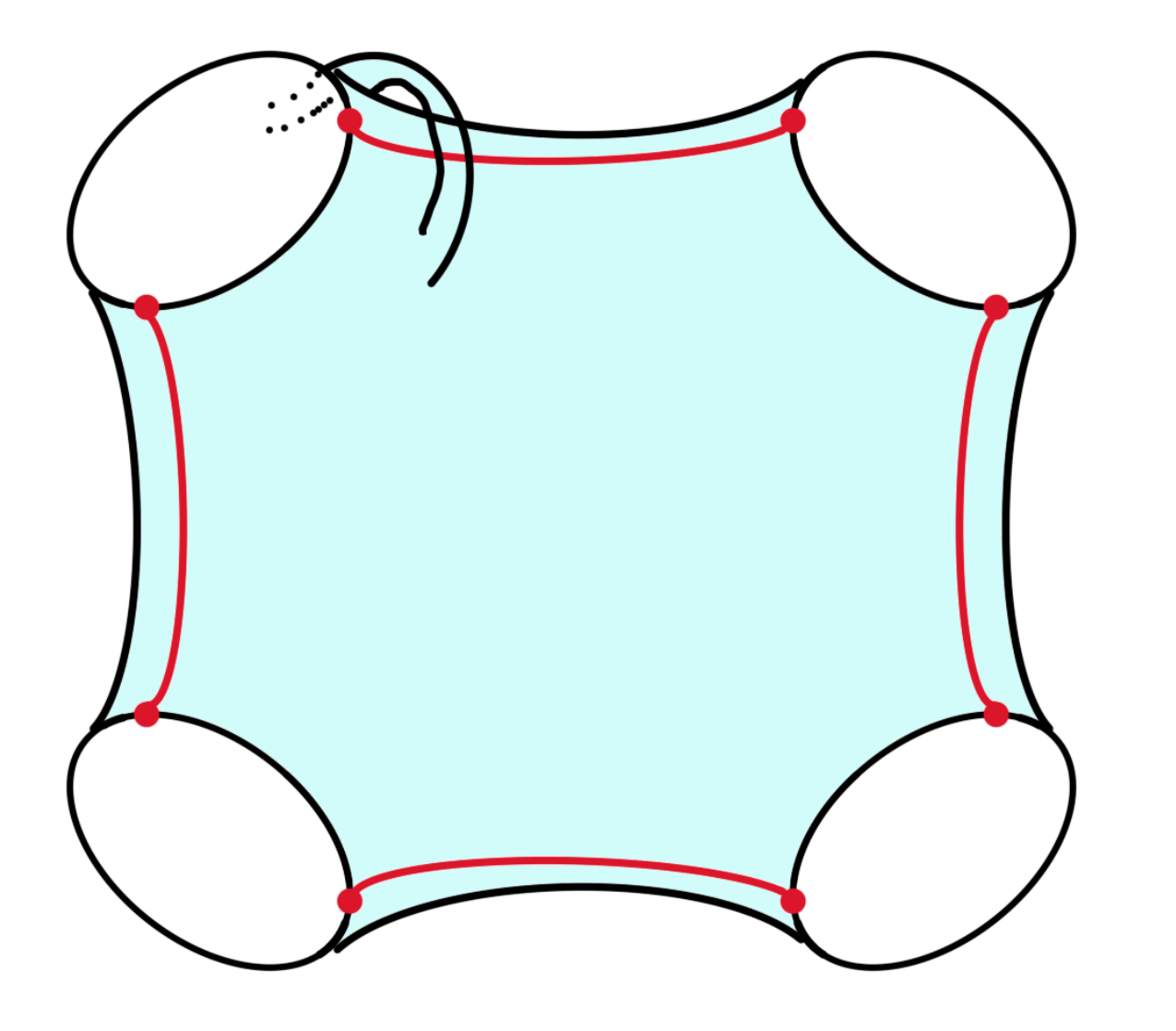}
    \caption{Example of a contribution breaking the semiclassical factorisation argument.}
    \label{fig:nonfactorcont}
\end{figure}

However, in the complete sum over topologies and geometries there are potential contributions to the path integral such as handles connecting the $L,R$ sides of the pinwheel (Fig.~\ref{fig:nonfactorcont}) that appear to destroy the factorisation in (\ref{eq:tracefac}).  Note that these contributions need not be saddlepoints; they need not even be continuously deformable to saddlepoints.   By contrast, adding handles to the $L,R$ sheets individually respects factorisation and simply leads to corrections to $\overline{ Z(\beta_1)}, \overline{ Z(\beta_2)}$ separately. To show (\ref{eq:tracefac}) we consider the full path integral equalities $\overline{\mathcal{A}}=\overline{\mathcal{B}}$ and $\overline{(\mathcal{A}-\mathcal{B})^2}=0$ for $\mathcal{A}=Tr_{\mathcal{H}_{LR}}(e^{-H_{L}\beta_1}e^{-H_{R}\beta_2})$ and $\mathcal{B}=Z(\beta_1)\times Z(\beta_2)$, not just in the sum over saddlepoints, but in the complete sum over topologies and geometries. 
We will achieve this in three  steps.

First, we express the trace over $\mathcal{H}_{LR}$ on the left hand side of (\ref{eq:tracefac}) in the $\kappa \to \infty$ shell basis:
\beq
Tr_{\mathcal{H}_{LR}}(e^{-\beta_1H_{L}}e^{-\beta_2H_{R}})= \lim_{n \to -1} G^n_{ij}\langle j| e^{-\beta_1H_{L}}e^{-\beta_2H_{R}} |i\rangle.
\eeq
Second, we make the equivalence of gravitational path integrals more apparent by inserting a resolution of the identity $\mathds{1}= G^{-1}_{ij} |i\rangle\langle j|$ in between the $Z(\beta)=\langle \beta|\beta\rangle$ bra-kets on the right hand side of (\ref{eq:tracefac}): 
\beq \label{eq:ZZallO}
\overline{ Z(\beta_1)\times Z(\beta_2)}=\overline{\langle \beta_1|\beta_1\rangle
\langle \beta_2|\beta_2\rangle}  =  \overline{\langle \beta_1|\mathds{1}|\beta_1\rangle
\langle \beta_2|\mathds{1}|\beta_2\rangle} = \overline{G^{-1}_{ij}G^{-1}_{mn}\langle \beta_1|i\rangle\langle j|\beta_1\rangle\langle \beta_2|m\rangle\langle n|\beta_2\rangle}.
\eeq
We also rewrite the trace as
\beq \label{eq:TrallO}
\overline{Tr_{\mathcal{H}_{LR}}(e^{-\beta_1H_{L}}e^{-H_{R}\beta_2})}=\overline{Tr_{\mathcal{H}_{LR}}(e^{\frac{-\beta_1H_L}{2}}e^{\frac{-\beta_2H_R}{2}}\mathds{1}e^{\frac{-\beta_1H_L}{2}}e^{\frac{-\beta_2H_R}{2}})}=
\\
\overline{G^{-1}_{ij}G^{-1}_{mn}\langle j|e^{\frac{-\beta_1H_{L}}{2}}e^{\frac{-\beta_2H_{R}}{2}} | m\rangle\langle n|e^{\frac{-\beta_1H_{L}}{2}}e^{\frac{-\beta_2H_{R}}{2}}|i\rangle}.
\eeq
The task is now to compare the maximal index loop geometries contributing to 
\beq \label{eq:ZZ}
\overline{G^{w}_{ij}G^{z}_{mn}\langle \beta_1|i\rangle\langle j|\beta_1\rangle\langle \beta_2|m\rangle\langle n|\beta_2\rangle},
\eeq
and 
\beq \label{eq:Tr}
\overline{G^{w}_{ij}G^{z}_{mn}\langle j|e^{\frac{\beta_1H_{L}}{2}}e^{\frac{-\beta_2H_{R}}{2}}|m\rangle\langle n|e^{\frac{-\beta_1H_{L}}{2}}e^{\frac{-\beta_2H_{R}}{2}}|i\rangle}
\eeq
which in the $w,z \to -1$ limit reduce to (\ref{eq:ZZallO}) and (\ref{eq:TrallO}) respectively.  Will refer to contributions to (\ref{eq:ZZ}) and (\ref{eq:Tr}) as as \textit{ZZ} and \textit{Tr} wormholes respectively.   The full path integral equality between (\ref{eq:ZZ}) and (\ref{eq:Tr}) can now be established by considering the corresponding sheet diagrams as we describe below.

\begin{figure}[h]
    
    \begin{subfigure}[b]{\linewidth}
    \centering
     \includegraphics[width=0.6\linewidth]{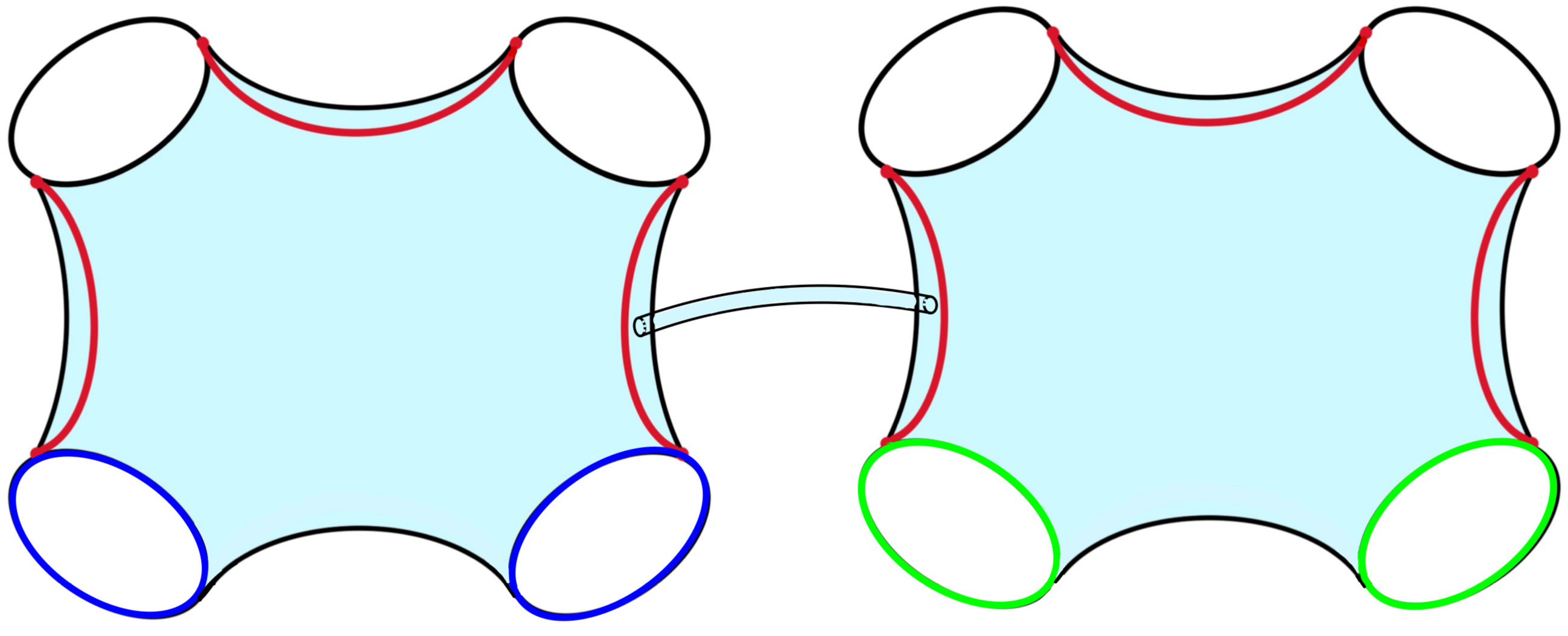}
     \caption{}
     \label{fig:connZZwheel}
    \end{subfigure}
    \hfill
     \begin{subfigure}[b]{\linewidth}
    \centering
    \includegraphics[width=0.8\linewidth]{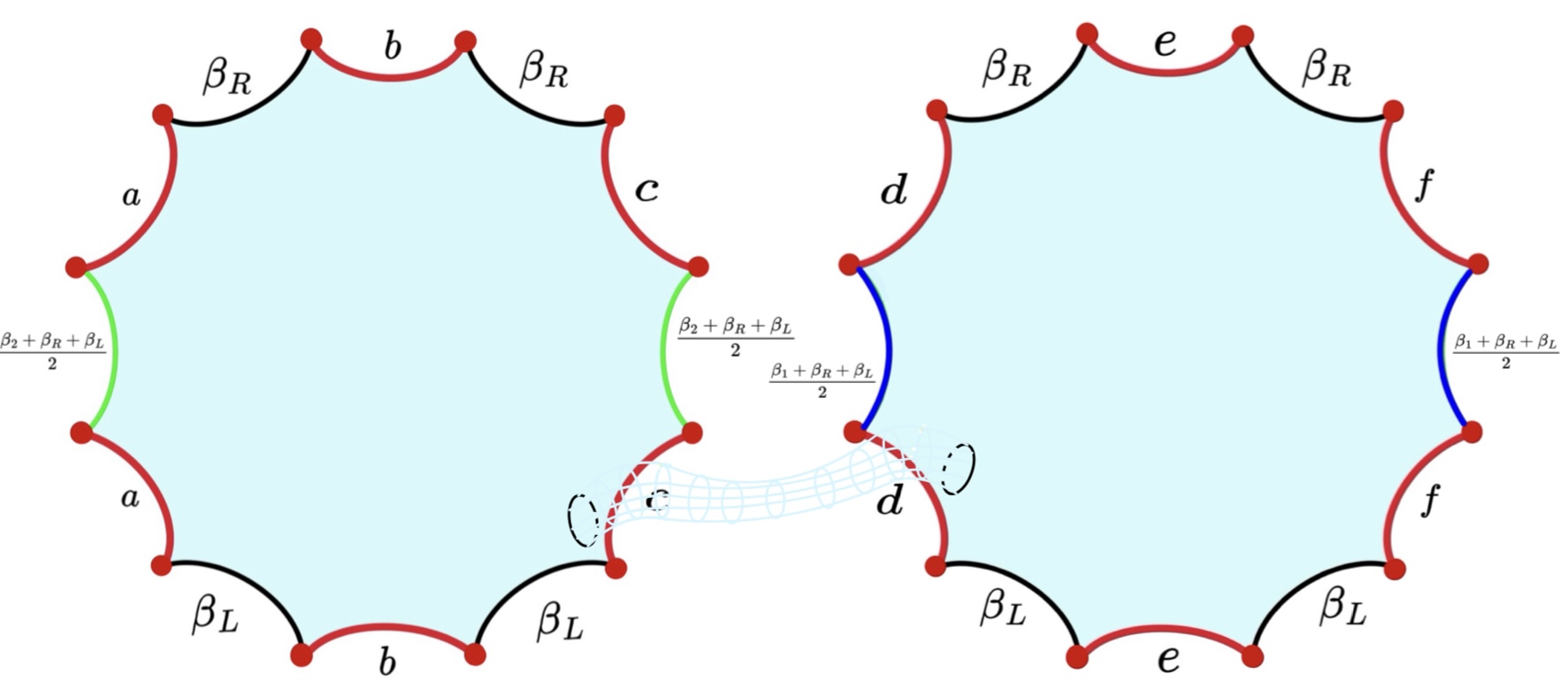}
    \caption{}
    \label{fig:zz corrections}
    \end{subfigure}
    \caption{ ({\bf a}) Wormhole contribution to (\ref{eq:ZZ}) including a non-factorizing handle, depicted for $(w,z)=(2,2)$. The pinwheel with the blue boundaries belongs to the $G^{w}_{ij}\langle \beta_1|i\rangle\langle j|\beta_1\rangle$ insertion and the green to $G^{z}_{mn}\langle \beta_2|m\rangle\langle n|\beta_2\rangle$ ({\bf b}) Corresponding sheet diagram, where $a,b,c \cdots$ denotes the shells to be identified to glue back into the wormhole. The wormhole is obtained by identifying $a \leftrightarrow a,b \leftrightarrow b \cdots i\leftrightarrow i$, and the handle is depicted by a wormhole connecting the two sheets. }
    \label{fig:ZZwheel}
\end{figure}

\paragraph{\textit{ZZ} sheets} The asymptotic boundary structure of the \textit{ZZ} wormholes contain two \textit{separate} maximal index loops corresponding to the $\langle \beta_1|\mathds{1}|\beta_1\rangle$ and $\langle \beta_2|\mathds{1}|\beta_2\rangle$ insertions. For example, the $G^{w}_{ij}\langle \beta_1|i\rangle\langle j|\beta_1\rangle$ boundary index loop consists of two boundaries of length $\frac{\beta_1+\beta_R+\beta_L}{2}$ from the $\langle \beta_1|i\rangle$, $\langle j|\beta_1\rangle$ insertions (green in Fig.~\ref{fig:ZZwheel}), separated by $w$ shell boundary insertions of length $\beta_{L}+\beta_{R}$. At a given $(w,z)$ the  saddles for (\ref{eq:ZZ}) consist of two disconnected folded wormholes, $\tilde{Z}_{w+2}(\beta_1)$ and $\tilde{Z}_{z+2}(\beta_2)$ discussed in Sec.~\ref{sec:Zid}. Hence, in the sum over saddlepoints (\ref{eq:ZZ}) evaluates to $\overline{Z}(\beta_1 + (w+1)\frac{\beta_L+\beta_R}{2})\times\overline{Z}(\beta_2 + (z+1)\frac{\beta_L+\beta_R}{2}) \times \Pi_{k=1}^{w+z+2}Z_{m_k} $. The non saddle geometries contributing to (\ref{eq:ZZ}) include all the contribution to  $\overline{\langle \beta_1|\mathds{1}|\beta_1\rangle}$ and  $\overline{\langle \beta_2|\mathds{1}|\beta_2\rangle}$ individually and all contributions connecting the two wormholes via bulk handles (Fig.~\ref{fig:connZZwheel}), which result in $\overline{Z(\beta_1)} \times\overline{Z(\beta_2)} \neq \overline{Z(\beta_1)Z(\beta_2)}$. The sheet diagrams for a generic contribution therefore consist of two sheets, one for each wormhole, connected by a number of handles, see Fig.~\ref{fig:zz corrections}.

\paragraph{\textit{Tr} sheets} 
\begin{figure}[h]
    \begin{subfigure}{\linewidth}
    \centering
    \includegraphics[width=0.4\linewidth]{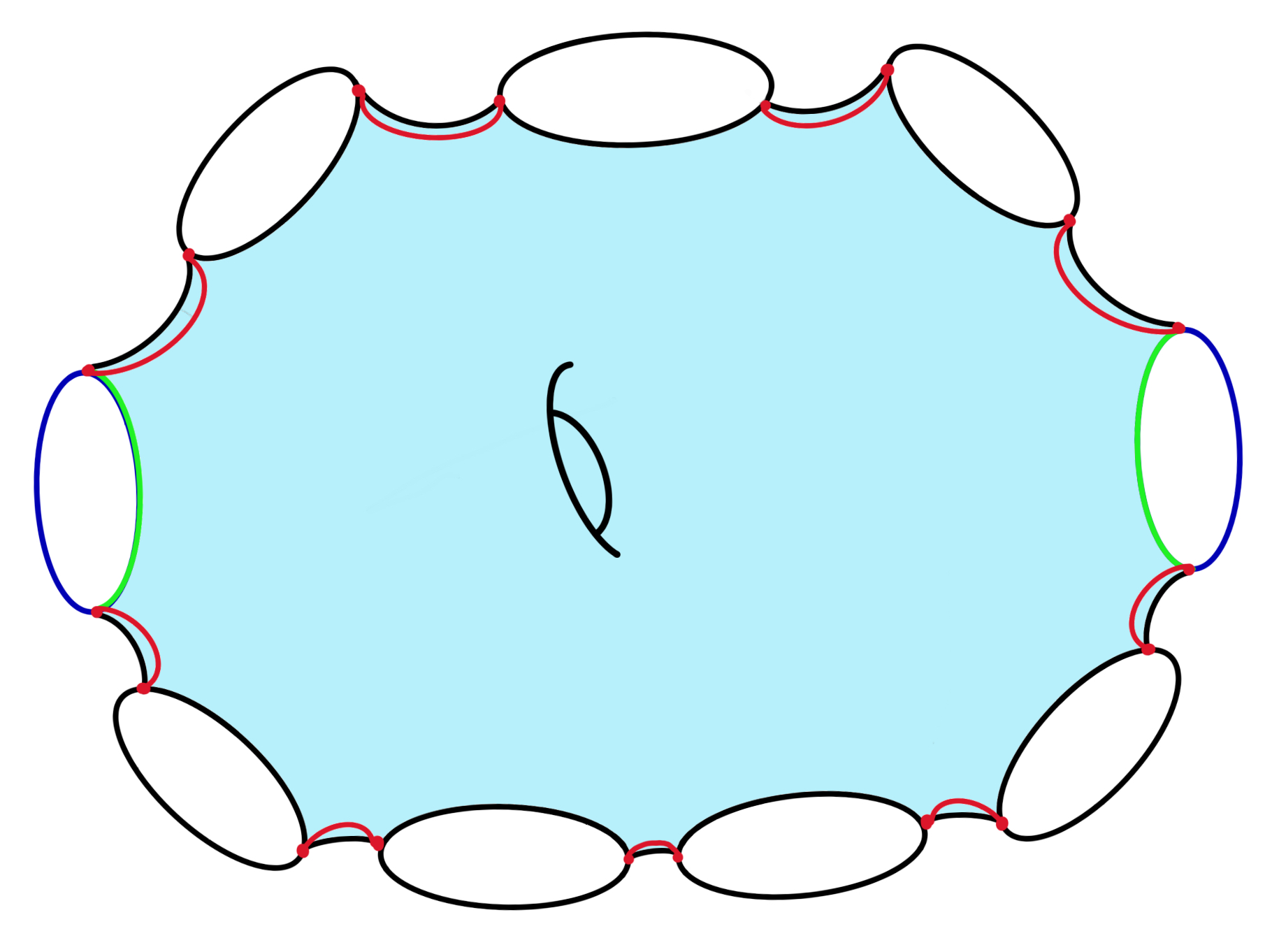}
    \caption{}
    \label{fig:TrWHnondiag}
     \end{subfigure}
     \hfill
     \begin{subfigure}[c]{\linewidth}
    \centering
    \includegraphics[width=0.9\linewidth]{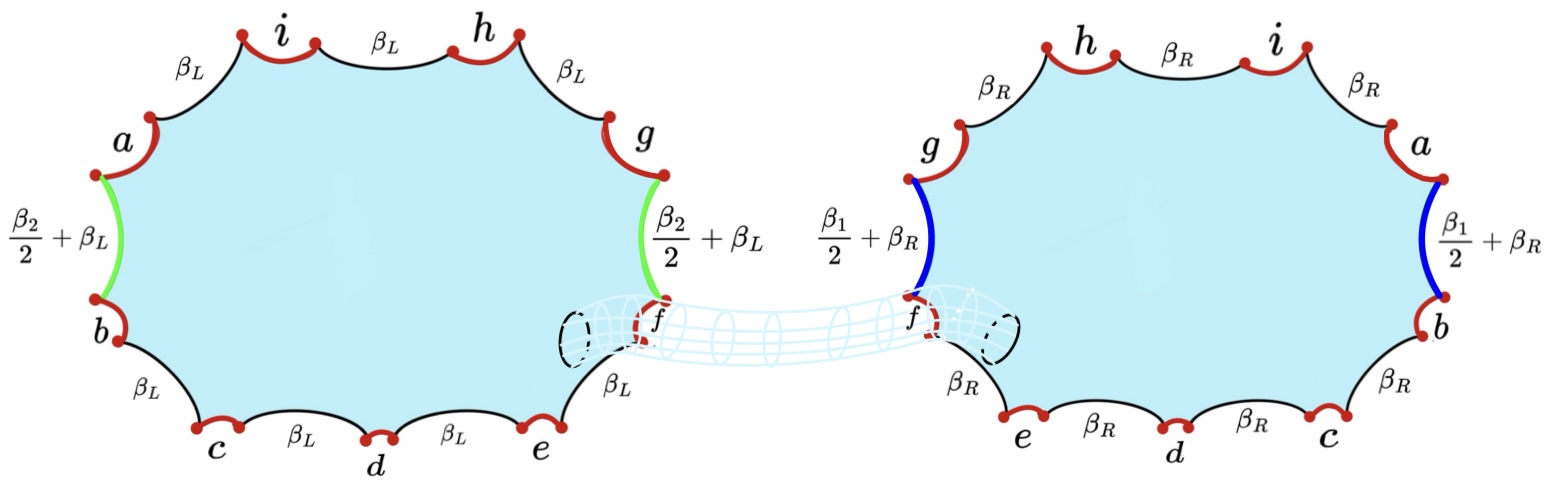}
    \caption{}
    \label{fig:correctTrwheel}

\end{subfigure}
\caption{({\bf a}) Wormhole contribution to (\ref{eq:Tr}), depicted for $(w,z)=(3,4)$. ({\bf b}) Corresponding sheet diagram, the wormhole is obtained by identifying $a \leftrightarrow a,b \leftrightarrow b \cdots i\leftrightarrow i$, the handle is depicted by the wormhole connecting the two sheets.}
     
\end{figure}
In contrast, the contributions to (\ref{eq:Tr}) consist of a single bulk wormhole as there is only a single maximal index loop. This loop consists of a boundary of length $\frac{\beta_1+\beta_2}{2}+\beta_R+\beta_L$ from the $\langle i|e^{\frac{-\beta_2H_{L}}{2}}e^{\frac{-\beta_1H_{R}}{2}}|j \rangle$ insertion followed by  $z$ shell boundaries of length $\beta_{L}+\beta_{R}$ from the resolution of the identity, followed by another boundary of length $\frac{\beta_1+\beta_2}{2}+\beta_R+\beta_L$ and lastly $w$ more shell boundaries from the resolution of the trace. The wormholes saddles are a trivial extension of those in  Sec.~\ref{sec:facSC} as they are constructed by identifying shells across two separate disks. Hence, in the sum over saddlepoints (\ref{eq:Tr}) evaluates to $\overline{Z}(\beta_1 + (w+z+2)\frac{\beta_R}{2})\times\overline{Z}(\beta_2 + (w+z+2)\frac{\beta_L}{2}) \times \Pi_{k=1}^{w+z+2}Z_{m_k} $. The sum over geometries for (\ref{eq:Tr}) includes non-saddle  higher topology pinwheels, see Fig.~\ref{fig:TrWHnondiag}. The sheet diagrams for a generic contribution to (\ref{eq:Tr}) therefore also consist of two sheets, corresponding to the $L,R$ side of the pinwheel, connected by a number of handles, see Fig.~\ref{fig:correctTrwheel}.

\paragraph{Diagonal terms}
There is some freedom left to fix in (\ref{eq:ZZ}) and (\ref{eq:Tr}) to make equality explicit. Apart from the shell identifications across/within the two sheets, the sheet diagrams (\ref{eq:ZZ}) and (\ref{eq:Tr}) at a given $(w,z)$ are identical for the {\it diagonal} $w=z$ terms if we choose the shell preparation temperatures to satisfy $\beta_L=\beta_R\equiv \beta_S $. As the shell states span $\mathcal{H}_{LR}$ at any preparation temperature this can always be done. Furthermore, for this choice of parameters for (\ref{eq:ZZ}) and (\ref{eq:Tr}) are equal in the saddle point approximation, as can be seen by expressions in the above two paragraphs. Schematically the path integrals (\ref{eq:ZZ}) and (\ref{eq:Tr}) for a given $(w,z)$ will evaluate to certain functions $g(w,z)$ and $f(w,z)$ where $\overline{Z(\beta_1)Z(\beta_2)}=\lim_{w,z \to -1} g(w,z)$ and $\overline{Tr_{\mathcal{H}_{LR}}(e^{-\beta_1H_{L}}e^{-\beta_2H_{R}})}=\lim_{w,z \to -1} f(w,z)$. This limit must either be independent of the contour in the $(w,z)$ plane or the contour must be chosen to recover $\overline{Z(\beta_1)Z(\beta_2)}$ and $\overline{Tr_{\mathcal{H}_{LR}}(e^{-\beta_1H_{L}}e^{-\beta_2H_{R}})}$. Either way, we are interested in the $ w,z \to -1$ limit of (\ref{eq:ZZallO}) and (\ref{eq:TrallO}) along the diagonal $w=z$ contour. We choose this contour, because the Gram matrix resolution of the trace ($Tr_{\mathcal{H}_{shell}}(\cdots) = G^{-1}_{ij} \langle j|\cdots| i\rangle$) and identity ($\mathds{1} = G^{-1}_{ij} |i\rangle \langle j|$) are related.  So from a physical standpoint we expect that they should be defined  by analytic continuation of the same powers of the Gram matrix in given calculation.   Alternatively we could think about this as an argument for how to choose the correct contour in analytic continuation to define the physical theory.  It would be useful to also have a strictly mathematical argument for why this is the sensible analytical continuation, but we leave that for future work.

\begin{figure}[h]
    \centering
    \includegraphics[width=0.5\linewidth]{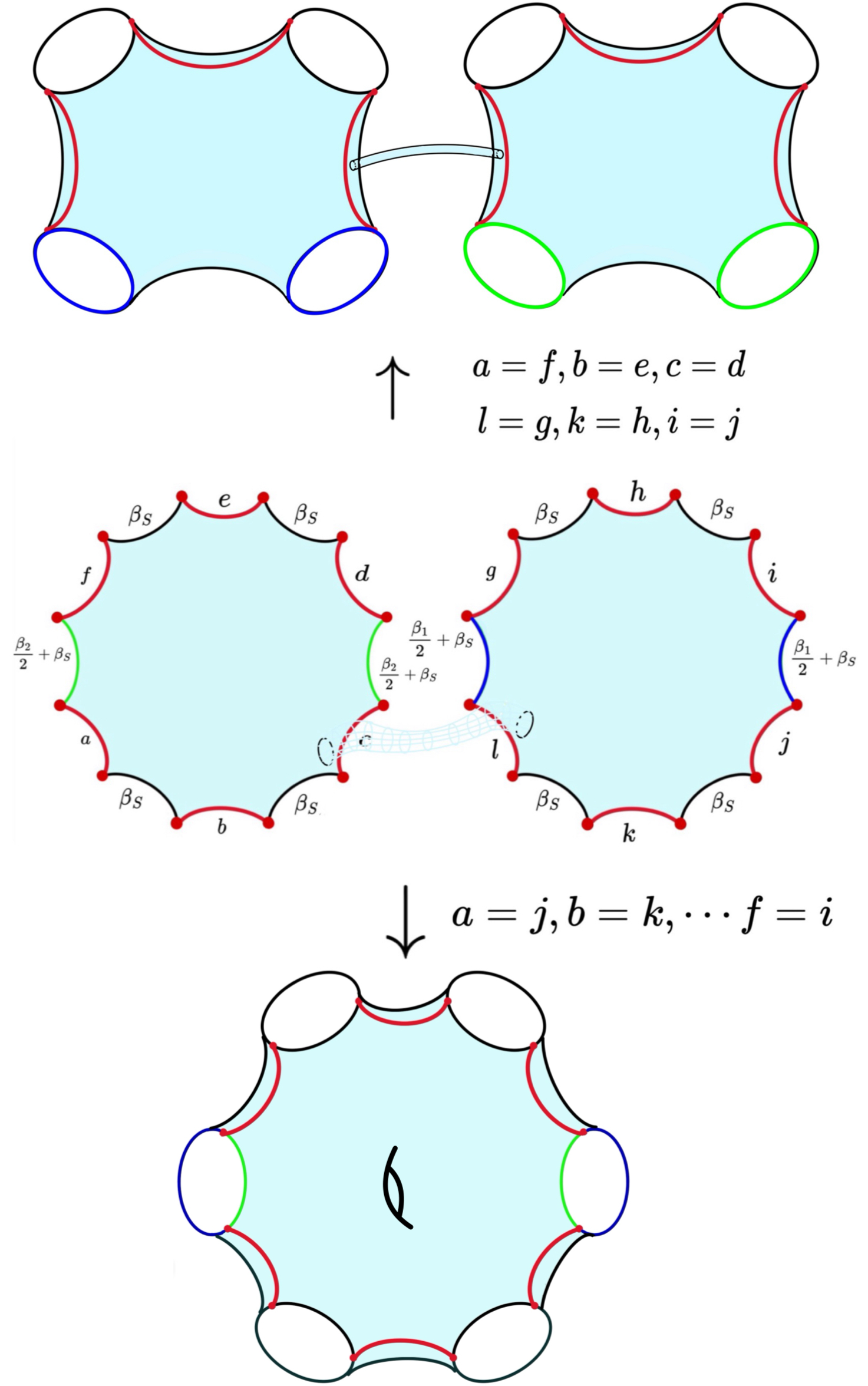}
    \caption{A given sheet geometry can be  re-glued into a \textit{Tr} wormhole via the identification pattern $a=j,b=k ,c\dots f=i$ whereas a \textit{ZZ} wormhole is obtained by $a=f, b=e , \dots k=h, j=i$.}
    \label{fig:re-glue}
\end{figure}

\subsubsection{Bijection by cutting and gluing}
Along the diagonal contour the semiclassical saddlepoint actions for (\ref{eq:ZZ}) and (\ref{eq:Tr}) are equal. To extend this equality beyond the sum over saddlepoints we will show that there is a bijection between  arbitrary  contributions with the same action for the full path integrals  (\ref{eq:ZZ}) and (\ref{eq:Tr}). Starting with a $L,R$ sheet geometry a \textit{Tr} wormhole is obtained by identifying the shells across the $L,R$ sheets in the pattern $a=g, d=j \cdots$, whereas identifying shells within the same sheet in the pattern $a=d,b=e \cdots$ and $g=j , h=k , \cdots$ results in a \textit{ZZ} wormhole, see Fig.~\ref{fig:re-glue}. If re-gluing any contribution to (\ref{eq:ZZ}) into a contribution to (\ref{eq:Tr}) and vice versa is possible without changing the action the full path integrals are the equal. To show this is the case we divide the action of a geometry $\mathcal{M}$ contributing to (\ref{eq:ZZ}) or (\ref{eq:Tr}) as:
\beq
I_{\mathcal{M}} =  I_{bulk}  + I_{GHY}
\eeq
where $I_{local}$ is the local bulk action 
and $I_{GHY}$ the Gibbons-Hawking-York boundary terms \cite{York:1972sj,Gibbons:1976ue}.

\paragraph{Re-gluing leaves the boundary action unchanged}
The contributions $I_{GHY}$ are determined by the asymptotic boundary structure of the glued wormholes. In gluing sheet diagrams into wormholes the asymptotic boundary sections of the sheet are compactified into circular boundaries as in Fig.~\ref{fig:re-glue}.  To compute the boundary terms the path integral should to be gauge fixed by choosing a conformal frame, say the Fefferman-Graham gauge \cite{Fefferman:2007rka} for the case of the asymptotically locally AdS  gravity path integral. Fixing a frame fixes the asymptotic behavior for any metric in the sum over geometries, and hence the boundary terms are universal and proportional to the length of the asymptotic boundary circle. Hence gluing a sheet into \textit{Tr} or \textit{ZZ} wormhole results in the same $I_{GHY}$.

The \textit{Tr} wormhole saddlepoints where constructed by gluing two Z-partition function saddles together along the shells. We can decorate this procedure with perturbative $\mathrm{G_N}$ corrections, for which the shell homology regions still pinch off in the large shell mass limit, and the shell worldvolumes trajectories are functions of the shell mass only, independent of the background geometry. This is because the heavy shells localize to the asymptotic region which does not fluctuate. Therefore the perturbatively corrected saddles \textit{Tr} wormholes correspond to the perturbative corrections to two Z-partition functions and some universal shell terms. Hence for every correction to (\ref{eq:ZZ}) we obtain a correction to (\ref{eq:Tr}) with the same action by simply changing the shell identification pattern, and vice versa, extending (\ref{eq:tracefac}) to include perturbative corrections around saddlepoints.  Extending the argument to all topologies, even those that do not support saddlepoints, and to complete sum over geometries, requires more careful consideration.

\begin{figure}
\begin{subfigure}{\linewidth}
    \centering
    \includegraphics[width=0.8\linewidth]{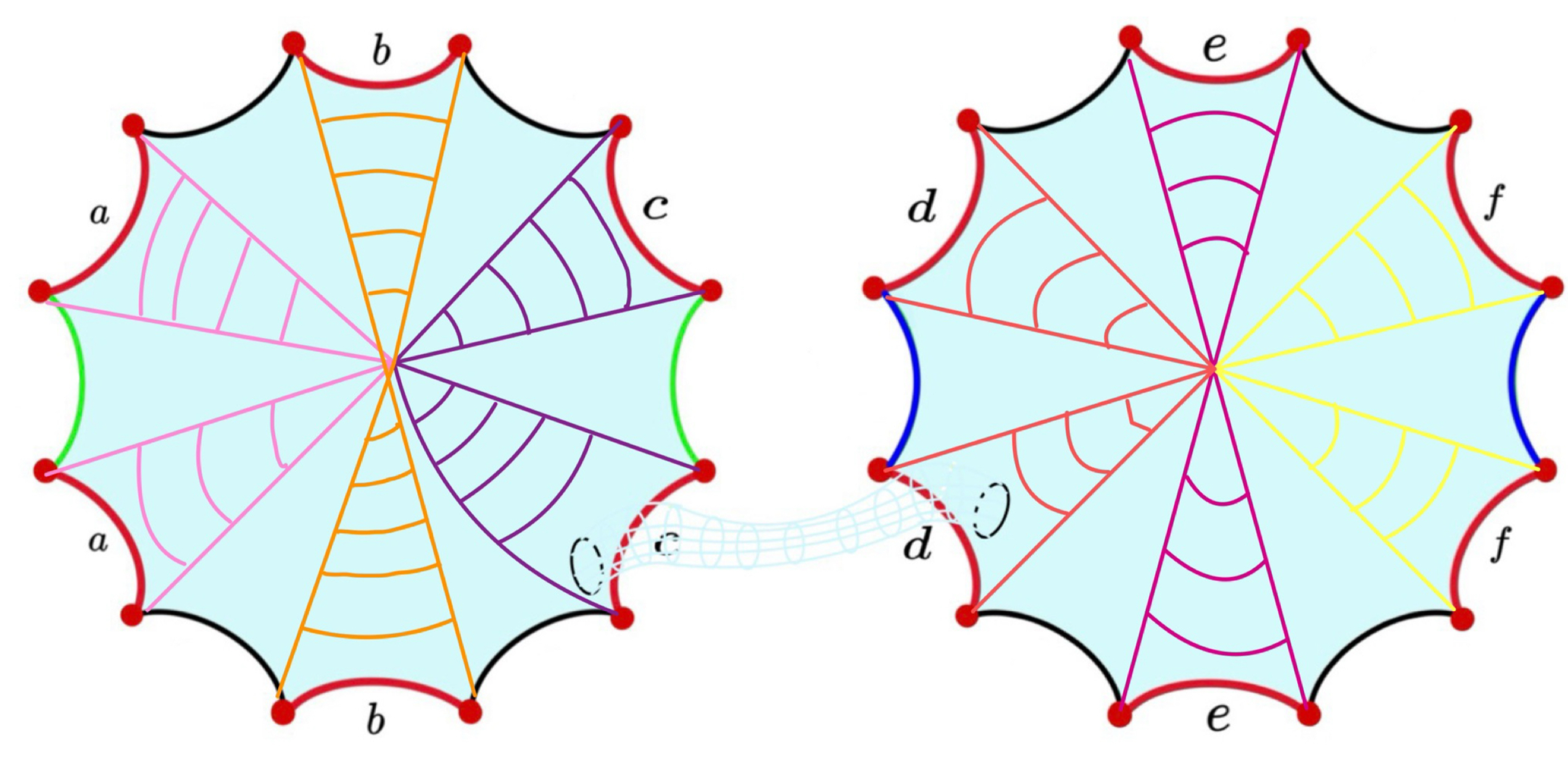}
    \caption{Sheet for a \textit{ZZ} wormhole geometry including a handle connecting the two sheets depicted in blue.}
    \label{fig:step1}
     \end{subfigure}
     \hfill
    \begin{subfigure}{\linewidth}
    \centering
    \includegraphics[width=0.8\linewidth]{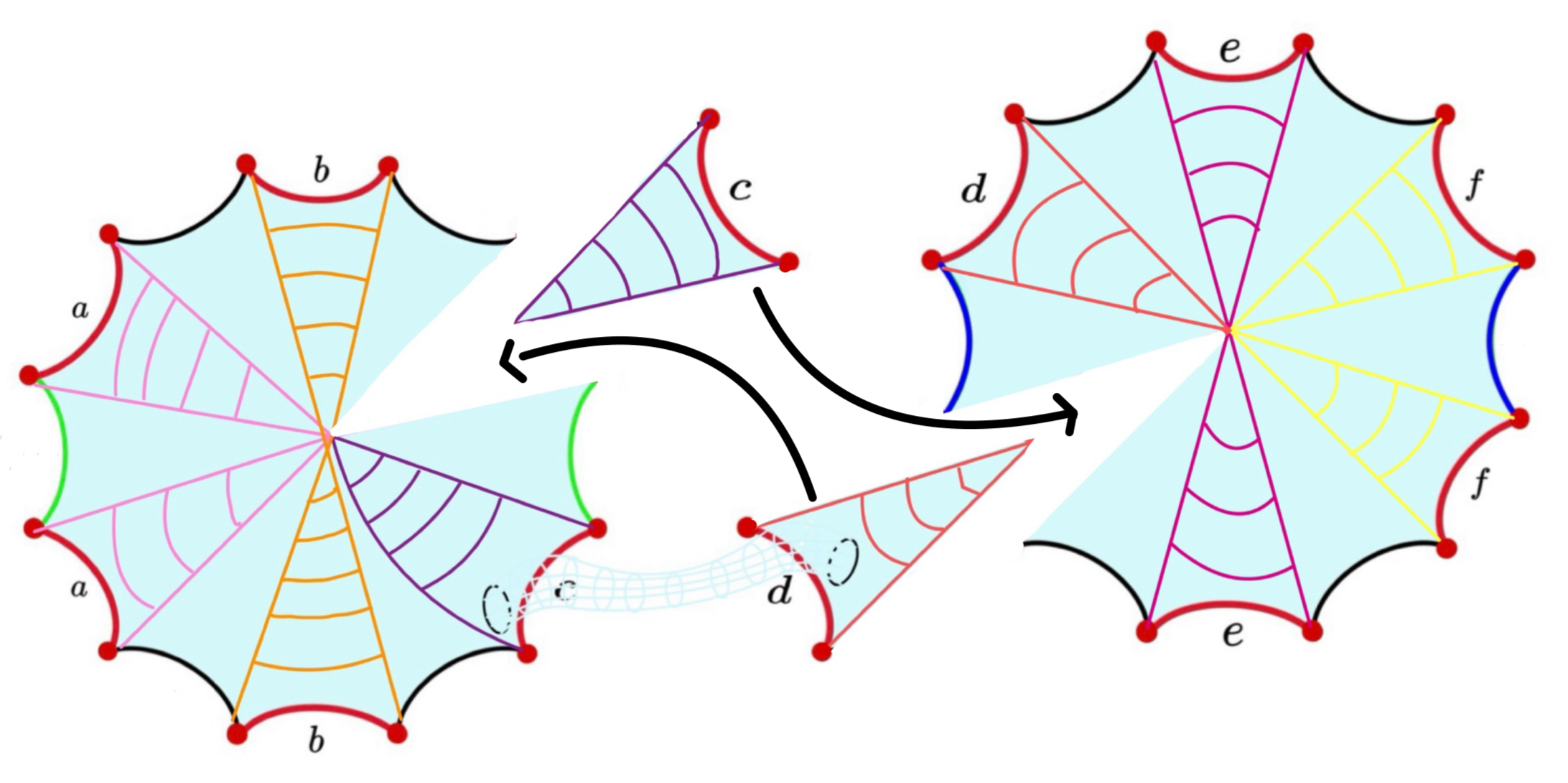}
    \caption{Swap the $d,e,f$ wedges with the $a,b,c$ wedges.}
    \label{fig:step2}
   
     \end{subfigure}
     \hfill
     \begin{subfigure}[c]{\linewidth}
    \centering
    \includegraphics[width=0.8\linewidth]{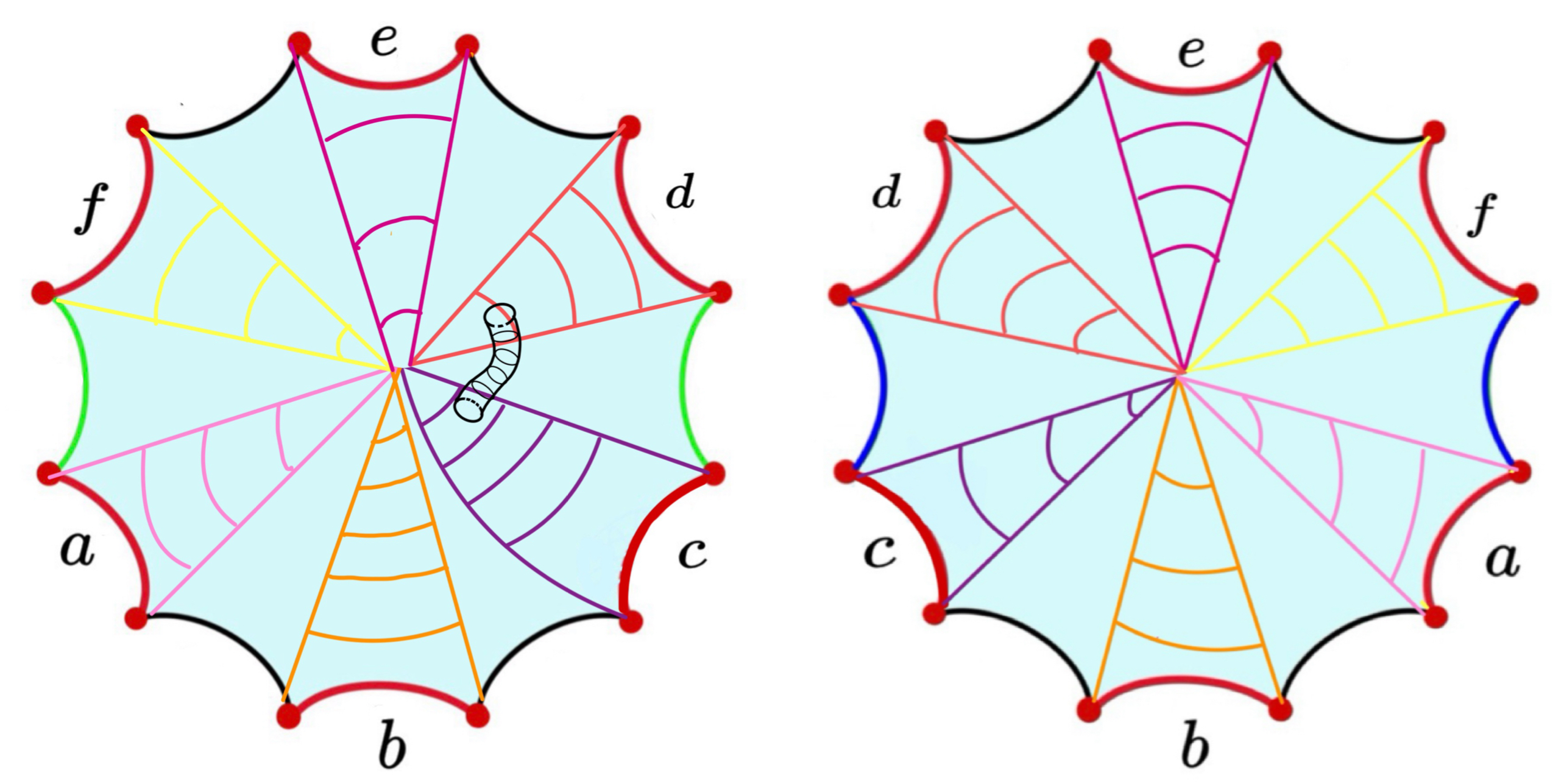}
    \caption{Sheet for a \textit{Tr} wormhole with the same action as the original \textit{ZZ} wormhole geometry, this geometry now includes a handle on only one of the sheets depicted in black.}
    \label{fig:step3}
\end{subfigure}
\caption{Diagrammatic sketch of the procedure mapping geometries contributing to (\ref{eq:ZZ}) to contributions to (\ref{eq:Tr}) with the same action.}
\label{pizza}
\end{figure}

\paragraph{Re-gluing leaves the bulk action unchanged} 
To do this extension we start with a \textit{ZZ} wormhole geometry $\mathcal{M}_{ZZ}$. We divide the $\mathcal{M}_{ZZ}$ sheet into alternating wedges, each including a certain shell or asymptotic boundary as in Fig.~\ref{fig:step1}. For higher topology contributions some of the resulting wedges might be connected by a number of handles. We can take the two wedges containing an identified shell, say the two $c$-wedges on the $L$ sheet and the $d$-wedges on the $R$ sheet, and move one of the $d$ wedges onto the $L$ sheet by swapping it with one of the $c$ wedges, see Fig.~\ref{fig:step2}.
Doing this for every pair of identified shell we are left with $L$ and $R$ sheets that each have the \textit{Tr} identification pattern, see Fig.~\ref{fig:step3}.  The rearrangement produces a \textit{Tr} geometry from the  original \textit{ZZ} geometry. The smooth metrics on the two spaces will match everywhere except possibly on the sets of measure zero where cut and re-joined them.  Hence their actions, computed from local integrals, will be the same. Since these smooth metrics constitute the finite contributions to the path integral we conclude that the path integrals over metrics on the  \textit{ZZ} and \textit{Tr} geometries should be the same. Here we assume that the re-glued geometry has the same measure factor as the original geometry. This requires more rigorous justification. It would be interesting to show this explicitly in lower dimensional models of gravity. 

In conclusion, by cutting and re-gluing any contribution to (\ref{eq:Tr}) results in a valid contribution to (\ref{eq:ZZ}) with the same action and vice versa. Hence the full path integrals (\ref{eq:Tr}) and (\ref{eq:ZZ}) are the same. Taking the $\lim_{w,z \to -1}$ limit this derives the full equality (\ref{eq:tracefac}). The above argument can trivially be extended to show $\overline{\left(Tr_{\mathcal{H}_{LR}}(e^{-H_{L}\beta_1}e^{-H_{R}\beta_2}) - Z(\beta_1)\times Z(\beta_2)\right)^2}=0$. Hence the full  fine-grained trace factorises (\ref{eq:tracefac}), even though the gravitational path integral includes explicitly non-factorizing off-shell geometries.

\paragraph{Direct limit}
Alternatively, (\ref{eq:tracefac}) can be derived by a slight modification of the $n \to -1$ limit argument of Sec.~\ref{sec:completebasis}. This approach avoids having  choosing the diagonal contour and $\beta_L=\beta_R$ restriction in the above argument. We do not repeat the full argument here, as it is a simple extension of the method of Sec.~\ref{sec:completebasis}, but briefly outline the key points. The wormhole contributions to $\overline{G^{n}_{ij}\langle i|e^{-H_{L}\beta_2}e^{-H_{R}\beta_1}|j\rangle}$ can be organized by topology and we can cut out small regions around the shells to split up the action of each contribution. This produces two sheets, possibly connected by handles, and we again sum over the completions of each of the sheets into an asymptotic boundary circle.  As in Sec.~\ref{sec:completebasis} in the ${n \to -1}$ limit the only part that survives is the path integral subject to the two circular asymptotic boundary conditions, the lengths of which limit to $\beta_1,\beta_2$ respectively.  Thus as $\lim_{n \to -1}$ the path integral computes $\overline{Z(\beta_1)Z(\beta_2)}$. The argument can be repeated for the cross terms in $\overline{(\mathcal{A}-\mathcal{B})^2}=0$ to show fine-grained factorisation.

\section{Summary and discussion} \label{sec:discussion}
We derived our results using only the rules for the gravitational path integral and the Einstein-Hilbert action treated as a universal effective field theory for gravitational systems. In particular, we did not assume any form of holographic duality.  While for concreteness we phrased some of the discussion in terms of asymptotically AdS gravity,  the main results carry over unchanged to the asymptotically flat space. Our saddlepoint calculations relies on the explicit form of the gravitational action, but the general method we proposed to showing equivalences between  path integrals only requires that the action consists of local bulk and boundary terms. We 
therefore expect the relational results to be insensitive to higher  derivative  corrections to the Einstein-Hilbert action arising from, e.g., its possible string theoretic completion at high energies.  Below we summarize the toolkit that we have developed and discuss some implications.

\subsection{Algorithm for deriving  relational equivalence} \label{sec:alg}
As discussed, the gravitational path integral seemingly has only access to coarse-grained data of an underlying fine grained theory, and is difficult to compute beyond the saddlepoint approximation. These difficulties can be overcome by collecting together the three tools we have developed into a general framework for deducing  equivalences between quantities in the fine-grained theory using the gravitational path integral. Suppose we want to show that two \textit{a priori} distinct quantities in the fine grained theory, $\mathcal{A}$ and $\mathcal{B}$, are equal.  We can do so by demonstrating the equivalence of the full path integrals $\overline{\mathcal{A}}{=}\overline{\mathcal{B}}$ and $\overline{(\mathcal{A}-\mathcal{B})^2}{=}0$. We achieve this by taking the following three steps:

\paragraph{Step 1: Express quantities in terms of shell states in the $\kappa \to \infty$ limit.}
In Secs.~\ref{sec:scspan} and \ref{sec:completebasis}  we argued that the shell states span the entire gravity Hilbert space $\mathcal{H}_{LR}$ in the $\kappa \to \infty$ limit. Use this fact to resolve the trace and identity on $\mathcal{H}_{LR}$  in terms of the shell states via Eq.~(\ref{eq:id}) and Eq.~(\ref{eq:shellTr}), which involve the inverse of the  shell Gram matrix defined as $\lim_{n \to -1} \overline{G^{n}_{ij}}$. In the $\kappa \to \infty$ limit the gravitational path integrals for these quantities include only the fully-connected geometries for each maximal index loop, leading to a highly simplified sum over topologies. 

\paragraph{Step 2: Insert the identity to obtain equality for positive integers $n$.}
It may be that the computation of $\overline{\mathcal{A}}$ requires an analytic continuation $\lim_{n \to -1}\overline{\mathcal{A}}(n)$ involving some insertions of $G^n_{ij}$, while this is not the case for $\mathcal{B}$. 
To show that $\lim_{n \to -1}\overline{\mathcal{A}}(n) = \overline{{\mathcal{B}}}$, reorganize the path integral by using the arguments in Sec.~\ref{sec:completebasis}. Alternatively, as in Sec.~\ref{sec:Trzz} we can establish equivalence by appropriately inserting a shell resolution of the identity $\mathds{1}_{\mathcal{H}_{LR}}=\lim_{n \to -1} G^{n}_{ij}|i\rangle\langle j|$ in the path integral for $\mathcal{B}$ so that  $\overline{\mathcal{B}} = \lim_{n \to -1}\overline{\mathcal{B}}(n)$, where by $\overline{\mathcal{B}}(n)$ we mean the path integral with $\mathds{1}_{\mathcal{H}_{LR}}$ inserted.  This procedure allows us to directly compare  $ \overline{\mathcal{A}}(n) \stackrel{?}{=} \overline{\mathcal{B}}(n)$ at fixed positive $n$. We then establish equality  by comparing the corresponding sheet diagrams as described in Step 3.

\paragraph{Step 3: Compare sheet diagrams to deduce a bijection of geometrical actions.}
Each of the  wormholes contributing to 
$\overline{\mathcal{A}}(n)$ and $\overline{\mathcal{B}}(n)$ can be cut along the shell worldvolumes to produce a set of sheet diagrams. These serve as a visual tool for organizing the sum over geometries in the $\kappa \to \infty$ limit, and hence are a geometrical analogue of Feynman diagrams.\footnote{Note that the sheet diagrams not literally gravitational Feynman diagrams representing graviton scattering.} Even if neither $\overline{\mathcal{A}}(n)$ or $\overline{\mathcal{B}}(n)$ can be explicitly calculated, their equivalence can be established by showing a bijection that maps a given $\mathcal{A}$-sheet to a $\mathcal{B}$-sheet with the same action and vice versa. We argued that such equalities can be naturally extended to include perturbative corrections around the saddle points. We also outlined how these equalities might be further generalized to incorporate geometries in the path integral whose topologies do not admit a classical saddle point. Although this outline relies on several heuristic assumptions regarding the integration measure and the analytic continuation of certain contributions, we expect the schematic to capture the essential qualitative features. It would be interesting to realize these ideas explicitly in lower-dimensional models of gravity, such as JT gravity or AdS$_3$ gravity.

\subsection{Macroscopic superpositions of semiclassical geometries}
The fact that the canonical shells states span $\mathcal{H}_{LR}$ at \textit{any} preparation temperature (\ref{Hshellspan}) has a striking consequence.  For example,  the basis change  
\beq \label{eq:basischange}
|i,\beta_L,\beta_R \rangle = \sum_{j} a_j(i,\beta_L,\beta_R) |j,\tilde{\beta}_L,\tilde{\beta}_R \rangle
\eeq
allows us to express a type 1 black hole state consisting of shells behind a horizon as a superposition of the over-complete basis states of type 2 horizon-less geometries which include a disconnected closed universe. This is evident after including the $\mathcal{O}(e^{-1/\mathrm{G_N}})$ wormhole corrections, while in the strict  $\mathrm{G_N} \to 0$  limit all the shell states at any temperature are linearly independent. 

Focusing on the AdS case, as $\overline{\langle i|i\rangle}=\overline{Z}(\beta_L) \overline{Z}(\beta_R)Z_{m_i}$ (see Sec.~\ref{sec:shellgeom}) a given shell state could be associated to nine possible saddles geometries, as there are three potential saddlepoints for each $Z(\beta_{L})$ and $Z(\beta_{R})$ -- the large and small AdS black holes which have horizons, and thermal AdS which does not. For any $\beta_L,\beta_R$ one of the saddles dominates  and this in the perturbative $\mathrm{G_N} \to 0$ limit only this one survives.  (Note that the small AdS black hole never dominates.) Hence in the $\mathrm{G_N} \to 0 $ limit we can associate a single geometry to a canonical shell state, but in general we should always think of it as a superposition which involves saddlepoint configurations with and without horizons. That said, the type 2 shell states with both preparation temperatures  below the black hole threshold $\beta_{L,R}> \frac{ 2\pi \ell }{\sqrt{2}}$, where $\ell$ is the AdS radius, are only associated to thermal AdS saddles.

In the microcanonical ensemble a black hole saddle can be selected out by imposing a  boundary condition fixing the ADM energy (see \cite{Chua:2023srl,Marolf:2018ldl}). Together (\ref{eq:SCTrfac}) and (\ref{Hshellspan}) imply that microcanonical shell states can also be obtained via the Laplace transform of the canonical states:
\beq \label{eq:MCstate}
|i\rangle_{E_L, E_R} \propto \int d\beta_{L}d\beta_{R} e^{\frac{\beta_{R}E_R+\beta_{L}E_L}{2}} \sqrt{Z(\beta_L)Z(\beta_R)}|i,\beta_L,\beta_R \rangle.
\eeq
Moreover, using the basis change (\ref{eq:basischange}) a microcanonical state (\ref{eq:MCstate}) can also be expanded as a superposition shell states at a \textit{fixed} temperature:
\beq \label{eq:BHastype2}
|i\rangle_{E_L, E_R} \propto \sum_{j}   \int d\beta_{L}d\beta_{R} e^{\frac{\beta_{R}E_R+\beta_{L}E_L}{2}} a_j(i,\beta_L,\beta_R) \sqrt{Z(\beta_L)Z(\beta_R)} |j,\tilde{\beta}_L,\tilde{\beta}_R \rangle \equiv
\\
\sum_{j} b_j(i,E_L, E_R) |j,\tilde{\beta}_L,\tilde{\beta}_R \rangle.
\eeq
Hence by choosing $\tilde{\beta}_L,\tilde{\beta}_R> \frac{ 2\pi \ell }{\sqrt{2}}$, a microcanonical black hole state, associated to a saddlpoint with a horizon, can be written as a superposition of type 2 geometries without horizons, but which have disconnected closed universe components instead (see Fig.~\ref{type2}).  Observers inside the compact disconnected universe in the latter cannot send signals to an asymptotic observer, much like observers in the black hole interior.  So like the black hole interior the events inside the closed universe component are hidden from infinity.  However, unlike for a black hole, the asymptotic observer cannot send a classical probe into this hidden region. That said, the Euclidean state preparation path integral entangles the quantum state in the compact component with the state in the asymptotic component.   This means that observers in either region might be able to use the entanglement of the state to send signals to each other, similar to recent proposals that the black hole interior lies within the entanglement wedge of the distant Hawking radiation, at least at late times \cite{Almheiri:2019qdq,Penington:2019kki}. See \cite{Antonini:2023hdh} for a detailed discussion of the encoding of the compact universe into the boundary theory.

\paragraph{Geometry must be emergent} \label{sec:holography}
One consequence of these observations is that bulk geometry cannot in any sense be a linear quantum observable. Consider a state $|\Psi\rangle $ for which the norm is associated to a single gravitational saddle with a metric $\mathbf{g}_{\Psi}$ and define this as the semiclassical geometry of the state:
\beq
\overline {\langle\Psi|\Psi\rangle} \approx e^{-I[\mathbf{g}_{\Psi}]} \rightarrow   |\Psi \rangle = |\mathbf{g}_{\Psi} \rangle .
\eeq
If this notion of the metric associated to a state is a well-defined quantum observable it should correspond to
the eigenvalue of some Hermitian ``geometry operator'' $\mathcal{G}$: 
\beq \label{eq:geom_operator}
\mathcal{G} |\mathbf{g}_{\Psi} \rangle = \mathbf{g}_{\Psi} |\mathbf{g}_{\Psi} \rangle
\eeq
See \cite{Balasubramanian:2007zt} for an example where such a notion can be defined reasonably precisely in the context of the half-BPS states of supersymmetric Yang-Mills and the dual AdS$_5$ gravity.  In our case, since the $\tilde{\beta}_L,\tilde{\beta}_R> \frac{ 2\pi \ell }{\sqrt{2}}$ type 2 shell states form a basis for the entire Hilbert space $\mathcal{H}_{LR}$, and have only a single horizonless saddle computing the norm, we can say that the latter geometric states form a complete basis of eigenvectors for $\mathcal{G}$. But since the microcanonical black hole states also have a single saddle computing the norm, but have a semiclassical geometry with horizons distinct from that of any type 2 state, (\ref{eq:BHastype2}) implies it is impossible for $\mathcal{G}$ to be hermitian with respect to the non-perturbative inner product. A version of this claim was made in \cite{Balasubramanian:2022gmo}, were it was suggested that micro-canonical type 1 shell with a very long wormhole can be understood as superpositions of short wormholes, suggesting that there is no linear operator that extracts the wormhole length from the underlying state. The nonexistence of a linear geometry operator is compatible with the conjecture that bulk connectedness emerges from quantum entanglement  \cite{Maldacena:2013xja,VanRaamsdonk:2010pw}, as entanglement cannot be measured by a linear operator.

In the $G_{N}\to 0 $ limit, shell states at any temperature are linearly independent, and hence the operator (\ref{eq:geom_operator}) can be Hermitian with respect to the perturbative inner product. This matches expectations from AdS/CFT that the semiclassical gravity description of the CFT emerges in the strict $N \to \infty$ limit, which translates to the $G_{N}\to 0 $ limit. Related ideas have been developed from the perspective of state-dependence of the description of black hole interiors \cite{Papadodimas:2015jra,Papadodimas:2015xma}, non-isometry in the holographic code \cite{Akers:2022qdl,Antonini:2024yif}, the nonexistence of a topology operator \cite{Jafferis:2017tiu}, the exterior description of a black hole \cite{Concepcion:2024nwv}  and the algebraic description of the semiclassical limit in gravity \cite{Leutheusser:2021qhd,Leutheusser:2022bgi,Chandrasekaran:2022eqq,Chandrasekaran:2022cip,Witten:2021unn}.

\paragraph{Nonperturbative definition of bulk geometry}
Despite the considerations described above, bulk observers can measure the spacetime they find themselves in. Thus, some meaningful non-perturbative definition of bulk geometry is necessary. In lower dimensional models of gravity a basis of states explicitly associated with bulk time-slice geometries can be constructed. In two dimensional Jackiw-Teitelboim (JT) gravity the perturbative Hilbert space is spanned by length basis $|\ell\rangle$, which are eigenstates of the operator $\hat{\ell}$ that measures the geodesic length between the two asymptotic boundaries (see for example \cite{Harlow:2018tqv,Yang:2018gdb}). However, as shown in \cite{Iliesiu:2024cnh} these states too are over-complete due to non-perturbative effects and therefore $\hat{\ell}$ is also not a well defined linear operator in the nonpertubative theory. 
To resolve this they proposed a nonperturbative definition of the length operator: $\mathcal{F}^{-1}\left(\int d\ell \mathcal{F}(\ell) |\ell\rangle \langle \ell | \right)$ for some real and invertible function $\mathcal{F}$. As this operator maps null states to null states it is a well defined linear operator in the quantum theory.  Several authors approached this problem from the alternative perspective of a dual Double Scaled SYK (DSSYK) model.  First, \cite{Berkooz:2018qkz} showed that that the  large system, low energy limit, of the DSSYK theory is a classical limit, and has an effective description that matches  JT gravity in two dimensions.  Then, \cite{Lin:2022rbf,Rabinovici:2023yex} pointed out that the length of wormholes in  JT gravity is precisely computed by the spread complexity of \cite{Balasubramanian:2022tpr} of the DSSYK model.  Finally, \cite{Balasubramanian:2024lqk} showed how, away from the classical limit, small quantum effects truncate the apparently infinite dimensional Hilbert space identified by \cite{Berkooz:2018qkz} to finite dimension. For similar reasons even if a bulk time-slice Hilbert space can be constructed in higher dimensions we expect it also to be overcomplete. Hence it again cannot support  a linear geometry operator.

\subsection{The gravity Hilbert space in the presence of an observer}
To find a way of defining what we mean by saying that a state has a definite geometry, it may help to understand how observers should be included in the gravitational path integral.
Recently, \cite{Abdalla:2025gzn} proposed a rule for calculating observables in the presence of a gravitating observer (see also \cite{Harlow:2025pvj,Akers:2025ahe} for discussion of rules for defining observers). The idea is that for every inner product inserted in the gravitational path integral, the observer's worldline should propagate between the boundaries defining the corresponding bra and ket, and not to any other.  Using this rule \cite{Abdalla:2025gzn}  showed that the Hilbert spaces of closed universes and the black hole interior are larger relative to an observer. These results suggest that we should imagine an additional holographic theory living on the world-line of the observer. It would interesting to use the simplified setting provided by the toolkit we have developed to probe this observer Hilbert space in more detail. For example, an observer should be able to probe its local patch of spacetime, and therefore it is unlikely that  geometries with a horizon can be expanded in terms of geometries without in the presence of an observer. Such an effect would reduce the linear dependence of  geometries shown in our results, perhaps providing a physical mechanism for the increase in the dimension of the Hilbert space when an observer is present.

Recently \cite{Antonini:2024mci} used AdS/CFT to argue that two distinct semiclassical geometries can be assigned to the type 2 shell states. One is the leading saddle in the gravitational path integral computing its norm, discussed in (\ref{sec:tool1}), and the other is obtained by the extrapolate dictionary, which results in two copies of empty AdS with some operator insertion, but without a compact big-crunch universe. It would be interesting to understand whether including an observer in the type 2 shell states and applying the rule from \cite{Abdalla:2025gzn} can reconcile these two views. Indeed \cite{Engelhardt:2025vsp} argued that there are asymptotic observables that can detect the presence or absence of a semiclassical compact universe.  In our work we have argued that in some circumstances a space with a horizon can be regarded as superposition of the type 2 shell states discussed in \cite{Antonini:2024mci} and \cite{Engelhardt:2025vsp}.  Our is not in tension with theirs because we are considering a very large quantum superstition of the type 2 states.  There is nothing semiclassical about it.

\subsection{Other possible bases and saturation of the spectral form factor}
By the argument in  Sec.~\ref{sec:tool3}, any set of $\kappa$ states that  satisfies $\overline{\braket{i|j}}\propto \delta_{ij}$ with  $\overline{\braket{i|j}\braket{j|i}}| \neq 0$ for any two states $|i\rangle, |j\rangle$ in the set should form a basis as  $\kappa \to \infty$. The shell states are useful because $\overline{\braket{i|j}\braket{j|i}} \neq 0$ already in the semiclassical saddlepoint approximation, and corresponding wormholes  are easily constructed. A potential alternative basis for  $\mathcal{H}_{LR}$ is the time-evolved TFD states recently discussed in \cite{Banerjee:2024fmh}. This family of states is obtained by evolving the  state $|\beta\rangle$ by a varying amount in Lorentzian boundary time. As discussed in \cite{Magan:2024nkr}, proving that these states span the full 
quantum gravity Hilbert space implies the saturation of the plateau of the gravity spectral form factor. In chaotic quantum systems, time evolved TFD states satisfy the two above properties as shown in \cite{Verlinde:2021kgt}. If this can be shown explicitly for the gravity construction of these states, a corollary is that the spectral form factor for gravity in any dimension has a late time plateau. However, the relevant gravitational wormholes have not been constructed, and may not exist as saddlepoints, because of the absence of matter stress energy supporting the wormhole topology.

\subsection{de Sitter space}
The arguments in this paper relied on cutting open an Euclidean asymptotic spatial boundary for spacetime to define states. However, Euclidean de Sitter space is compact and has no spatial boundary. So the Euclidean de Sitter gravitational path integral computes no-boundary wave functionals $\Psi[h_{ij{}}]$ of the bulk data on a time slice $h_{ij}$.  However, it is not clear that these wavefunctional can be expanded in states with definite geometries because there is no privileged asymptotic observer relative to whose observables such an expansion can be defined. Although Lorentzian de Sitter space does have boundaries at future and past infinity, probes of the bulk geometry sent out by observers located on these boundaries would need to travel backwards in time, so these would not constitute a measurement {\it per se} of the spacetime in the conventional sense. 

One way of achieving greater control is to create a bubble of Euclidean de Sitter  within Euclidean asymptotically anti-de Sitter spacetime. To achieve this, \cite{Balasubramanian:2023xyd} imposed an additional constraint on the Euclidean AdS path integral that the relevant saddlepoints must contain an asymptotic space-like boundary at future infinity after continuation to Lorentzian signature. This procedure can be seen as imposing a mixed Euclidean/Lorentzian boundary condition that requires  the saddle geometries  to be locally dS in the infinite future. This rule could be used in conjunction with the highly simplified AdS path integral setting provided by our toolkit to probe de Sitter quantum gravity.

\acknowledgments 
 TY would like to thank Juan Hernandez, Maria Knysh, Simon Ross and Gabriel Wong for useful discussion. TY would like to Charlie Cummings and Krishnendu Ray for comments on an early version of this draft.  TY is funded by the Peter Davies Scholarship and wishes to thank them for their continued support.   VB is supported in part by the DOE
through DE-SC0013528 and QuantISED grant DE-SC0020360, and in part by the Eastman
Professorship at Balliol College, University of Oxford.
 
\appendix

\section{Appendix: Gram matrix of overlaps} \label{LA}
In this appendix we detail the Gram matrix formalism for obtaining an orthonormal basis from an overcomplete set of states. We consider a $d$ dimensional Hilbert space $\mathcal{H}$ spanned by a set of  $\kappa \geq d$, generally non-orthogonal, vectors $ \mathcal{U} = \{ |u_i\rangle \}$.  The Gram matrix of overlaps
\beq
G_{ij} = \langle u_i | u_j \rangle\,
\eeq
 is Hermitian and can therefore be diagonalized by a unitary matrix $U$,
\beq \label{eq:diagonalGram}
G=\ U D U^{\dagger}
\eeq
where $D$ is a diagonal matrix with real, positive semi-definite eigenvalues $\lambda_i$. The rank of $G$ is given by the number of nonzero eigenvalues of $D$, which must equal the dimension of $d$ of $\mathcal{H}$, and the $\kappa -d$ dimensional kernel of $G$ is the space of null states. Using the decomposition (\ref{eq:diagonalGram}) powers of the Gram matrix can be defined for any real $n$ as
\beq \label{eq:grampowers}
G^n \equiv U D^n U^{\dagger} \, ,
\eeq
 where $D^n \equiv {\rm diag}(\lambda_1^n, \lambda_2^n,\ldots \lambda_d^n, 0,0,\ldots 0)$. For positive non-integer powers $k$ this definition is understood as an analytic continuation of the positive integer $n$ result $G^n$: $G^{k}=\lim_{n \to k} G^n$. For example, the generalized inverse $G^{-1} = U D^{-1} U^\dagger$ is defined by inverting all non-zero eigenvalues and leaving the rest untouched, resulting in $GG^{-1}G\equiv \lim_{n \to -1} GG^{n}G=\lim_{n \to -1} UD^{n+2}U^{\dagger}=G$. As it is intractable to explicitly diagonalize the Gram matrix in the cases of interest in this paper, we write expressions like $G^{-1}$ directly by their analytic continuation $G^{-1} \equiv \lim_{n\to -1} G^{n}$ from $G^{n}$ at positive integer $n$ without reference to $D$.

\paragraph{Orthonormal basis: } This formalism can also be used to obtain an orthonormal basis for $\mathcal{H}$ in terms of the overcomplete set $ \mathcal{U}$. Consider the states $\{|v_a\rangle\}$ defined as
\beq \label{eq:appendixproject}
|v_a\rangle = \sum_{ij}  G^{-{1\over2}}_{ji} U_{ia} |u_j\rangle  = \sum_{j b}  U_{jb}D_{ba}^{- {1\over 2}} |u_j\rangle.
\eeq
The first $d$ states in this set are orthonormal and span $\mathcal{H}$, while the remaining states are null
\begin{equation}
\langle v_a | v_b \rangle =
\sum_{jkcd}D_{ac}^{-{1\over 2}} U^{\dagger}_{cj} \langle u_j| u_k \rangle U_{kd} D_{db}^{-{1\over 2}}
=
\sum_{cd}D_{ac}^{-{1\over 2}} D_{cd} D_{db}^{-{1\over 2}} = \left(D^0\right)_{ab} \, ,
\end{equation}
where by the above defintion $D^0 = {\rm diag}(1,1,\ldots,1_d,0,0,\ldots,0)$. As $\mathcal{H}$ is a Hilbert space the null states must all be the zero vector, hence the states $\{|v_a\rangle\}$ with $a,b,\ldots$ running from $1$ to $d$ provide an orthonormal basis for $\mathcal{H}$.

\paragraph{Identity, inner products and traces:} 
The resolution of the identity in the orthonormal basis is given by $\sum_a|u_a\rangle \langle u_a|$ and can be expressed in terms of the overcomplete basis using (\ref{eq:appendixproject}):

\beq
\mathds{1}_\mathcal{H} = \sum_a|v_a\rangle \langle v_a| = \sum_{abcjk}  |u_j\rangle U_{jb} D^{-{1\over 2}}_{ba} D^{-{1\over 2}}_{ac} U^{\dagger}_{ck}\langle u_k| =\sum_{ij} G^{-1}_{ij} |u_i\rangle \langle u_j|\equiv \lim_{n \to -1}\sum_{ij} G^{n}_{ij} |u_i\rangle \langle u_j|\ .
\eeq
Inserting this identity into the inner product we obtain
\beq
\langle A | B \rangle =\langle A |\mathds{1}_\mathcal{H}  |B \rangle
= \sum_{jk}   
\langle A |u_j\rangle G^{-1}_{jk} \langle u_k| B \rangle \, .
\eeq

We can similarly derive a formula for the trace of an operator $\mathcal{O}$:
\begin{equation}\label{eq:traceformula}
Tr_{\mathcal{H}}({\mathcal{O}}) = 
\sum_a \langle v_a | \mathcal{O} | v_a \rangle =
\sum_{abcjk} D^{-{1\over 2}}_{ab} U^{\dagger}_{bj} \langle u_j| {\cal O}| u_k \rangle U_{kc} D^{-{1\over 2}}_{ca} =
\sum_{jk} G^{-1}_{kj}  \langle u_j | \mathcal{O}  
 | u_k \rangle \, .
\end{equation}

Hence this trace is formally defined by the analytic continuation
\beq
Tr_{\mathcal{H}}({\mathcal{O}}) = \lim_{n \to -1}\sum_{jk} G^{n}_{kj}\langle u_j | \mathcal{O}| u_k \rangle.
\eeq

The trace can be used to extract the dimension of $\mathcal{H}$:

\beq
 d=Tr_{\mathcal{H}}(\mathds{1}_\mathcal{H})=\sum_{jk} G^{-1}_{kj}  \langle u_j | \mathds{1}_\mathcal{H}
 | u_k \rangle = \sum_{jk,lm} G^{-1}_{kj}  G^{-1}_{lm} \langle u_j  |u_l\rangle \langle u_m|u_k\rangle
\\
= \sum_{jk,lm} G^{-1}_{kj}  G^{-1}_{lm} G_{jl}G_{mk} = \sum_{i} ({GG^{-1}GG^{-1}})_{ii}  = \sum_{i} ({GG^{-1}})_{ii},
\eeq

where the sixth equality follows from the definition of the generalized inverse. Hence, the dimension can be extracted directly from the over-complete basis by the analytic continuation:
\beq
d=  \lim_{n\to -1} \sum_{i,j} G^{n}_{ij}G_{ji}.
\eeq

The fact that the identity, trace and inner product (although not matrix elements) can be expressed in terms of the over-complete set $\mathcal{U}$  while being independent of $U$ will be very useful in the gravity constructions considered in this paper, as finding $U$ explicitly is general intractable in this setting.

\bibliographystyle{jhep}
\bibliography{Main_final}
\end{document}